\documentclass[a4paper,noarxiv,twocolumn,accepted=2020-06-24]{quantumarticle}
\usepackage[utf8]{inputenc}
\usepackage[english]{babel}
\usepackage[T1]{fontenc}

\usepackage[numbers,sort&compress]{natbib}
\usepackage{color}
\usepackage{bm}
\usepackage{xcolor}
\usepackage{amsmath}
\usepackage{mathtools}
\usepackage{amssymb}
\usepackage{graphicx}
\usepackage{braket}
\usepackage{psfrag}
\usepackage{tikz}
\usepackage[normalem]{ulem}
\usepackage{qcircuit}
\usepackage{hyperref}

\begin{document}


\title{Option Pricing using Quantum Computers}

\author{Nikitas Stamatopoulos}
\orcid{0000-0002-0906-5340}
\thanks{Current Address: Goldman Sachs \& Co., New York, NY, 10282}
\affiliation{%
Quantitative Research, JPMorgan Chase \& Co., New York, NY, 10017
}%

\author{Daniel J. Egger}
\orcid{0000-0002-5523-9807}
\affiliation{%
IBM Quantum, IBM Research -- Zurich
}%

\author{Yue Sun}
\orcid{0000-0002-0756-164X}
\affiliation{%
Quantitative Research, JPMorgan Chase \& Co., New York, NY, 10017
}%

\author{Christa Zoufal}
\orcid{0000-0003-4126-3141}
\affiliation{%
IBM Quantum, IBM Research -- Zurich
}%
\affiliation{%
ETH Zurich
}%

\author{Raban Iten}
\orcid{0000-0001-5332-5093}
\affiliation{%
IBM Quantum, IBM Research -- Zurich
}%
\affiliation{%
ETH Zurich
}%

\author{Ning Shen}
\orcid{0000-0003-4059-5187}
\affiliation{%
Quantitative Research, JPMorgan Chase \& Co., New York, NY, 10017
}%

\author{Stefan Woerner}
\orcid{0000-0002-5945-4707}
\affiliation{%
IBM Quantum, IBM Research -- Zurich
}%



\begin{abstract}
We present a methodology to price options and portfolios of options on a gate-based quantum computer using amplitude estimation, an algorithm which provides a quadratic speedup compared to classical Monte Carlo methods. The options that we cover include vanilla options, multi-asset options and path-dependent options such as barrier options. We put an emphasis on the implementation of the quantum circuits required to build the input states and operators needed by amplitude estimation to price the different option types. Additionally, we show simulation results to highlight how the circuits that we implement price the different option contracts. Finally, we examine the performance of option pricing circuits on quantum hardware using the IBM Q Tokyo quantum device. We employ a simple, yet effective, error mitigation scheme that allows us to significantly reduce the errors arising from noisy two-qubit gates.
\end{abstract}

\maketitle


\section{\label{sec:introduction} Introduction}

Options are financial derivative contracts that give the buyer the right, but not the obligation, to buy (call option) or sell (put option) an underlying asset at an agreed-upon price (strike) and timeframe (exercise window). In their simplest form, the strike price is a fixed value and the timeframe is a single point in time, but exotic variants may be defined on more than one underlying asset, the strike price can be a function of several market parameters and could allow for multiple exercise dates. As well as providing investors with a vehicle to profit by taking a view on the market or exploit arbitrage opportunities, options are core to various hedging strategies and as such, understanding their properties is a fundamental objective of financial engineering. For an overview of option types, features and uses, we refer the reader to Ref.~\cite{Hull}.

Due to the stochastic nature of the parameters options are defined on, calculating their fair value can be an arduous task and while analytical models exist for the simplest types of options \cite{BlackScholes}, the simplifying assumptions on the market dynamics required for the models to provide closed-form solutions often limit their applicability \cite{Dupire1994}. Hence, more often than not, numerical methods have to be employed for option pricing, with Monte Carlo being one of the most popular due to its flexibility and ability to generically handle stochastic parameters \cite{Boyle1977, Glasserman2003}. However, despite their attractive features in option pricing, classical Monte Carlo methods generally require extensive computational resources to provide accurate option price estimates, particularly for complex options. Because of the widespread use of options in the finance industry, accelerating their convergence can have a significant impact in the operations of a financial institution.

By leveraging the laws of quantum mechanics a quantum computer \cite{Nielsen2010} may provide novel ways to solve computationally intensive problems such as quantum chemistry \cite{Kandala2017, Kandala2019, Moll2018, Ganzhorn2019}, solving linear systems of equations \cite{Harrow2009b}, and machine learning \cite{Lloyd2014, Biamonte2017, Havlicek2019}. Quantitative finance, a field with many computationally hard problems, may benefit from quantum computing. Recently developed applications of gate-based quantum computing for use in finance \cite{Orus2019} include portfolio optimization \cite{Rebentrost2018b}, the calculation of risk measures \cite{Woerner2019} and pricing derivatives \cite{Rebentrost2018, Zoufal2019, Martin2019}. Several of these applications are based on the Amplitude Estimation algorithm \cite{Brassard2000} which can estimate a parameter with a convergence rate of $1/M$, where $M$ is the number of quantum samples used. This represents a theoretical quadratic speed-up compared to classical Monte Carlo methods.

In this paper we extend the pricing methodology presented in \cite{Woerner2019, Rebentrost2018} and place a strong emphasis on the implementation of the algorithms in a gate-based quantum computer. We first classify options according to their features and show how to take the different features into account in a quantum computing setting. In Sec.~\ref{sec:tools}, we review the quantum methodology to price options and discuss how to represent relevant probability distributions in a quantum computer. In Sec.~\ref{sec:option_pricing}, we show a framework to price vanilla options and portfolios of vanilla options, options with path-dependent dynamics and options on several underlying assets. In Sec.~\ref{sec:hardware} we show results from evaluating our option circuits on quantum hardware, and describe the error mitigation scheme we employ to increase the accuracy of the estimated option prices. In particular, we employ the maximum likelihood estimation method introduced in \cite{Suzuki2020} to perform amplitude estimation without phase estimation in option pricing using three qubits of a real quantum device.

\section{Review of option types and their challenges}

Option contracts are valid for a pre-determined period of time, and their value at the expiration date is called the \emph{payoff}. 
The goal of option pricing is to estimate the option payoff at the expiration date in the future and then \emph{discount} that value to determine its worth today. 
The discounted payoff is also called the \emph{fair value} and indicates the amount of money one should pay to enter the option contract today, making it worthwile receiving the payoff value at the expiration date.

In practice, we price complex options numerically using Monte Carlo methods by following these steps:

\begin{enumerate}
 \item Model the asset price of the option's underlying(s) and any other sources of uncertainty as random variables $\mathbf{X} = \{X_1, X_2, \dots, X_N\}$ following a stochastic process.
 \item Generate a large number $M$ of random price paths $\{\mathbf{X}_1, \mathbf{X}_2, \dots, \mathbf{X}_M\}$ for the underlying(s) from the probability distribution $\mathbb{P}$ implied by the stochastic process.
 \item Calculate the option's payoff $f(\mathbf{X}_i)$ on each generated price path and compute an estimator for the expectation value of the payoff $\mathbb{E}_{\mathbb{P}}[f(\mathbf{X})]$ as an average across all paths
   \begin{equation*}
    \hat{\mathbb{E}}_{\mathbb{P}}[f(\mathbf{X})] = \frac{1}{M} \sum_{i=1}^M{f(\mathbf{X}_i)}
   \end{equation*}
By the Central Limit Theorem, the estimator $\hat{\mathbb{E}}_{\mathbb{P}}$ converges to the expectation value $\mathbb{E}_{\mathbb{P}}$ as the number of paths goes to infinity, with convergence  $\mathcal{O}\left(M^{-1/2}\right)$ \cite{Rubinstein1981}.
 \item Discount the calculated expectation value to get the option's fair value.
\end{enumerate}

The discounting process requires knowledge of interest rates at future dates which is itself an important question from a financial modelling perspective.
However, for the types of options we consider, this process is not computationally challenging and can be performed classically after the payoff calculation. 
We therefore do not discount the expected payoff for simplicity.

We classify options according to two categories: path-independent vs path-dependent and options on a single asset or on multiple assets. Path-independent options have a payoff function that depends on an underlying asset at a single point in time. Therefore, the price of the asset up to the exercise date of the option is irrelevant for the option price. By contrast, the payoff of path-dependent options depends on the evolution of the price of the asset and its history up to the exercise date. Table~\ref{Tab:option_types} exemplifies this classification. Options that are path-independent and rely on a single asset are the easiest to price, and in most cases numerical calculation is straightforward and would likely not benefit by the use of a quantum computer. Path-independent options on multiple assets are only slightly harder to price since more than one asset is now involved and the probability distributions must account for correlations between the assets, but usually these can be priced quite efficiently on classical computers as well. Path-dependent options on the other hand are significantly harder to price than path-independent options since they require an often expensive payoff calculation at multiple time points on each path, therefore minimizing the number of paths required for this step would lead to a significant benefit in the pricing process. It is this last case where we envision the largest impact of quantum computing.

\begin{table*}[t]
 \begin{center}
  \caption{\label{Tab:option_types} Example of the different option types.}
  \begin{tabular}{l c c} \hline \hline
                    & Single-asset & Multi-asset \\ \hline
   Path-independent & European put/call & Basket option \\
   Path-dependent   & Barrier \& Asian options & Multi-asset \\
                    &                          & barrier options \\ \hline \hline
  \end{tabular}
 \end{center}
\end{table*}

\section{Quantum Methodology \label{sec:tools}}

Here we outline the building blocks needed to price options on a gate-based quantum computer. As discussed in the previous section, the critical components are 1) represent the probability distribution $\mathbb{P}$ describing the evolution of random variables $\mathbf{X} = \{X_1, X_2, \dots, X_N\}$ on the quantum computer, 2) construct the circuit which computes the payoff $f(\mathbf{X})$ and 3) calculate the expectation value of the payoff $\mathbb{E}_{\mathbb{P}}[f(\mathbf{X})]$. 
In Sec.~\ref{sec:ae} we show how to use Amplitude Estimation to calculate the expectation value of a function of random variables.
In Sec.~\ref{sec:dist_loading} we describe the process of loading the relevant probability distributions to a quantum register, and in Sec.~\ref{sec:payoff} we construct the circuits to compute the payoff and set up Amplitude Estimation to estimate the expectation value of the payoff.
We then have all the ingredients to price options on a quantum computer.

\subsection{Amplitude Estimation \label{sec:ae}}

\begin{figure}
\centering
\includegraphics[width=0.45\textwidth]{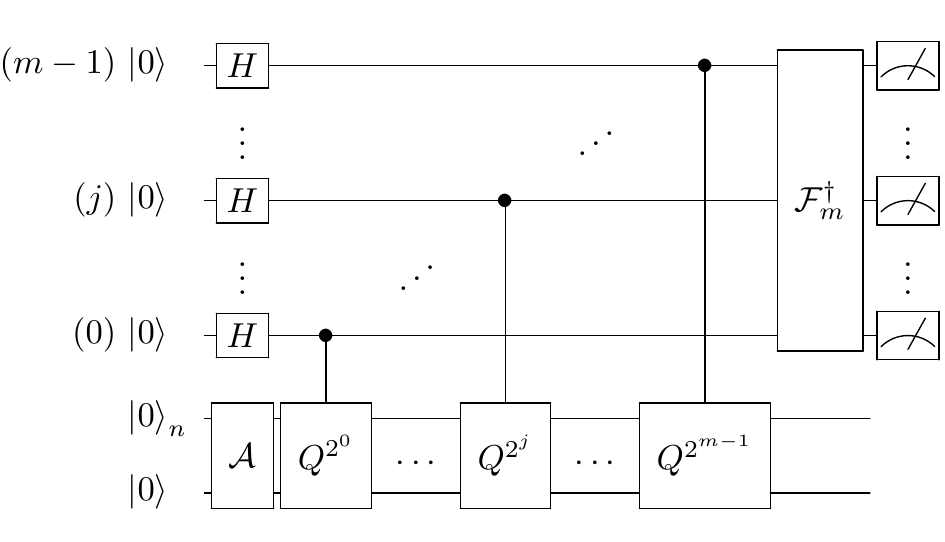}
\caption{\label{fig:ae} The quantum circuit of amplitude estimation, where $H$ denotes a Hadamard gate and $\mathcal{F}^{\dagger}$ the inverse QFT.}
\end{figure}

The advantage of pricing options on a quantum computer comes from the Amplitude Estimation (AE) algorithm \cite{Brassard2000} which provides a quadratic speed-up over classical Monte Carlo simulations \cite{Abrams1999, Montanaro2017}.
Suppose a unitary operator $\mathcal{A}$ acting on a register of $(n+1)$ qubits such that 
\begin{align} \label{eqn:psiAE}
\mathcal{A} \ket{0}_{n+1} = \sqrt{1 - a}\ket{\psi_0}_n\ket{0} + \sqrt{a}\ket{\psi_1}_n\ket{1}
\end{align} 
for some normalized states $\ket{\psi_0}_n$ and $\ket{\psi_1}_n$, where $a \in [0, 1]$ is unknown.
AE allows the efficient estimation of $a$, i.e., the probability of measuring $\ket{1}$ in the last qubit.
This estimation is obtained with an operator $\mathcal{Q} = \mathcal{A}\mathcal{S}_0\mathcal{A}^\dagger\mathcal{S}_{\psi_0}$, where $\mathcal{S}_0 = 1 - 2\ket{0}\bra{0}$ and $\mathcal{S}_{\psi_0} = 1 - 2\ket{\psi_0}\ket{0}\bra{\psi_0}\bra{0}$, which is a rotation of angle $2\theta_a$ in the two-dimensional space spanned by $\ket{\psi_0}_n\ket{0}$ and $\ket{\psi_1}_n\ket{1}$.
From Eq.~(\ref{eqn:psiAE}) we know that $a = \sin^2(\theta_a)$.
To obtain an approximation for $\theta_a$, AE applies Quantum Phase Estimation \cite{Kitaev1995,Cleve1998} to approximate certain eigenvalues of $\mathcal{Q}$, which are classically mapped to an estimator for $a$.
The Quantum Phase Estimation uses $m$ additional sampling qubits to represent result and $M = 2^m$ applications of $\mathcal{Q}$, i.e., $M$ quantum samples.
The $m$ qubits, initialized to an equal superposition state by Hadamard gates, are used to control different powers of $\mathcal{Q}$.
After applying an inverse Quantum Fourier Transform (QFT), their state is measured resulting in an integer $y \in \{0, ..., M-1\}$, which is classically mapped to the estimator for $a$, i.e.
\begin{equation}
\label{eqn:estimator}
\tilde{a} = \sin^2(y\pi/M) \in [0, 1].
\end{equation}
The full circuit for AE is shown in Fig.~\ref{fig:ae}.
The estimator $\tilde{a}$ satisfies
\begin{eqnarray}
\label{eqn:estimation_error}
| a - \tilde{a} |\leq& \frac{\pi}{M} + \frac{\pi^2}{M^2} \;=\; \mathcal{O}\left(M^{-1}\right),
\end{eqnarray}
with probability of at least $8/\pi^2$.
This represents a quadratic speedup compared to the $\mathcal{O}\left(M^{-1/2}\right)$ convergence rate of classical Monte Carlo methods \cite{Glasserman2000}.

To reduce the required number of qubits and the resulting circuit depth, Suzuki \emph{et al.} have shown that AE can be performed without requiring quantum phase estimation while still maintaining a quadratic speed-up \cite{Suzuki2020}.
To this extent, they exploit that
\begin{eqnarray}
\label{eqn:QkA}
\mathcal{Q}^k \mathcal{A}\ket{0}_n\ket{0} &=&
\cos \left( (2k+1)\theta_a \right)\ket{\psi_0}_n\ket{0} + \notag \\
&&\sin \left( (2k+1)\theta_a \right)\ket{\psi_1}_n\ket{1},
\end{eqnarray}
and by measuring $\mathcal{Q}^k\mathcal{A}\ket{0}$ for $k=2^0, ..., 2^{m-1}$ for a given $m$ and applying a maximum likelihood estimation, an approximation for $\theta_a$ (and hence $a$) can be recovered.
If we define $M = 2^m-1$, i.e. the total number of $\mathcal{Q}$-applications, and we consider $N$ shots for each experiment, it has been shown empirically that the resulting estimation error scales as $\mathcal{O}(1/(M\sqrt{N}))$, i.e., the algorithm achieves the quadratic speed-up in terms of $M$.
We will use this approach to demontrate results from real quantum hardware in Sec.~\ref{sec:hardware}.

For the option contracts we consider, the random variables involved represent the possible values $S_i$ the underlying asset can take, and the corresponding probabilities $p_i$ that those values will be realized.
For an option with payoff $f$, the $\mathcal{A}$ operator will create the state

\begin{equation}
\label{eqn:explicit_option_A}
\sum_{i=0}^{2^n-1}\sqrt{1-f(S_i)} \sqrt{p_i}\ket{S_i}\ket{0} + \sum_{i=0}^{2^n-1}\sqrt{f(S_i)} \sqrt{p_i}\ket{S_i}\ket{1}.
\end{equation}
Comparing Eq.~(\ref{eqn:psiAE}) and Eq.~(\ref{eqn:explicit_option_A}), we can see that 

\begin{equation}
\label{eqn:expectation_value}
a = \sum_{i=0}^{2^n-1}f(S_i) p_i = \mathbb{E}[f(S)],
\end{equation}
meaning AE allows us to compute the undiscounted price of an option given a way to represent the option's payoff as a quantum circuit and  create the state of Eq.~(\ref{eqn:explicit_option_A}). 
In the following sections, we describe the necessary components to achieve that.

\subsection{Distribution loading \label{sec:dist_loading}}
The first component of our option pricing model is the circuit that takes a probability distribution implied for possible asset prices in the future and loads it into a quantum register such that each basis state represents a possible value and its amplitude the corresponding probability. In other words, given an $n$-qubit register, asset prices $\{S_i\}$ for $i \in \{0, ..., 2^n-1\}$ and corresponding probabilities $\{p_i\}$, the distribution loading module creates the state:

\begin{equation}
\label{eqn:prob_distribution}
\ket{\psi}_n = \sum\limits_{i=0}^{2^n-1}\sqrt{p_i}\ket{S_i}_n.
\end{equation} 

The analytical formulas used to price options in the Black-Scholes-Merton (BSM) model \cite{BlackScholes, Merton1973} assume that the underlying stock prices at maturity follow a log-normal distribution with constant volatility. 
In \cite{Grover2002}, the authors show that log-concave probability distributions (such as the log-normal distribution of the BSM model) can be efficiently loaded in a gate-based quantum computer. 
The option types considered in this paper are modeled using the underlying BSM dynamics and thus loading the relevant probability distributions onto quantum registers does not require prohibitive complexity. 

However, in option models of practical interest, the simplified assumptions in the BSM model fail to capture important market dynamics, limiting the model's applicability in real-life scenarios.
As such, the market-implied probability distribution of the underlying needs to be captured properly in order for valuation models to accurately estimate the intrinsic value of option contracts.
To address these shortcomings, Artificial Neural Networks (ANN) are becoming increasingly more popular as a means to capture the real-life dynamics of the relevant market parameters, without the need to assume a simplified underlying model \cite{Koshiyama2019, Horvath2019}.
It is thus important to be able to efficiently represent distributions of financial parameters on a quantum computer which might not have explicit analytical representations.

The loading of arbitrary states into quantum systems requires exponentially many gates \cite{Plesch2010}, making it inefficient to model arbitrary distributions as quantum gates. 
Since the distributions of interest are often of a special form, the limitation may be overcome by using quantum Generative Adverserial Networks (qGAN).
These networks allow us to load a distribution using a polynomial number of gates \cite{Zoufal2019}. 
A qGAN can learn the random distribution $X$ underlying the observed data samples $\set{x^0, \ldots, x^{k-1}}$ and load it directly into a quantum state. 
This generative model employs the interplay of a classical discriminator, a neural network \cite{Goodfellow2014}, and a quantum generator (a parametrized quantum circuit). 
More specifically, the qGAN training consists of alternating optimization steps of the discriminator's parameters $\phi$ and the generator's parameters $\theta$. 
After the training, the output of the generator is a quantum state
\begin{equation}
\ket{\psi(\theta)}_n = \sum\limits_{i=0}^{2^n-1}\sqrt{p_i(\theta)}\ket{i}_n,
\end{equation}
that represents the target distribution.
The $n$-qubit state $\ket{i}_n=\ket{i_{n-1}...i_{0}}$ encodes the integer $i=2^{n-1}i_{n-1}+...+2i_1+i_0\in\{0,...,2^n-1\}$ with $i_k\in\{0,1\}$ and $k=0,...,n-1$.
The probabilities $p_i(\theta)$ approximate the random distribution underlying the training data.
We note that the outcomes of a random variable $X$ can be mapped to the integer set $\{0,...,2^n-1\}$ using an affine mapping.
Furthermore, the approach can be easily extended to multivariate data, where we use a separate register of qubits for each dimension \cite{Zoufal2019}.

Another useful feature in the use of qGANs for loading probability distributions, is the fact that we can tailor the qGAN variational form to construct short-depth circuits for an acceptable degree of accuracy. 
This in turn allows us to evaluate the performance of option pricing quantum circuits in near-term quantum hardware where resources are still quite limited.

The use of ANNs to represent probability distributions inevitably imposes a cost associated with the training component, in both classical and quantum models.
However, during common business practices such as overnight risk assessment of large portfolios which may consist of millions of option contracts, the same probability distributions can be used across a large number of pricing calls defined on the same underlyings.
For example, it is quite common during risk assessment valuations to require pricing of several thousands of option contracts defined on the same underlying.
As such, the effective training cost per pricing call can be efficiently amortized to represent only a small fraction of the total individual option pricing cost.

It is also noteworthy that the qGAN training performs better if the initial distribution is close to the target distribution \cite{Zoufal2019}. Therefore, as new market data comes in which needs to be incorporated into the probability distribution (e.g. for overnight risk, a lot of the same options as the day before need to be priced, but there is one more day's worth of market data) the previously trained qGAN can be used as the initial distribution, leading to faster convergence.

\subsection{Computing the payoff \label{sec:payoff}}

We obtain the expectation value of a linear function $f$ of a random variable $X$ with AE by creating the operator $\mathcal{A}$ such that $a=\mathbb{E}[f(X)]$, see Eq.~(\ref{eqn:expectation_value}).
Once $\mathcal{A}$ is implemented we can prepare the state in Eq.~(\ref{eqn:psiAE}), and the $\mathcal{Q}$ operator, which only requires $\mathcal{A}$ and generic quantum operations \cite{Kitaev1995,Cleve1998}.
In this section, we show how to create a close relative of $\mathcal{A}$ and how to combine it with AE to calculate the expectation value of $f$.

\begin{figure}
\centering
\includegraphics[width=0.48\textwidth, clip, trim=25 37 0 10]{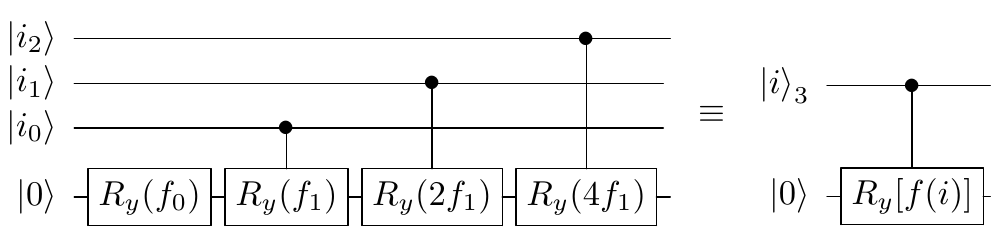}
\caption{\label{fig:controlledY} Quantum circuit that creates the state in Eq.~(\ref{eqn:yrotstate}). Here, the independent variable $i=4i_2+2i_1+i_0\in\{0,...,7\}$ is encoded by three qubits in the state $\ket{i}_3=\ket{i_2i_1i_0}$ with $i_k\in\{0,1\}$. Therefore, the linear function $f(i)=f_1i+f_0$ is given by $4f_1i_2+2f_1i_1+f_1i_0+f_0$. After applying this circuit the quantum state is $\ket{i}_3[\cos(f_1 i+f_0)\ket{0}+\sin(f_1 i+f_0)\ket{1}]$. The circuit on the right shows an abbreviated notation.}
\end{figure}

The payoff function for the option contracts of interest is piece-wise linear and as such we only need to consider linear functions $f:\{0,...,2^n-1\}\to[0,1]$ which we write $f(i)=f_1i+f_0$.
We can efficiently create an operator that performs
\begin{align} \label{eqn:yrotstate}
\ket{i}_n\ket{0} \to \ket{i}_n 
\left(
\cos\left[f(i)\right]\ket{0}+\sin\left[f(i)\right]\ket{1}
\right)
\end{align}
using controlled Y-rotations \cite{Woerner2019}.  
To implement the linear term of $f(i)$ each qubit $j$ (where $j\in \{0, \dotsc n-1\}$) in the $\ket{i}_n$ register acts as a control for a Y-rotation with angle $2^jf_1$ of the ancilla qubit. The constant term $f_0$ is implemented by a rotation of the ancilla qubit without any controls, see Fig.~\ref{fig:controlledY}. The controlled Y-rotations can be implemented with CNOT and single-qubit gates \cite{Barenco1995}.

We now describe how to obtain $\mathbb{E}[f(X)]$ for a linear function $f$ of a random variable $X$ which is mapped to integer values $i\in\{0,...,2^n-1\}$ that occur with probability $p_i$ respectively.
To do this we use the procedure outlined immediately above to create the operator that maps $\sum_i\sqrt{p_i}\ket{i}_n\ket{0}$ to
\begin{align}
\label{eqn:yrotstate_scaled}
\sum_{i=0}^{2^n-1}\sqrt{p_i}\ket{i}_n\left[\cos\left(c\tilde f(i)+\frac{\pi}{4}\right)\ket{0}+\sin\left(c\tilde f(i)+\frac{\pi}{4}\right)\ket{1}\right],
\end{align}
where $\tilde f(i)$ is a scaled version of $f(i)$ given by
\begin{align}\label{eqn:scaling}
\tilde f(i)=2\frac{f(i)-f_\text{min}}{f_\text{max}-f_\text{min}}-1,
\end{align}
with $f_\text{min}=\min_if(i)$ and $f_\text{max}=\max_if(i)$,
and $c\in[0,1]$ is an additional scaling parameter.
The relation in Eq.~(\ref{eqn:scaling}) is chosen so that $\tilde f(i)\in[-1,1]$.
Consequently, $\sin^2[c\tilde f(i)+\pi/4]$ is an anti-symmetric function around $\tilde f(i)=0$.
With these definitions, the probability to find the ancilla qubit in state $\ket{1}$, namely
\begin{align}\notag
P_1=\sum_{i=0}^{2^n-1}p_i\sin^2\left(c\tilde f(i)+\frac{\pi}{4}\right),
\end{align}
is well approximated by
\begin{align}\label{eqn:approxP1}
P_1 \approx \sum_{i=0}^{2^n-1} p_i \left(c\tilde f (i) + \frac{1}{2}\right)
= c\frac{2\mathbb{E}[f(X)]-f_\text{min}}{f_\text{max}-f_\text{min}}-c+\frac{1}{2}.
\end{align}
To obtain this result we made use of the approximation
\begin{align}\label{eqn:approx}
\sin^2\left(c\tilde f(i)+\frac{\pi}{4}\right)=c\tilde f(i)+\frac{1}{2}+\mathcal{O}(c^3\tilde f^3(i))
\end{align}
which is valid for small values of $c\tilde f(i)$.
With this first order approximation the convergence rate of AE is $\mathcal{O}(M^{-2/3})$ when $c$ is properly chosen which is already faster than classical Monte Carlo methods \cite{Woerner2019}.
We can recover the $\mathcal{O}(M^{-1})$ convergence rate of AE by using higher orders implemented with quantum arithmetic. The resulting circuits, however, have more gates.
This trade-off, discussed in Ref.~\cite{Woerner2019}, also gives a formula that specifies which value of $c$ to use to minimize the estimation error made when using AE.
From Eq.~(\ref{eqn:approxP1}) we can recover $\mathbb{E}[f(X)]$ since AE allows us to efficiently retrieve $P_1$ and because we know the values of $f_\text{min}$, $f_\text{max}$ and $c$.

\section{Option pricing on a quantum computer \label{sec:option_pricing}}

In this section we show how to price the different options shown in Tab.~\ref{Tab:option_types}.
We put an emphasis on the implementation of the quantum circuits that prepare the states needed by AE.
We use the different building blocks reviewed in Sec.~\ref{sec:tools}.

\subsection{Path-independent options}
\label{sec:path_independent_options}

\begin{figure}[h!]
\centering
\includegraphics[width=0.48\textwidth]{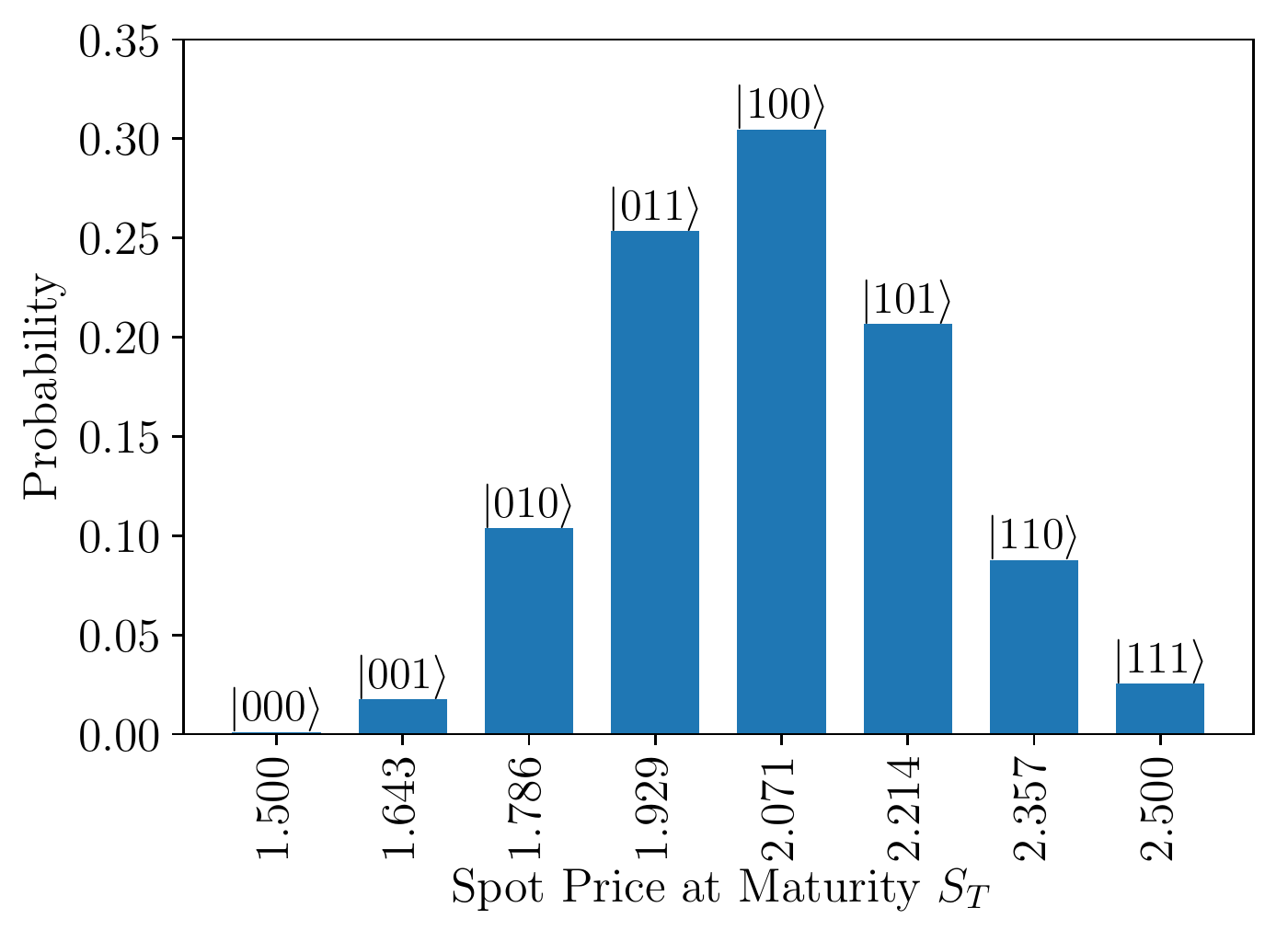}
\caption{\label{fig:distribution}Example price distribution at maturity loaded in a three-qubit register. In this example we followed the Black-Scholes-Merton model which implies a log-normal distribution of the asset price at maturity $T$ with probability density function $P(S_T)=\frac{1}{S_T\sigma \sqrt{2\pi T}} \text{exp}\left(-\frac{(\ln{S_T} -\mu)^2}{2\sigma^2T}\right)$. $\sigma$ is the volatility of the asset and $\mu=\left(r-0.5\sigma^2\right)T + \ln(S_0)$, with $r$ the risk-free market rate and $S_0$ the asset's spot at $t=0$. In this figure we used $S_0=2$, $\sigma=10\%$, $r=4\%$ and $T=300/365$.}
\end{figure}

The price of path-independent \emph{vanilla} options (e.g. European call and put options) depends only on the distribution of the underlying asset price $S_T$ at the option maturity $T$ and the payoff function $f(S_T)$ of the option.
To encode the distribution of $S_T$ in a quantum state, we truncate it to the range $[S_{T,\text{min}}, S_{T,\text{max}}]$ and discretize this interval to $\{0,..., 2^n-1\}$ using $n$ qubits.
In the quantum computer, the distribution loading operator $\mathcal{P}_X$ creates a state
\begin{align} \label{eqn:prob_state}
\ket{0}_n \xrightarrow{\mathcal{P}_X} \ket{\psi}_n=\sum\limits_{i=0}^{2^n-1}\sqrt{p_i}\ket{i}_n,
\end{align}
with $i\in\{0,...,2^n-1\}$ to represent $S_T$.
This state, exemplified in Fig.~\ref{fig:distribution}, may be created using the methods discussed in Sec.~\ref{sec:dist_loading}.

We start by showing how to price vanilla call or put options, and then generalize our method to capture the payoff structure of portfolios containing more than one vanilla option.

\subsubsection{Vanilla options \label{sec:vanilla}}
Vanilla call options are structured so that if the underlying asset price is larger than a fixed value $K$ (the strike price) at the expiration date, the contract pays the difference between the realized price and the strike. As such, the call option payoff $f_C(S_T)$ can be written as

\begin{equation}
\label{eqn:call_payoff}
f_C(S_T) = \max(0, S_T-K).
\end{equation}
Equivalently, the corresponding put option has a similar payoff but it pays if the asset price at expiry is smaller than the strike. That is, the put option payoff is

\begin{equation}
\label{eqn:put_payoff}
f_P(S_T) = \max(0, K-S_T).
\end{equation}
The linear part of the payoff can be computed as a quantum circuit with the approach outlined in Sec.~\ref{sec:payoff}, but we need a way to express the $\max$ operation as a quantum circuit as well.
We show how to achieve that for both call and put option types by implementing a comparison between the values encoded in the basis states of Eq.~(\ref{eqn:prob_state}) and $K$.

\begin{figure*}
\centering
\includegraphics[width=0.7\textwidth]{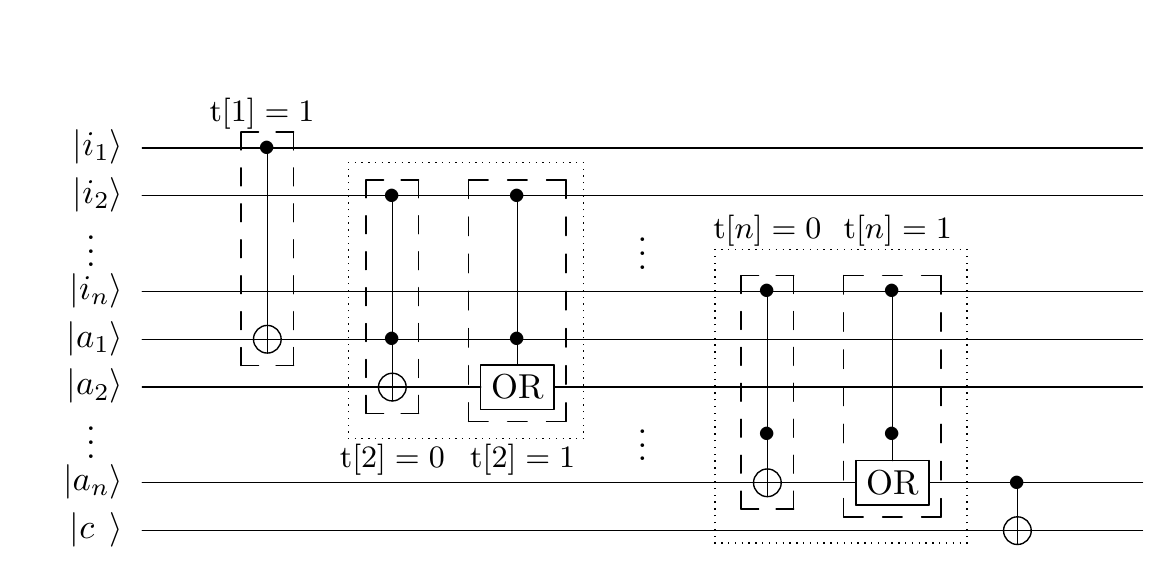}
\caption{\label{fig:comparator_circ_general} Circuit that compares the value represented by an $n$-qubit register $\ket{i}_n$, to a fixed value $K$. We use $n$ ancilla qubits $\ket{a_1,...,a_n}$, a classical array $t[n]$ holding the precomputed binary value of $K$'s two's complement and a qubit $\ket{c}$ which will hold the result of the comparison with $\ket{c}=1$ if $\ket{i} \geq K$. For each qubit $\ket{i_k}$, with $k \in \{1,..., n\}$, we use a Toffoli gate to compute the carry at position $k$ if $t[k]=1$ and a logical OR, see Fig.~\ref{fig:logical_or}, if $t[k]=0$. For $k=1$, we only need to use a CNOT on $\ket{i_1}$ if $t[1]=1$. In the circuit above, only one of two unitaries in a dotted box needs to be added to the circuit, depending on the value of $t[k]$ at each qubit. The last carry qubit $\ket{a_n}$ is then used to compute the final result of the comparison in qubit $\ket{c}$.  }
\end{figure*}

\begin{figure}
\centering
\includegraphics[width=0.48\textwidth]{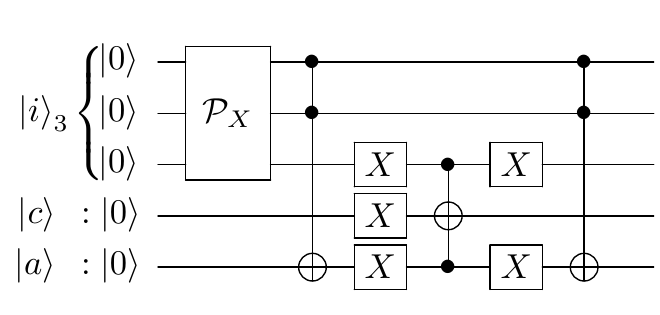}
\caption{\label{fig:comparator_circ} Quantum circuit that sets a comparator qubit $\ket{c}$ to $\ket{1}$ if the value represented by $\ket{i}_3$ is larger than a strike $K=1.9$, for the spot distribution in Fig.~\ref{fig:distribution}. The unitary $\mathcal{P}_X$ represents the set of gates that load the probability distribution in Eq.~(\ref{eqn:prob_state}). An ancilla qubit $\ket{a}$ is needed to perform the comparison. It is uncomputed at the end of the circuit.}
\end{figure}

A quantum comparator circuit sets an ancilla qubit $\ket{c}$, initially in state $\ket{0}$, to the state $\ket{1}$ if $i \geq K$ and $\ket{0}$ otherwise.
The state $\ket{\psi}_n$ in the quantum computer therefore undergoes the transformation
\begin{align}
\notag
\ket{\psi}_n\ket{0}\to\ket{\phi_1}=\sum\limits_{i < K}\sqrt{p_i}\ket{i}_n\ket{0}+\sum\limits_{i \geq K}\sqrt{p_i}\ket{i}_n\ket{1}.
\end{align}
This operation can be implemented by a quantum comparator \cite{Cuccaro2004} based on CNOT and Toffoli gates.
Since we know the value of the strike, we can implement a circuit tailored to the specific strike price. We use $n$ ancilla qubits $\ket{a_1,...,a_n}$ and compute the two's complement of the strike price $K$ in binary using $n$ bits, storing the digits in a (classical) array $t[n]$. For each qubit $\ket{i_k}$ in the $\ket{i}_n$ register, with $k \in \{0,..., n-1\}$, we compute the possible carry bit of the bitwise addition of $\ket{i_k}$ and $t[k]$ into $\ket{a_k}$. If $t[k]=0$, there is a carry qubit at position $k$ only if there is a carry at position $k-1$ \emph{and} $\ket{i_k}=1$. If $t[k]=1$, there is a carry qubit at position $k$ if there is a carry at position $k-1$ \emph{or} $\ket{i_k}=1$.  After going through all $n$ qubits from least to most significant, $\ket{i}_n$ will be greater or equal to the strike price, only if there is a carry at the last (most significant) qubit. This procedure along with the necessary gate operations is illustrated in  Fig.~\ref{fig:comparator_circ_general}.
An implementation for $K=1.9$ and a three-qubit register is shown in Fig.~\ref{fig:comparator_circ}.

\begin{figure}
\centering
\includegraphics[width=0.48\textwidth]{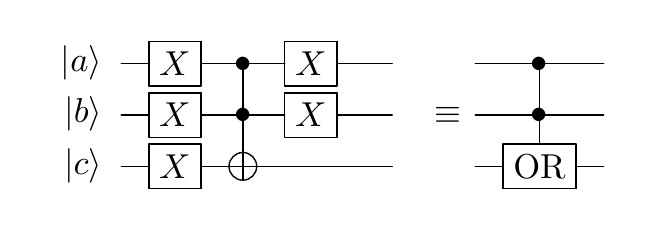}
\caption{\label{fig:logical_or} Circuit that computes the logical OR between qubits $\ket{a}$ and $\ket{b}$ into qubit $\ket{c}$. The circuit on the right shows the abbreviated notation used in Fig.~\ref{fig:comparator_circ_general}.}
\end{figure}

\begin{figure}
\centering
\includegraphics{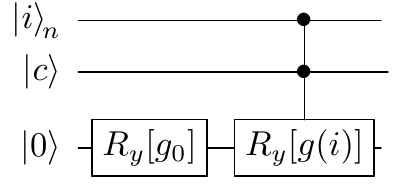}
\caption{\label{fig:phi2} Circuit that creates the state in Eq.~(\ref{eqn:aeState}). We apply this circuit directly after the comparator circuit shown in Fig.~\ref{fig:comparator_circ}. The multi-controlled $y$-rotation is the gate shown in Fig.~\ref{fig:controlledY} controlled by the ancilla qubit $\ket{c}$ that contains the result of the comparison between $i$ and $K$.}
\end{figure}

\begin{figure}
\centering
\includegraphics[width=0.48\textwidth, height=0.48\textheight]{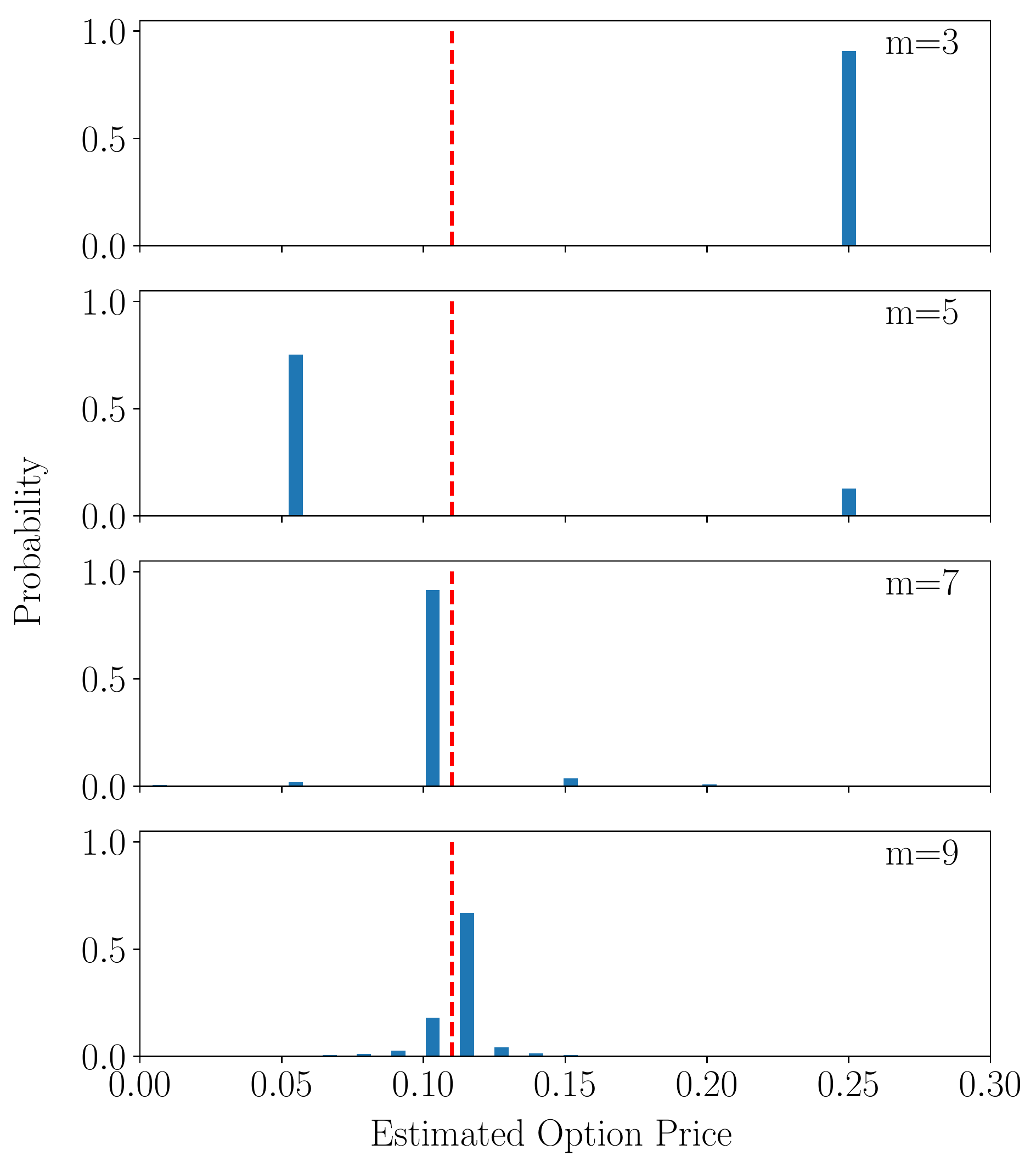}
\caption{\label{fig:european_call_ae} Results from applying amplitude estimation (Sec.~\ref{sec:ae}) on a European call option with spot price distribution as given in Fig.~\ref{fig:distribution} and a strike price $K=2.0$, on a simulated quantum device with $m \in \{3,5,7,9\}$ sampling qubits, i.e., $M \in \{8, 32, 128, 512\}$ quantum samples. The red dashed line corresponds to the (undiscounted) analytical value for this option, calculated using the Black-Scholes-Merton model. We limit the range of possible option values shown to $[0, 0.3]$ to illustrate the convergence of the estimation, as the cumulative probability in the windows shown exceeds $90\%$ in each case.}
\end{figure}

To represent the payoff function $f(i)$, we add to $\ket{\phi_1}$ a second ancilla qubit, which corresponds to the last qubit in Eq.~(\ref{eqn:yrotstate_scaled}).
The payoff function of vanilla options is piece-wise linear
\begin{align}
\label{eqn:general_payoff}
f(i)=
\begin{cases}
a_<\cdot i+b_< & i<K, \\
a_\geq\cdot i+b_\geq & i\geq K.
\end{cases}
\end{align}
We now focus on a European call option with payoff $f(i) = \text{max}(0, i-K)$, i.e., $a_<=b_<=0$, $a_\geq=1$, and $b_\geq=-K$.
To prepare the operator that calculates the payoff in the form of Eq.~(\ref{eqn:yrotstate_scaled}) for use with AE, we set
\begin{align}
\notag
c \tilde f(i) + \frac{\pi}{4} =
\begin{cases}
g_0 & i<K, \\
g_0+g(i) & i\geq K,
\end{cases}
\end{align}
where $g(i)$ is a linear function of $i$, and $g_0$ is an angle whose value we will carefully select.
With this setup, the payoff in Eq.~(\ref{eqn:yrotstate_scaled}) can be constructed, starting from the state $\ket{\phi_1}\ket{0}$, by first initializing the last ancilla qubit to the state $\cos(g_0)\ket{0}+\sin(g_0)\ket{1}$, and then performing a rotation of the last ancilla qubit controlled by the comparator qubit $\ket{c}$ and the qubits in $\ket{\psi}_n$.
This rotation operation, implemented by the quantum circuit in Fig.~\ref{fig:phi2}, applies a rotation with an angle $g(i)$ only if $i\geq K$.
The state of the $n+2$ qubits after this operation becomes
\begin{align}\label{eqn:aeState} 
&\sum_{i<K}\sqrt{p_i}\ket{i}_n\ket{0}\left[\cos(g_0)\ket{0}+\sin(g_0)\ket{1}\right] + \\
&\sum_{i\geq K}\sqrt{p_i}\ket{i}_n\ket{1}\left\{\cos[g_0+g(i)]\ket{0}+\sin[g_0+g(i)]\ket{1}\right\}. \notag
\end{align}
The probability to find the second ancilla in state $\ket{1}$, efficiently measurable using AE, is
\begin{align} \label{eqn:PsingleStrike}
P_1=\sum_{i< K}p_i\sin^2(g_0)+\sum_{i\geq K}p_i\sin^2[g_0+g(i)].
\end{align}
Now, we must carefully choose the angle $g_0$ and the function $g(i)$ to recover the expected payoff $\mathbb{E}[f(X)]$ of the option from $P_1$ using the approximation in Eq.~(\ref{eqn:approxP1}).
To reproduce $f(i)=i-K$ for $i\geq K$ and simultaneously satisfy $c \tilde f(i) = g_0+g(i)-\pi/4 \in [-c,c]$ (see Eq.~(\ref{eqn:scaling})), we must set
\begin{align} \label{eqn:gofi}
g(i)=\frac{2c(i-K)}{i_\text{max}-K},
\end{align}
where $i_\text{max}=2^n-1$. This choice of $g(i)$ forces us to choose
\begin{align} \label{eqn:g0}
g_0=\frac{\pi}{4}-c.
\end{align}
To see why, we substitute Eqs.~(\ref{eqn:gofi}) and (\ref{eqn:g0}) into Eq.~(\ref{eqn:PsingleStrike}) and use the approximation in Eq.~(\ref{eqn:approx}), which leads to
\begin{align} \label{eqn:option_scaling} \notag
P_1 &\approx \sum_{i<K}p_i\left(\frac{1}{2}-c\right)+\sum_{i\geq K}p_i\left(\frac{2c(i-K)}{i_\text{max}-K}+\frac{1}{2}-c\right) \\
&= \frac{1}{2}-c+\frac{2c}{i_\text{max}-K}\sum_{i\geq K}p_i(i-K),
\end{align}
where we have used $\sum_ip_i=1$ in the last equality.
Eq.~(\ref{eqn:option_scaling}) shows that by setting $g_0=\pi/4-c$, we could recover $\mathbb{E}[\max(0,i-K)]$ from $P_1$ up to a scaling factor and a constant,
from which we can subsequently recover the expected payoff $\mathbb{E}[f(i)]$ of the option given the probability distribution of the underlying asset.
We should note that the fair value of the option requires appropriately discounting the expected payoff of the option to today, but as the discounting can be performed after the expectation value has been calculated, we omit it from our discussion for simplicity.
We demonstrate the performance of our approach by running amplitude estimation using Qiskit \cite{Qiskit} on the overall circuit produced by the elements described in this section, and verifying the convergence to the analytically computed value or classical Monte Carlo estimate. An illustration of the convergence of a European call option with increasing evaluation qubits is shown in Fig.~\ref{fig:european_call_ae}.

A straightforward extension of the analysis above yields a pricing model for a European put option, whose payoff $f(i) = \text{max}(0, K-i)$ is equivalent to Eq.~(\ref{eqn:general_payoff}) with $a_>=b_>=0$, $a_\leq=-1$, and $b_\leq=K$.

\subsubsection{Portfolios of options}
Various popular trading and hedging strategies rely on entering multiple option contracts at the same time instead of individual call or put options and as such, these strategies allow an investor to effectively construct a payoff that is more complex than that of vanilla options. For example, an investor who wants to profit from a volatile asset without picking a direction of where the volatility may drive the asset's price, may choose to enter a \emph{straddle} option strategy, by buying both a call and a put option on the asset with the same expiration date and strike. If the underlying asset moves sharply up to expiration date, the investor can make a profit regardless of whether it moves higher or lower in value. Alternatively, the investor may opt for a \emph{butterfly} option strategy by entering four appropriately structured option contracts with different strikes simultaneously. Because these option strategies give rise to piecewise linear payoff functions, the methodology described in the previous section can be extended to calculate the fair values of these option portfolios.

\begin{figure}
\centering
\includegraphics[width=0.48\textwidth]{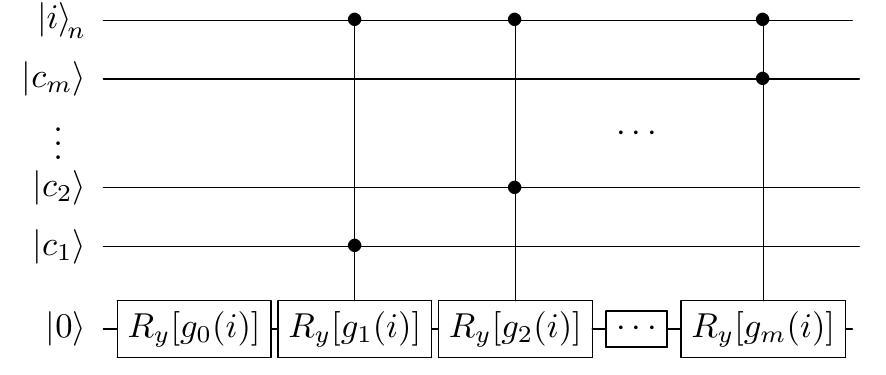}
\caption{\label{fig:multi_strike} Quantum circuit that implements the multi-controlled Y-rotations for a portfolio of options with $m$ strike prices.}
\end{figure}

In order to capture the structure of such option strategies, we can think of the individual options as defining one or more effective strike prices $K_j$ and add a linear function $f_j(S)=a_jS+b_j$ between each of these strikes. For example, to price an option strategy with the payoff function
\begin{align} \label{eqn:call_spread1}
f_s(S) = \max(0, S-K_1)-\max(0, S-K_2),
\end{align}
which corresponds to a call spread (the option holder has purchased a call with strike $K_1$ and sold a call with strike $K_2$), we use functions $f_0$, $f_1$, and $f_2$ such that
\begin{align} \label{eqn:call_spread2}
f_s(S)=
\begin{cases}
f_0(S) & S<K_1, \\
f_0(S) + f_1(S) & K_1\leq S < K_2, \\
f_0(S) + f_1(S)+f_2(S) & K_2\leq S.
\end{cases}
\end{align}
To match Eq.~(\ref{eqn:call_spread1}) with Eq.~(\ref{eqn:call_spread2}), we set $f_0(S)=0$, $f_1(S)=S-K_1$ and $f_2(S)=-S+K_2$. 
In general, to price a portfolio of options with $m$ effective-strike prices $K_1$, ..., $K_m$ and $m+1$ functions $f_0(S)$, ..., $f_{m}(S)$, we need an ancilla qubit per strike to indicate if the underlying has reached the strike. This allows us to generalize the discussion from Sec.~\ref{sec:vanilla}. We apply a multi-controlled Y-rotation with angle $g_j(i)$ if $i\geq K_j$ for each strike $K_j$ with $j\in\{1,...,m\}$. The rotation $g_0(i)$ is always applied, see the circuit in Fig.~\ref{fig:multi_strike}. The functions $g_j(i)$ are determined using the same procedure as in Sec.~\ref{sec:vanilla}.

\subsection{Multi-asset and path-dependent options}

In this section we show how to price options with path-dependent payoffs as well as options on more than one underlying asset.
In these cases, the payoff function depends on a multivariate distribution of random variables $\{S_j\}$ with $j\in\{1,..., d\}$. 
The ${S_j}$'s may represent one or several assets at discrete moments in time or a basket of assets at the option maturity.
In both cases, the probability distribution of the random variables $S_j$ are truncated to the interval $[S_{j,\text{min}},S_{j,\text{max}}]$ and discretized using $2^{n_j}$ points so that they can be represented by $d$ quantum registers where register $j$ has $n_j$ qubits.
Thus, the multivariate distribution is represented by the probabilities $p_{i_1,...,i_d}$ that the underlying has taken the values $i_1$, ..., $i_d$ with $i_j\in\{0,...,2^{n_j}-1\}$.
The quantum state that represents this probability distribution, a generalization of Eq.~(\ref{eqn:prob_state}), is
\begin{align}
\label{eqn:multi_register_state}
\ket{\psi}_n=\sum_{i_1,...,i_d}\sqrt{p_{i_1,...,i_d}}\ket{i_1}_{n_1}\otimes ... \otimes \ket{i_d}_{n_d},
\end{align}
with $n=\sum_j n_j$.
Various types of options, such as Asian options or basket options, require us to compute the sum of the random variables $S_j$.
The addition of the values in two quantum registers $\ket{a,b}\to\ket{a,a+b}$ may be calculated in quantum computers with adder circuits based on CNOT and Toffoli gates \cite{Vedral1995, Draper2000, Draper2004}. 
To this end we add an extra qubit register with $n'$ qubits to serve as an accumulator.
By recursively applying adder circuits we perform the transformation $\ket{\psi}_{n}\ket{0}_{n'}\to\ket{\phi}_{n+n'}$ where $\ket{\phi}_{n+n'}$ is given by
\begin{align}
\sum_{i_1,...,i_d}\!\sqrt{p_{i_1,...,i_d}}\ket{i_1}_{n_1}\otimes ... \otimes \ket{i_d}_{n_d}\otimes\ket{i_1+...+i_d}_{n'}.
\end{align}
Here circuit optimization may allow us to perform this computation in-place to minimize the number of qubit registers needed.
Now, we use the methods discussed in the previous section to encode the option payoffs into the quantum circuit.

\subsubsection{Basket Options}

A European style basket option is an extension of the single asset European option discussed in Sec.~\ref{sec:path_independent_options}, only now the payoff depends on a weighted sum of $d$ underlying assets. A call option on a basket has the payoff profile

\begin{equation}
f(S_{\text{basket}}) = \text{max}(0, S_{\text{basket}} -K)
\end{equation}
where $S_{\text{basket}}=\vec{w}\cdot\vec{S}$, for basket weights $\vec{w}=[w_1, w_2, \dotsc, w_{d}]$, $w_i \in [0,1]$, underlying asset prices at option maturity $\vec{S}=[S_1, S_2, \dotsc S_d]$ and strike $K$. In the BSM model, the underlying asset prices are described by a multivariate log-normal distribution with probability density function \cite{Tarmast2001}

\begin{equation}
\label{eqn:multivariate_lognormal_pdf}
P(\vec{S}) = \frac{\text{exp}\left(-\frac{1}{2}(\ln S -\mu)^T \Sigma^{-1}(\ln S -\mu) \right)}{(2\pi)^{d/2}(\text{det}\Sigma)^{1/2}\prod_{i=1}^d S_i},
\end{equation}
where $\ln S = (\ln S_1, \ln S_2 \dotsc, \ln S_d)^T$ and $\mu=(\mu_1, \mu_2 \dotsc \mu_d)^T$, where each $\mu_i$ is the log-normal distribution parameter for each asset defined in the caption of Fig.~\ref{fig:distribution}. $\Sigma$ is the $d\times d$ positive-definite covariance matrix of the $d$ underlyings 
\begin{equation}
\label{eqn:covariance_matrix}
\Sigma = T \begin{bmatrix} 
    \sigma_1^2 & \rho_{12} \sigma_1\sigma_2 & \dots  & \rho_{1d}\sigma_1\sigma_d \\
    \rho_{21} \sigma_2\sigma_1 & \sigma_2^2 & \dots  & \rho_{2d}\sigma_2\sigma_d\\
    \vdots & \vdots & \ddots & \vdots \\
    \rho_{d1} \sigma_d\sigma_1 & \dots & \dots  & \sigma_d^2 
    \end{bmatrix}
\end{equation}
with $\sigma_i$ the volatility of the $i$th asset, and $-1\leq \rho_{ij} \leq 1$ the correlation between assets $i$ and $j$ and $T$ the time to maturity.

\begin{figure}
\centering
\includegraphics[width=0.48\textwidth]{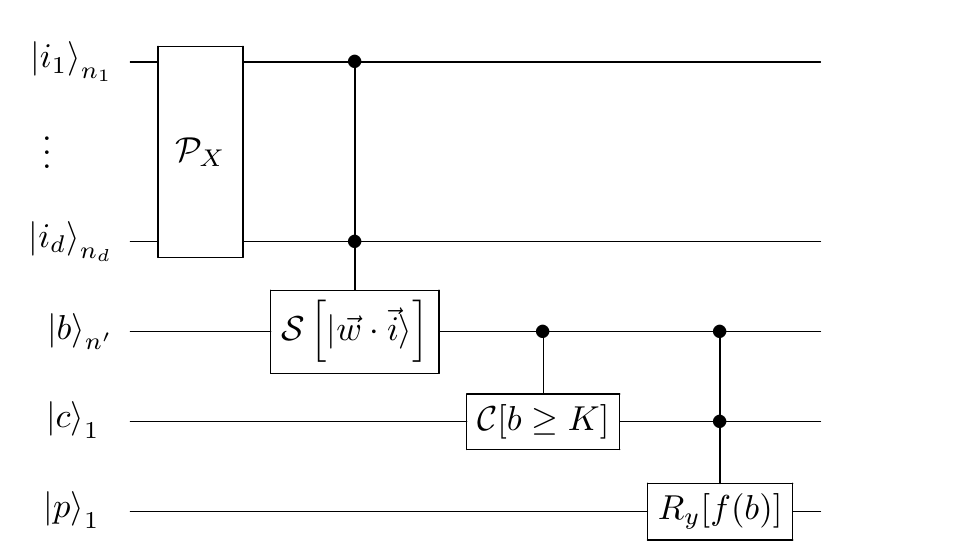}
\caption{\label{fig:basket_option_circuit} Schematic of the circuit that encodes the payoff of a basket call option of $d$ underlying assets into the amplitude of a payoff qubit $\ket{p}$. First, a unitary $\mathcal{P}_X$ loads the multivariate distribution of Eq.~(\ref{eqn:multivariate_lognormal_pdf}) into $d$ registers $\ket{i_1}_{n_1} \dotsc \ket{i_d}_{n_d}$ using the methods described in Sec.~\ref{sec:dist_loading}. The weighted sum operator $\mathcal{S}$, see Appendix~\ref{sec:weighted_sum}, calculates the weighted sum $\ket{w_1\cdot i_1+\dotsc+w_d \cdot i_d}$ into a register $\ket{b}_{n'}$ with $n'$ qubits, where $n'$ is large enough to hold the maximum possible sum. The comparator circuit $\mathcal{C}$ sets a comparator qubit $\ket{c}$ to $\ket{1}$ if $b\geq K$. Lastly, controlled-Y rotations are used to encode the option payoff $f(b) = \text{max}(0, b - K)$ into the payoff qubit using the method shown in Fig.~\ref{fig:phi2}, controlled by the comparator qubit $\ket{c}$. }
\end{figure}

The quantum circuit for pricing a European style basket call option is analogous to the single asset case, with an additional unitary to compute the weighted sum of the uncertainty registers $\ket{i_1}_{n_1} \dotsc \ket{i_d}_{n_d}$ before applying the comparator and payoff circuits, controlled by the accumulator register $\ket{b}_{n'}=\ket{i_1+...+i_d}_{n'}$. A schematic of these components is shown in Fig.~\ref{fig:basket_option_circuit}. The implementation details of the circuit that performs the weighted sum operator is discussed in Appendix~\ref{sec:weighted_sum}. 

\begin{figure}
\centering
\includegraphics[width=0.48\textwidth]{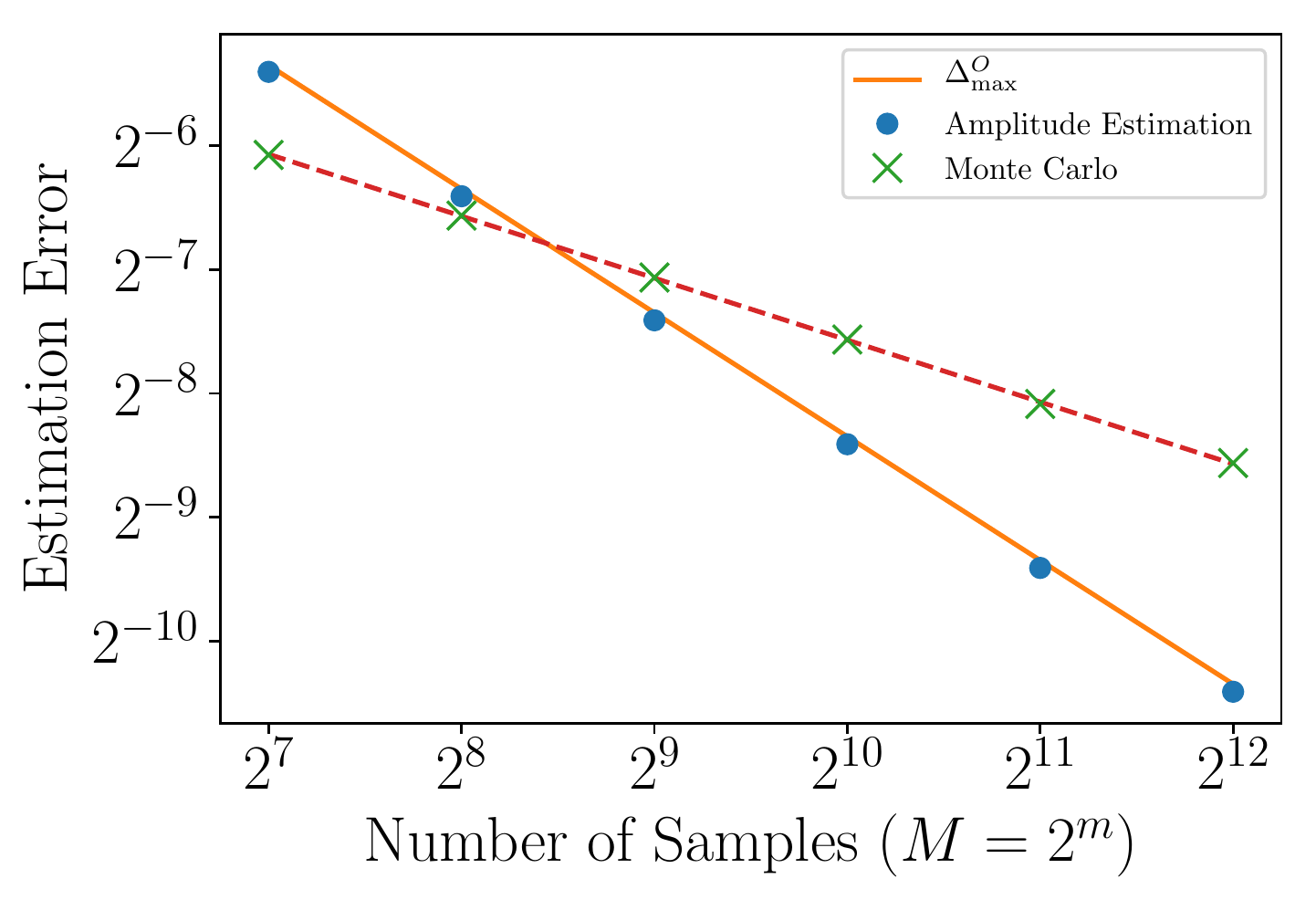}
\caption{\label{fig:estimation_error_QC_MC} Comparison of the estimation error between Amplitude Estimation and Monte Carlo at the $8/\pi^2 \approx 81\%$ confidence interval for a basket option consisting of 3 identical, equally weighted assets with the parameters of Fig.~\ref{fig:distribution}, strike price $K=2.0$ and asset correlations $\rho_{12}=\rho_{13}=\rho_{23}=0.8$. The approximation error for amplitude estimation is plotted against the maximum expected error given by Eq.~(\ref{eqn:max_AE_error}), illustrating the $\mathcal{O}(M^{-1})$ convergence. We calculate the equivalent Monte Carlo error at the same $81\%$ confidence interval over 10,000 simulations for each sample number $2^m$. The red dashed line shows a linear fit across the Monte Carlo errors, displaying the expected $\mathcal{O}(M^{-1/2})$ scaling. }
\end{figure}

We use a basket option to compare the estimation accuracy between AE and classical Monte Carlo.
From Eq.~(\ref{eqn:estimator}), we know that for $M$ applications of the $\mathcal{Q}$ operator (see Fig.~\ref{fig:ae}), the possible values returned by AE will be of the form $\sin^2(y\pi/M)$ for $y \in \{0, ..., M-1\}$ and the maximum distance between two consecutive values is 

\begin{equation}
\Delta_{\text{max}} = \sin^2\left(\frac{\pi}{4} + \frac{2\pi}{4M}\right) - \sin^2\left(\frac{\pi}{4} - \frac{2\pi}{4M}\right).
\end{equation}
This quantity determines how close $M$ operations of $\mathcal{Q}$ can get us to the amplitude which encodes the option price. Using $\sin^2(\pi/4 + x) = x + 1/2 + \mathcal{O}(x^3)$ for $x\ll 1$, we get $\Delta_{\text{max}} = \pi/M + \mathcal{O}(M^{-3})$ for $\pi/M \ll 1$. From Eq.~(\ref{eqn:estimation_error}) and Eq.~(\ref{eqn:option_scaling}), we can determine that with probability of at least $8/\pi^2$, our estimated option price using AE will be within

\begin{equation}
\label{eqn:max_AE_error}
\Delta_{\text{max}}^O = \frac{\pi/M}{2c} \times (i_{\text{max}} - K) + \mathcal{O}(M^{-3})
\end{equation}
of the exact option price, where $c$, $i_{\text{max}}$ and $K$ are the parameters used to encode the option payoff into our quantum circuit, discussed in Sec.~\ref{sec:vanilla}. To compare this estimation error to Monte Carlo, we use the same number of  samples to price an option classically, and determine the approximation error at the same $8/\pi^2 \approx 81\%$ confidence interval by repeated simulations. The comparison for a basket option on three underlying assets shows that AE provides a quadratic speed-up, see Fig.~\ref{fig:estimation_error_QC_MC}. It is worth noting that for typical business cases, the number of paths required for acceptable pricing accuracy goes from tens of thousands to millions (depending on the option underlyings) \cite{Gobet2009, Shevchenko2014}, so they are well within the range where amplitude estimation becomes more efficient than Monte Carlo, as shown in Fig.~\ref{fig:estimation_error_QC_MC}.

\subsubsection{Asian Options}
We now examine arithmetic average Asian options which are single-asset, path-dependent options whose payoff depends on the price of the underlying asset at multiple time points before the option's expiration date. Specifically, the payoff of an Asian call option is given by

\begin{equation}
\label{eqn:asian_payoff}
f(\bar{S}) = \text{max}(0, \bar{S}-K)
\end{equation}
where $K$ is the strike price, $\bar{S}$ is the arithmetic average of the asset's value over a pre-defined number of points $d$ between $0$ and the option maturity $T$

\begin{equation}
\label{eqn:asian_avg}
\bar{S} = \frac{1}{d}\sum_{t=1}^{d}S_t.
\end{equation}

\begin{figure}
\centering
\includegraphics[width=0.48\textwidth]{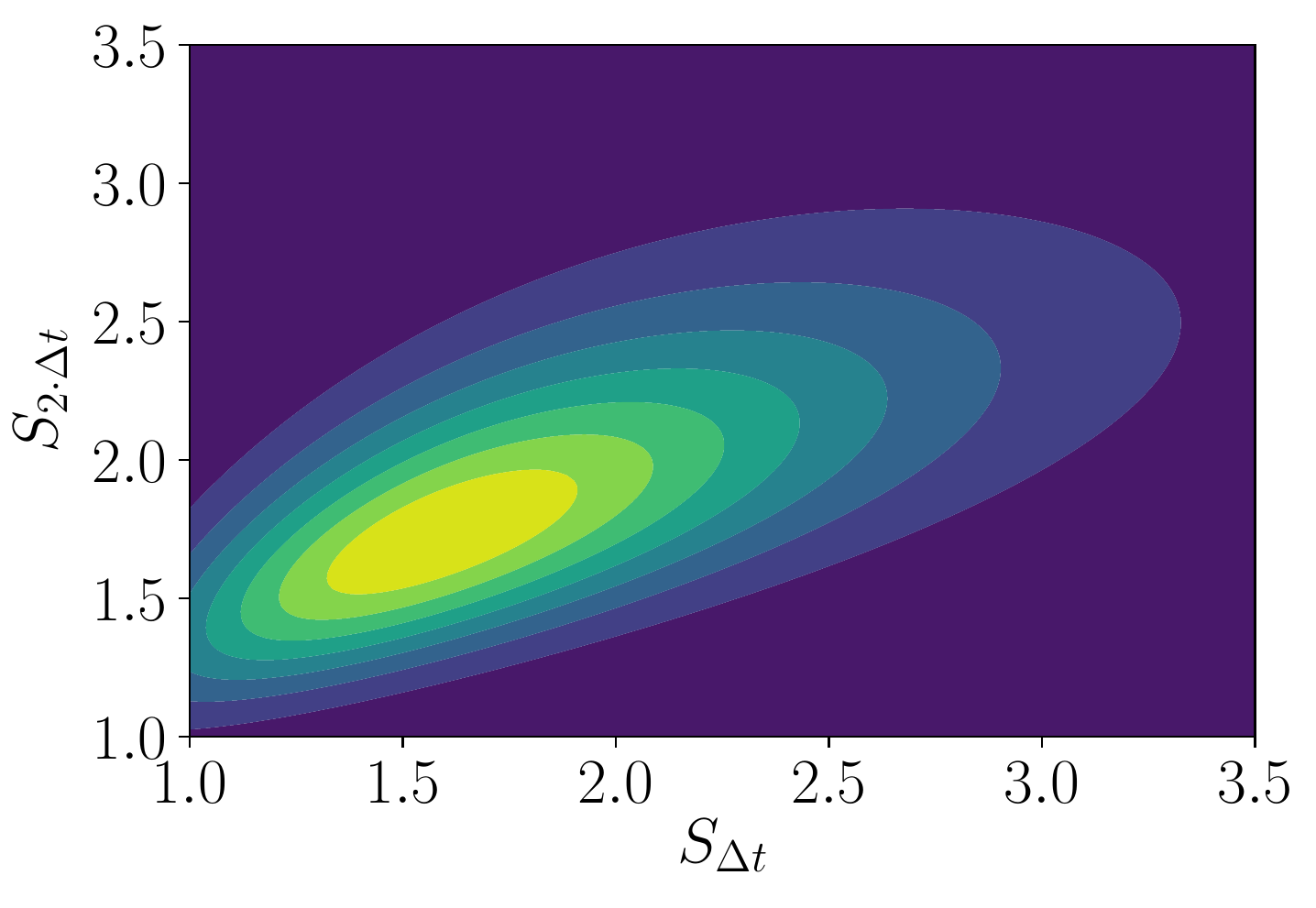}
\caption{\label{fig:multi_step_pdf} Probability density function of a multivariate log-normal distribution, see Eq.~(\ref{eqn:multivariate_lognormal_pdf}), for the asset shown in Fig.~\ref{fig:distribution} defined on two time steps $t=\Delta t = T/2$ and $t=2\Delta t = T$ }
\end{figure}

The probability distribution of the asset price at time $t$ will again be log-normal with probability density function

\begin{equation}
\label{eqn:lognormal_t}
P(S_t)=\frac{1}{S_t\sigma \sqrt{2\pi \Delta t}} e^{-\frac{(\ln{S_t} -\mu_t)^2}{2\sigma^2 \Delta t}},
\end{equation}
with $\mu_t = \left(r-0.5\sigma^2\right)\Delta t + \ln(S_{t-1})$ and $\Delta t = T/d$. We can then use the multivariate distribution in Eq.~(\ref{eqn:multivariate_lognormal_pdf}), with $\vec{S}$ now a $d$-dimensional vector of asset prices at time points $[t_1 \dotsc t_d]$, instead of distinct underlying prices at maturity $T$. As we are not considering multiple underlying assets that could be correlated, the covariance matrix is diagonal $\Sigma=\Delta t [ \text{diag}(\sigma^2,...,\sigma^2)]$. An illustration of the probability density function used for an asset defined on two time steps is shown in Fig.~\ref{fig:multi_step_pdf}.

We now prepare the state $\ket{\psi}_n$, see Eq.~(\ref{eqn:multi_register_state}), where each register represents the asset price at each time step up to maturity.  Using the weighted sum operator of Appendix~\ref{sec:weighted_sum} with equal weights $1/d$, we then calculate the average value of the asset until maturity $T$, see Eq.~(\ref{eqn:asian_avg}), into a register $\ket{\bar{S}}$

\begin{equation}
\ket{\underbrace{i_1}_{\Delta t}\underbrace{i_2}_{2\Delta t}...\underbrace{i_d}_{T}} \mapsto \ket{\bar{S}} = \ket{ \frac{1}{d}\sum_{t=1}^{d}S_t}.
\end{equation}
Finally, we use the same comparator and rotation circuits that we employed for the basket option illustrated in Fig.~\ref{fig:basket_option_circuit} to load the payoff of an arithmetic average Asian option into the payoff qubit $\ket{p}$.

\subsubsection{Barrier Options}
Barrier options are another class of popular option types whose payoff is similar to vanilla European options, but they become activated or extinguished if the underlying asset crosses a pre-determined level called the \emph{barrier}. In their simplest form, there are two general categories of barrier options

\begin{description}
\item[$\bullet$ Knock-Out] The option expires worthless if the underlying asset crosses a certain price level before the option's maturity.
\item[$\bullet$ Knock-In] The option has no value unless the underlying asset crosses a certain price level before maturity.
\end{description}
If the required barrier event for the option to have value at maturity occurs, the payoff then depends only on the value of the underlying asset at maturity and not on the path of the asset until then. If we consider a Knock-In barrier option and label the barrier level $B$, we can write the option's payoff as

\begin{equation}
\label{eqn:barrier_payoff}
f(S) = \begin{cases}\max(0, S_T - K) &\quad \text{if } \exists t \text{ s.t. } S_t \geq B\\
                      0 &\quad\text{otherwise}\\
            \end{cases}
\end{equation}
where $T$ is the time to maturity, $S_t$ the asset price at time $t$ with $0<t\leq T$ and $K$ the option strike.

To construct a quantum circuit to price a Knock-In barrier option, we use the same method as for the Asian option where $T$ is divided into $d$ equidistant time intervals with $\Delta t=T/d$, and use registers $\ket{i_1}_{n_1}\ket{i_2}_{n_2}\dotsc \ket{i_d}_{n_d}$ to represent the discretized range of asset prices at time $t \in \{\Delta t, 2\Delta t, \dotsc, d\cdot \Delta t = T\}$. The probability distribution of Eq.~(\ref{eqn:lognormal_t}) is used again to create the state $\ket{\psi}_n$ in Eq.~(\ref{eqn:multi_register_state}).

\begin{figure*}[htbp!]
\centering
\includegraphics[width=0.7\textwidth]{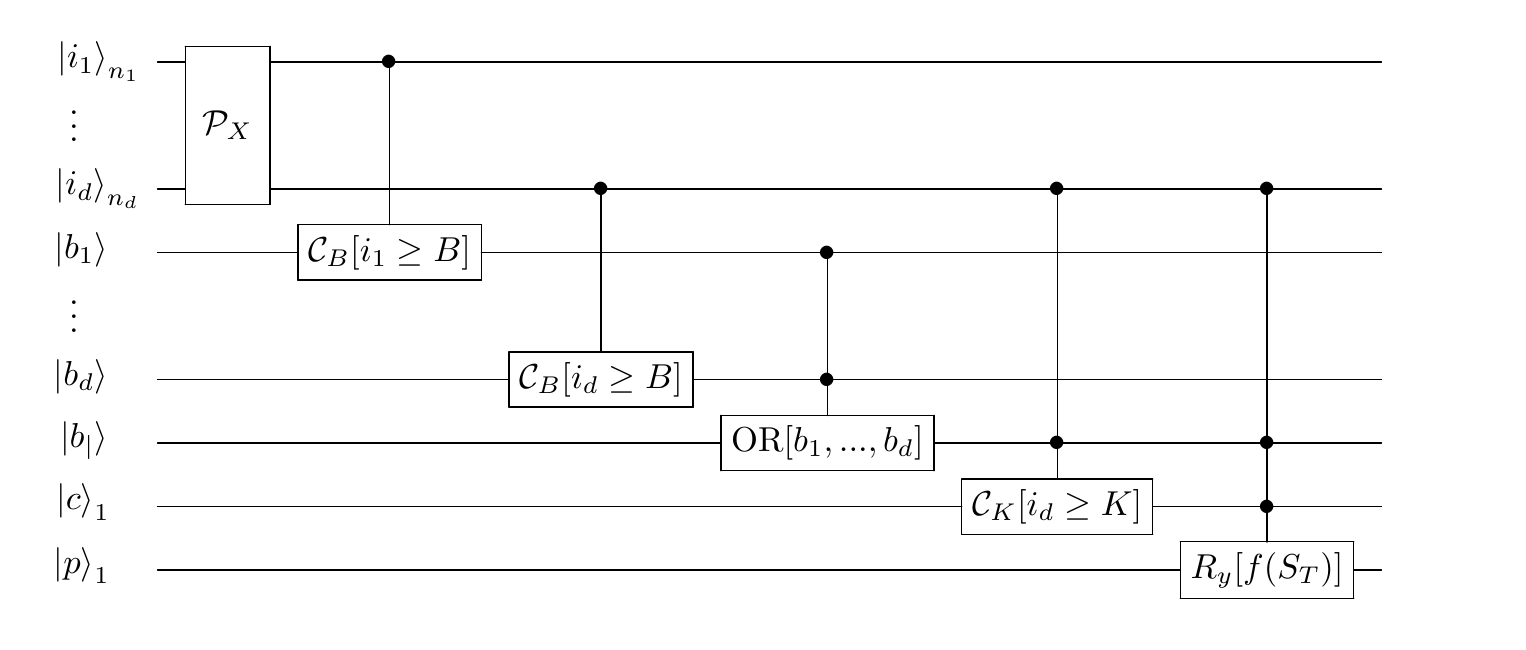}
\caption{\label{fig:barrier_option_circuit} Circuit that encodes the payoff of a Knock-In barrier option in the state of an ancilla qubit $\ket{p}_1$. The unitary operator $\mathcal{P}_X$ is used to initialize the state of Eq.~(\ref{eqn:multi_register_state}). Comparator circuits $\mathcal{C}_B$ are used to set a barrier qubit $b_j$ for all $j \in [1, d]$  if the asset price represented by $\ket{i_j}$ crosses the barrier $B$. The logical OR of all $b_j$ qubits is computed into ancilla $\ket{b_|}$. The strike comparator circuit $\mathcal{C}_K$ sets the comparator qubit $\ket{c}_1$ to $\ket{1}$ if the asset price at maturity $\ket{i_d} \geq K$. Finally, Y-rotations encode the payoff qubit $\ket{p}_1$, controlled on $\ket{i_d}$, the strike qubit $\ket{c}_1$ and the barrier qubit $\ket{b_|}$ which is $\ket{1}$ only if the asset price has crossed the barrier at least once before maturity. }
\end{figure*}

\begin{figure}
\centering
\includegraphics[width=0.48\textwidth]{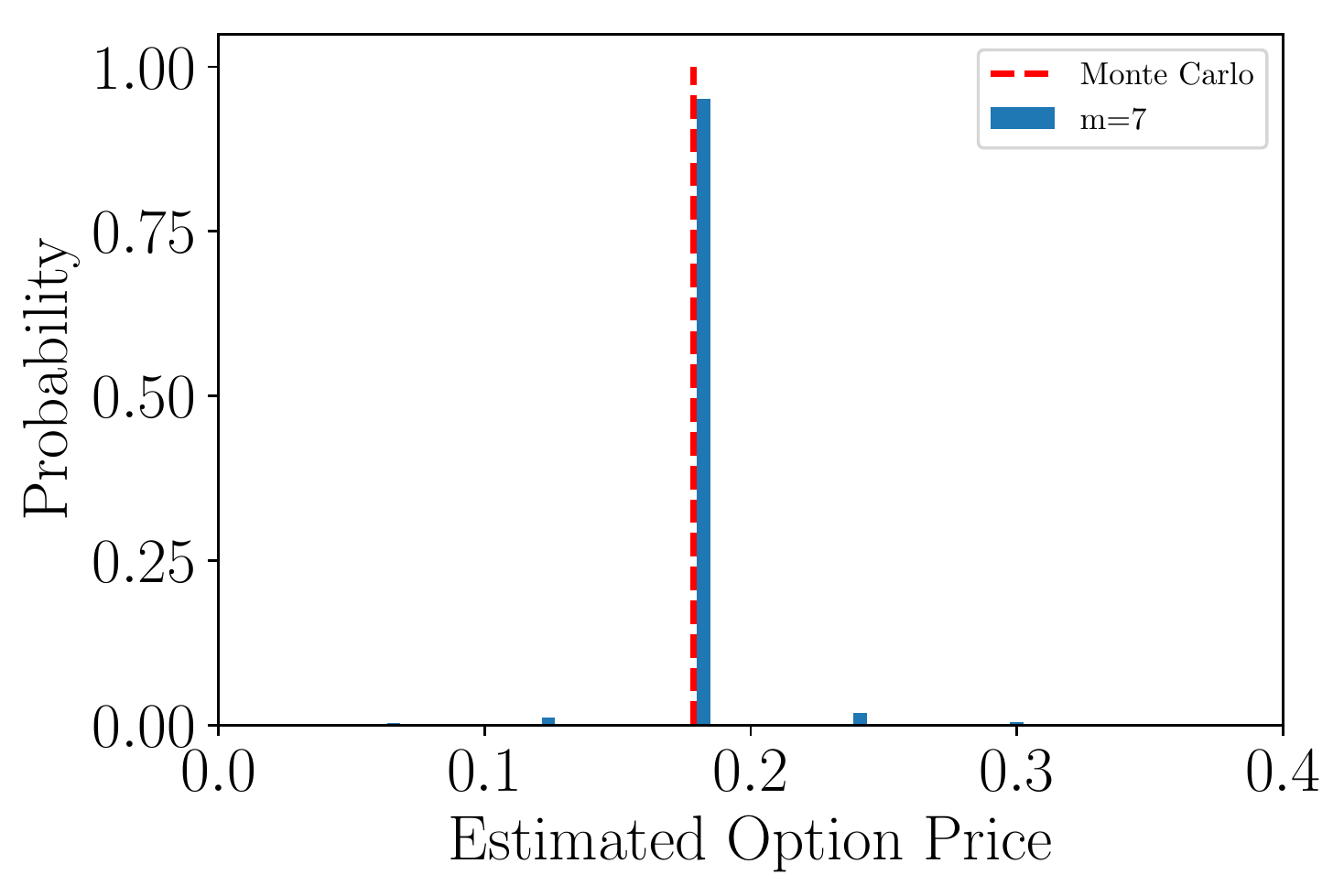}
\caption{\label{fig:barrier_results} Estimated option price for a barrier option using amplitude estimation on a quantum simulator. The option is defined on the asset of Fig.~\ref{fig:distribution} with two timesteps $T/2$ and $T=300/365$ and $2$ qubits used to represent the uncertainty per timestep. The option strike is $K=1.9$ and a barrier was added at $B=2.0$ on both timesteps. The red dotted line is the (undiscounted) value of the option calculated with classical Monte Carlo and $100,000$ paths and the blue bars show the estimated option values using amplitude estimation with $m=7$ sampling qubits.}
\end{figure}

To capture the path dependence introduced by the barrier, we use an additional $d$-qubit register $\ket{b}_d$ to monitor if the barrier is crossed. Each qubit $\ket{b_t}$ in $\ket{b}_d$ is set to $\ket{1}$ if $\ket{i_t}_{n_t} \geq B$. An ancilla qubit $\ket{b_|}$ is set to $\ket{1}$ if the barrier has been crossed in at least one time step. This is done by computing the logical OR, see Fig.~\ref{fig:logical_or}, of every qubit in $\ket{b}_d$ and storing the result in the ancilla

\begin{equation}
\label{eqn:barrier_calculation}
\ket{b_1 b_2\dotsc b_d}\ket{0} \mapsto \ket{b_1 b_2\dotsc b_d} \ket{b_1 \text{ || } b_2  \dotsc\text{ || } b_d }.
\end{equation}
This is computed with $X$ (NOT) and Toffoli gates and $d-2$ ancilla qubits. The ancilla qubit $\ket{b_|}$ is then used as a control for the payoff rotation into the payoff qubit, effectively knocking the option \emph{in}. For Knock-Out barrier options, we can follow the same steps and apply an $X$ gate to the ancilla barrier qubit before using it as control, in this manner knocking the option \emph{out} if the barrier level has been crossed. A circuit displaying all the components required to price a Knock-In barrier option is shown in Fig.~\ref{fig:barrier_option_circuit}. Results from amplitude estimation on a barrier option circuit using a quantum simulator are shown in Fig.~\ref{fig:barrier_results}.

Even though we have focused our attention on barrier options where the barrier event is the underlying asset crossing a barrier from below, we can use the same method to price barrier options where barrier events are defined as the asset crossing the value from above. This only requires changing the comparator circuits to compute $S_t \leq B$ in the barrier register $\ket{b}_d$.

For all path-dependent options, including the Asian and barrier options we have examined in this section, we note that the choice of time intervals on which we need to represent the probability distribution on quantum registers depends on the type of option in question and the type of underlying(s). For instance, when pricing barrier options, we only need to represent the probability distribution at the barrier dates and at maturity. However, as the choice of which time intervals need to be represented for option pricing is independent of whether we employ a quantum or a classical pricing model, a detailed analysis of this choice is beyond the scope of this work.

\section{Quantum hardware results \label{sec:hardware}}

\begin{table}
 \begin{center}
  \begin{tabular}{l | c | c | c | c} \hline \hline
    \#                & Single-qubit  & CX & CCX & Depth \\ \hline
   $m=3$ & 2,091 & 2,056 & 90 & 3,927\\
   $m=5$  & 12,768  & 9,078 & 378 & 17,332 \\
   $m=7$  & 52,275  & 37,132 & 1,530 & 70,916 \\
   $m=9$  & 210,144  & 149,290 & 6,138  & 285,204 \\
  \end{tabular}
 \end{center}
  \caption{\label{Tab:gate_counts} Single-qubit, CNOT, Toffoli gate counts and overall circuit depth required for the full amplitude estimation circuits for each instance in Fig.~\ref{fig:european_call_ae}, as a function of the number of sampling qubits $m$.
These figures assume all-to-all connectivity across qubits. }
\end{table}

In this section we examine the performance of European call option circuits evaluated on quantum hardware.
Quantum circuits based on standard amplitude estimation are not promising candidates for near-term devices, given that they require extra qubits to control the accuracy of the calculation, and multi-controlled gate operations.
Tab.~\ref{Tab:gate_counts} lists the number of single and multi-qubit gates required for the European call option examples in  Fig.~\ref{fig:european_call_ae}, all of which are far from the reach of current and near-term quantum hardware.

However, a quadratic speed-up is also possible by performing AE without quantum phase estimation (see Sec.~\ref{sec:ae}).
Even though removing phase estimation from amplitude estimation does not make the overall algorithm immediately compatible with noisy quantum computers, it does lead to significantly shorter circuits than the original implementation of AE, allowing us to examine how it performs on noisy quantum hardware.

We hence focus on the circuits required to perform AE without phase estimation and show results for a European call option, using three qubits, two of which represent the uncertainty and one encodes the payoff.
We consider a log-normal random distribution with $S_0 = 2$, $\sigma = 40\%$, $r = 5\%$, and $T = 40/365$, see Fig.~\ref{fig:distribution}, and truncate the distribution to the interval defined by three standard deviations around the mean. With two qubits encoding this distribution, the possible values are $[1.21, 1.74, 2.28, 2.81]$, represented by $\ket{00}, \ldots, \ket{11}$, with corresponding probabilities $0.1\%, 55.4\%, 42.5\%$, and $1.9\%$.
We set the strike price to $K = 1.74$.

To examine the behavior of the circuit for different input probability distributions, we run eight experiments that differ by the initial spot price $S_0$ and all other parameters are kept constant. The spot price is varied from $1.8$ to $2.5$ in increments of $0.1$. This way we can use the same circuit for all experiments and only vary the Y-rotation angles used to map the initial probability distribution onto the qubit register. This choice of input parameters allows us to evaluate our circuits for expected option prices in the range $[0.0754, 0.7338]$. 

Following the procedure detailed in Sec.~\ref{sec:ae}, we construct the circuits for $\mathcal{A}\ket{0}_3$ and $\mathcal{QA}\ket{0}_3$, which correspond to $k=0,1$ (i.e. $m=1$) in Eq.~(\ref{eqn:QkA}), with $n=2$.
We then perform repeated measurements of the circuits, and by combining the measured probabilities for all $k$ in a single likelihood function, we can perform a maximum likelihood estimation for $\theta_a$, hence obtaining an estimate for $a=\mathbb{E}[f(S)]$ (see Eq.~(\ref{eqn:expectation_value})), i.e. the expected payoff.
Each experiment is evaluated on the \emph{IBM Q Tokyo} 20-qubit device with 8192 shots. We repeat each 8192-shot experiment 20 times and average over the 20 measured probabilities in order to limit the effect of any one-off issues with the device. The standard deviation of the measured probabilities across the 20 runs was always $<2\%$.
The connectivity of \emph{IBM Q Tokyo} allows to choose three fully connected qubits for the experiments, and thus, no swaps are required to implement any two-qubit gate in our circuits \cite{Qiskit}. For all circuits described in the following sections, we used qubits 6, 10 and 11. 

Note that even though we are only interested in the result of a single qubit, we always measure all three qubits to be able to apply readout error mitigation as discussed later in Sec.~\ref{sec:hardware_error_mitigation}.

\subsection{Algorithm and Operators}

\begin{figure*}[htbp!]
\centering
\includegraphics[width=0.7\textwidth]{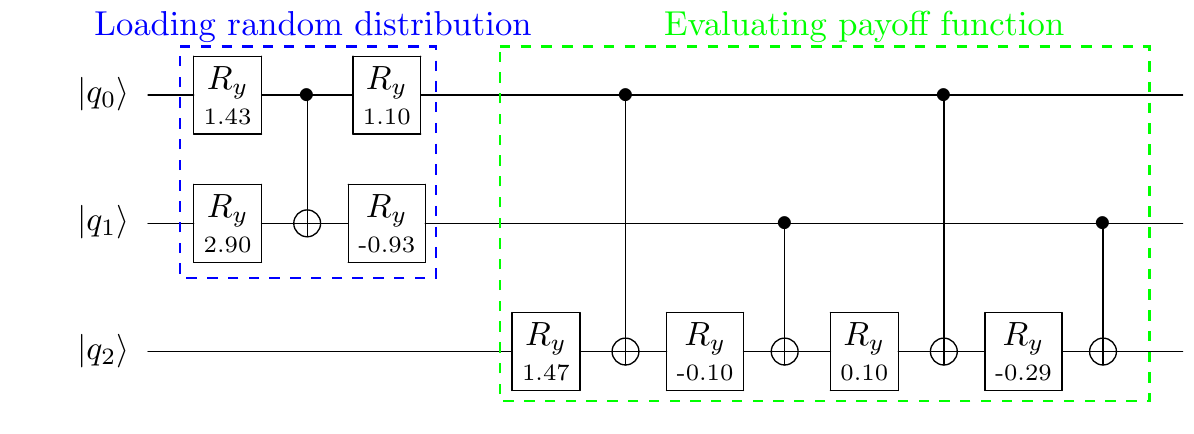}
\caption{\label{fig:european_call_a_operator} The $\mathcal{A}$ operator of the considered European call option: first, the 2-qubit approximation of a log-normal distribution is loaded, and second, the piecewise linear payoff function is applied to last qubit controlled by the first two. This operator can be used within amplitude estimation to evaluate the expected payoff of the corresponding option.}
\end{figure*}

We now show how to construct the operator $\mathcal{A}$ that is required for AE. 
The log-normal distribution on two qubits can be loaded using a single CNOT gate and four single-qubit rotations \cite{Znidaric2008}.
To encode the payoff function, we also exploit the small number of qubits and apply a uniformly controlled Y-rotation instead of the generic construction using comparators introduced in Sec.~\ref{sec:option_pricing}.
A uniformly controlled Y-rotation, i.e.
\begin{align}
\ket{i}_n\ket{0}\to\ket{i}_nR_y(\theta_i)\ket{0},
\end{align}implements a different rotation angle $\theta_i$, $i=0,...,2^n-1$ for each state of the $n$-control qubits. 
For $n=2$, this operation can be efficiently implemented using four CNOT gates and four single qubit Y-rotations \cite{shende2006, iten2019}.
The resulting circuit implementing $\mathcal{A}$ is shown in Fig.~\ref{fig:european_call_a_operator}.
Note that in our setup, different initial distributions only lead to different angles of the first four Y-rotations and do not affect the rest of the circuit.
Although we use a uniformly controlled rotation, the rotation angles are constructed in the same way described in Sec.~\ref{sec:payoff}.
We use an approximation scaling of $c = 0.25$ and the resulting angles are $[\theta_0, \ldots, \theta_3] = [1.1781, 1.1781, 1.5708, 1.9635]$, which shows the piecewise linear structure of the payoff function.

The total resulting circuit requires five CNOT gates and eight single-qubit Y-rotations, see Fig.~\ref{fig:european_call_a_operator}.
Since we use uniformly controlled rotations, we do not need any ancilla qubit.
Note that if we want to evaluate the circuit for $\mathcal{A}$ alone, we can replace the last CNOT gate in Fig.~\ref{fig:european_call_a_operator} by classical post-processing of the measurement result: if $q_1$ is measured as $\ket{1}$, we flip $q_2$ and otherwise we do nothing. 
This further reduces the overall CNOT gate count to four.

In the remainder of this section, we focus on $\mathcal{Q}\mathcal{A}\ket{0}_3$, i.e., the outlined algorithm for $m=1$, to examine the reach of today's quantum hardware in evaluating AE option pricing circuits which do not require phase estimation. We note that this evaluation is relevant to any quantum algorithm realizing AE without phase
estimation and is independent of the approach described in \cite{Suzuki2020} or any other particular implementation.

After optimzing the gate count, the resulting circuit for $\mathcal{QA}\ket{0}_3$ consists of $18$ CNOT gates and 33 single-qubit gates.
The detailed circuit diagram and applied circuit optimization steps are provided in Appendix \ref{sec:ga_circuit}.

\subsection{Error mitigation and results \label{sec:hardware_error_mitigation}}
We run the circuits for $\mathcal{A}\ket{0}_3$ and $\mathcal{QA}\ket{0}_3$ on noisy quantum hardware. The results are affected by readout errors and errors that occur during the execution of the circuits. 

To mitigate readout errors we run a calibration sequence in which we individually prepare and measure all eight basis states \cite{Dewes2012, Qiskit}. The result is a $8\times8$ readout-matrix $\mathcal{R}$ that holds the probability of measuring each basis state as function of the basis state in which the system was prepared. We use $\mathcal{R}$ to correct all subsequent measurements. The error we measure on $P_1$ for $\mathcal{A}\ket{0}_3$ was reduced from $\sim 6\%$ to $\sim 4\%$ using readout error mitigation.

Errors occuring during the quantum circuit can be mitigated using Richardson extrapolation \cite{Temme2017}. First, the quantum circuit is run using a rescaled Hamiltonian to amplify the effect of the noise. Second, a Richardson extrapolation is used to extract the result of the quantum circuit at the zero noise limit. In hardware, error mitigation is done by stretching the duration of the gates. For each stretch factor the qubit gates need to be recalibrated \cite{Kandala2019}. Here, we use a simplified error mitigation protocol that circumvents the need to recalibrate the gates but still allows us to increase the accuracy of the quantum hardware \cite{Dumitrescu2018}. Since the single-qubit and CNOT gates have an average randomized benchmarking fidelity of $99.7\%$ and $97.8$\%, respectively, we restrict our error mitigation to the CNOT gates. Furthermore, because the optimized circuit for $\mathcal{A}\ket{0}_3$ contains only 4 CNOT gates, we employ the error mitigation protocol only when evaluating $\mathcal{QA}\ket{0}_3$ which consists of 18 CNOT gates. 

\begin{figure*}[htbp!]
\centering
\begin{tabular}{ccc}
\includegraphics[width=0.315\textwidth]{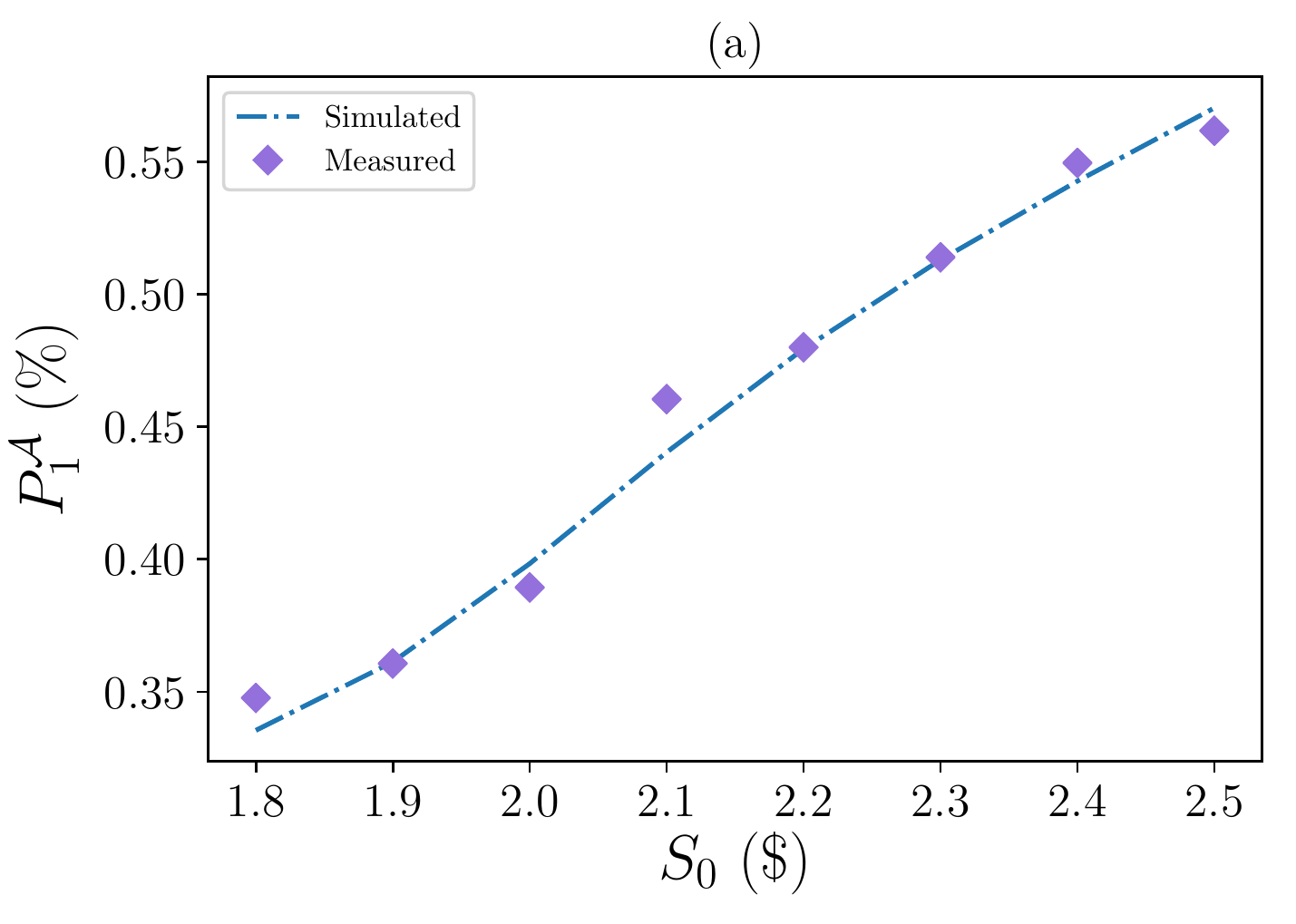}&
\includegraphics[width=0.315\textwidth]{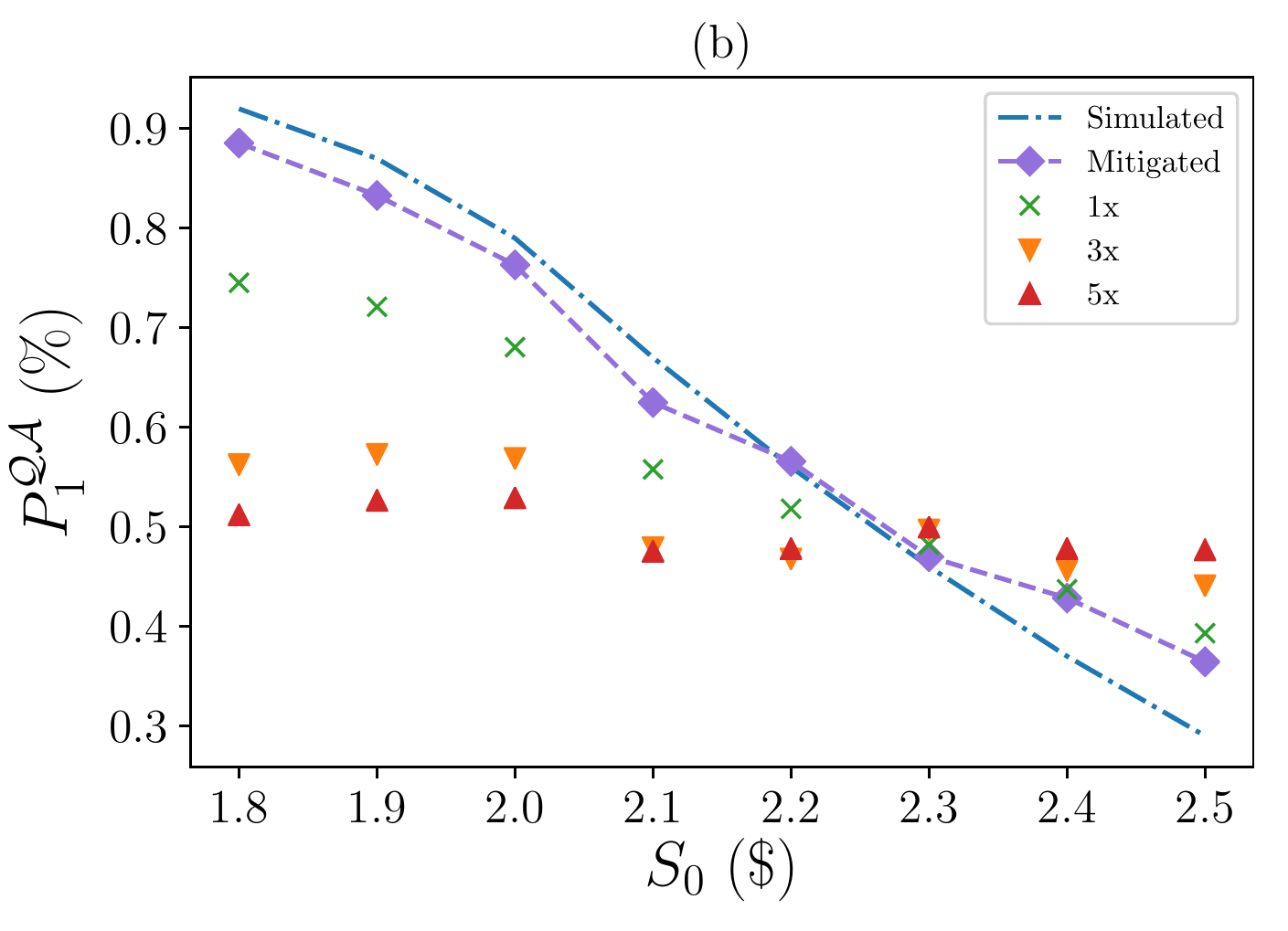}&
\includegraphics[width=0.315\textwidth]{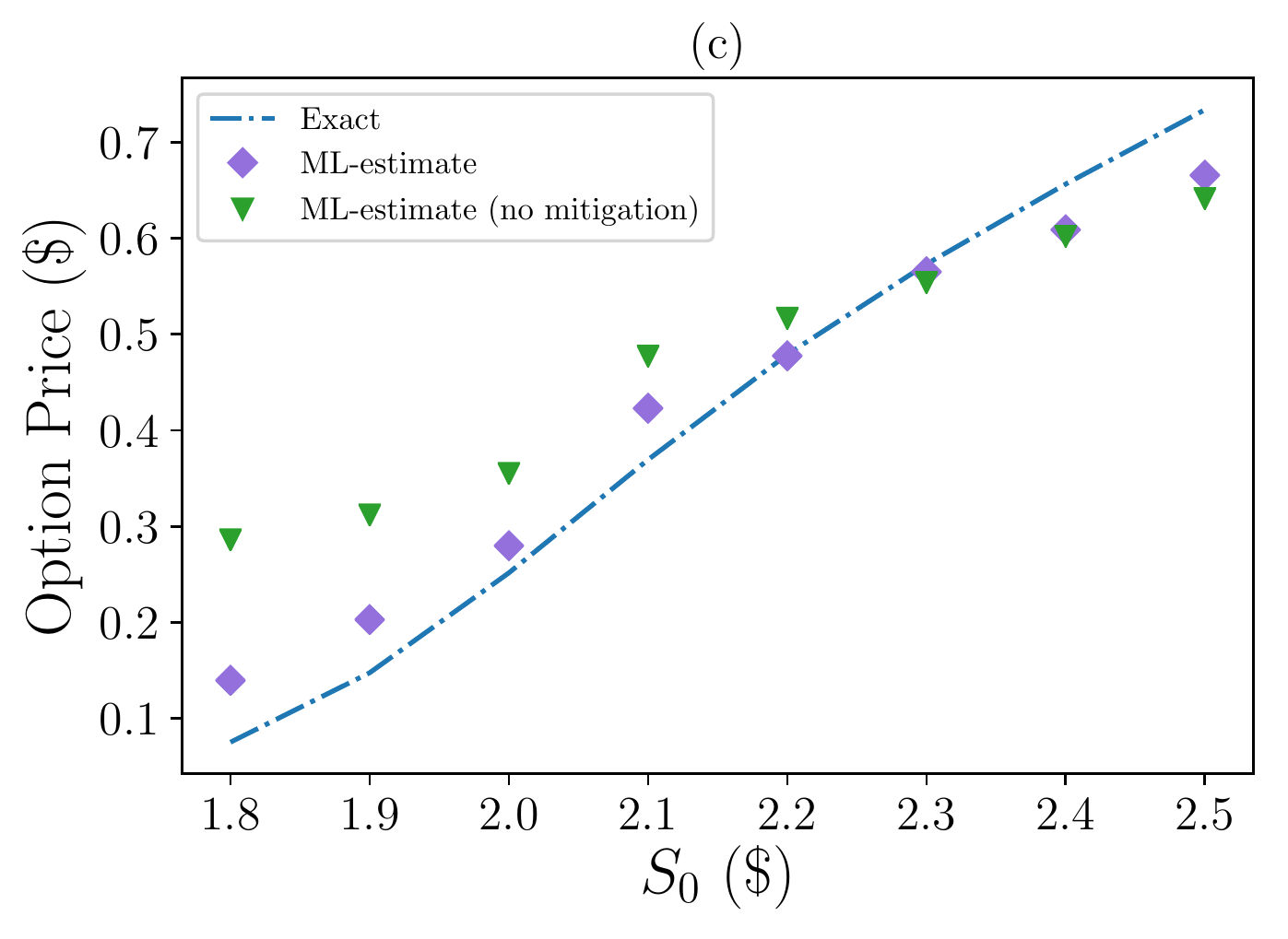}
\end{tabular}
\caption{\label{fig:qa_hardware}{Error-mitigated hardware results for $\mathcal{A}\ket{0}_3$, $\mathcal{QA}\ket{0}_3$ and the estimated option price after applying maximum likelihood estimation as a function of the initial spot price $S_0$. (a) Probability of measuring $\ket{1}$ for the $\mathcal{QA}\ket{0}_3$ circuit (see Appendix \ref{sec:ga_circuit}}, Fig.~\ref{fig:european_call_a_operator}) (b) Probability of measuring $\ket{1}$ for the $\mathcal{QA}\ket{0}_3$ circuit (see Fig.~\ref{fig:qa_circuit}). We show the measured probabilities when replacing each CNOT by one, three and five CNOT gates (green, orange, red, respectively), the zero-noise limit calculated using a second-order Richardson extrapolation method (purple), and the probability measured using the simulator (blue). (c) Option price estimated with maximum likelihood estimation from measurements of $\mathcal{QA}\ket{0}_3$ and $\mathcal{A}\ket{0}_3$ with error mitigation (purple) and without (green). The exact option price for each initial spot price $S_0$ is shown in blue.}
\end{figure*}

We run the circuit for $\mathcal{QA}\ket{0}_3$ three times. In each run we replace the CNOT gates of the original circuit by one, three and five CNOT gates for a total of 18, 54, and 90 CNOT gates, respectively. Since a pair of perfect CNOT gates simplifies to the identity these extra gates allow us to amplify the error of the CNOT gate without having to stretch the gate duration, thus, avoiding the need to recalibrate the gate parameters. As the number of CNOT gates is increased the probability of measuring $\ket{1}$ tends towards 0.5 for all initial spot prices, see Fig.~\ref{fig:qa_hardware}(b). After applying a second-order Richardson extrapolation, i.e quadratic extrapolation, we recover the same behavior as the option price obtained from classical simulations, see Fig.~\ref{fig:qa_hardware}(c). Our simple error mitigation scheme dramatically increased the accuracy of the calculated option price: it reduced the error, averaged over the initial spot price, from $62\%$ to $21\%$.

\section{Conclusion}

We have presented a methodology and the quantum circuits to price options and option portfolios on a gate-based quantum computer. We showed how to account for some of the more complex features present in exotic options such as path-dependency with barriers and averages. The results that we show are available in the finance module in \emph{Qiskit} \cite{Qiskit}. Future work may involve calculating the price derivatives \cite{Broadie1996} with a quantum computer. Pricing options relies on AE. This quantum algorithm allows a quadratic speed-up compared to traditional Monte Carlo simulations, but will most likely require a universal fault-tolerant quantum computer \cite{Fowler2012}. However, research to improve the algorithms is ongoing \cite{Dobsicek2006, OLoan2010, Svore2013}. Here we have used a new algorithm \cite{Suzuki2020} that retains the AE speed-up but that uses less gates to measure the price of an option. Furthermore, we employed a simple error mitigation scheme that allowed us to greatly reduce the errors from the noisy quantum hardware. However, larger quantum hardware capable of running deeper quantum circuits with more qubits than the currently available quantum computers is needed to price the typical portfolios seen in the financial industry. Future work could focus on reducing the number of quantum registers in our implementation by performing some of the computation in-place.

\section{Acknowledgments}
The authors want to thank Abhinav Kandala for the very constructive discussions on error mitigation and real quantum hardware experiments.
C.Z. and R.I. acknowledge the support of the National Centre of Competence in \textit{Research Quantum Science and Technology} (QSIT).

Opinions and estimates constitute our judgment as of the date of this Material, are for informational purposes only and are subject to change without notice. This Material is not the product of J.P. Morgan's Research Department and therefore, has not been prepared in accordance with legal requirements to promote the independence of research, including but not limited to, the prohibition on the dealing ahead of the dissemination of investment research. This Material is not intended as research, a recommendation, advice, offer or solicitation for the purchase or sale of any financial product or service, or to be used in any way for evaluating the merits of participating in any transaction. It is not a research report and is not intended as such. Past performance is not indicative of future results. Please consult your own advisors regarding legal, tax, accounting or any other aspects including suitability implications for your particular circumstances. J.P. Morgan disclaims any responsibility or liability whatsoever for the quality, accuracy or completeness of the information herein, and for any reliance on, or use of this material in any way. Important disclosures at: www.jpmorgan.com/disclosures

IBM, IBM Q, Qiskit are trademarks of International Business Machines Corporation, registered in many jurisdictions worldwide. Other product or service names may be trademarks or service marks of IBM or other companies.

%
%

\appendix

\section{\label{sec:weighted_sum} Circuit implementation of weighted sum operator}

\subsection{Weighted sum of single qubits}

In this appendix, we demonstrate an implementation of the weighted sum operator on a quantum circuit. 
The weighted sum operator $\mathcal{S}$ computes the arithmetic sum of the values of $n$ qubits 
$\ket{a}_n = \ket{a_1 \dotsc a_n}$
weighted by $n$ classically defined non-negative integer weights $\mathbf{\omega} = \left(\omega_1, \omega_2, \dotsc, \omega_n \right)$, and stores the result into another $m$-qubit register $\ket{s}_m = \ket{s_1 \dotsm s_m}$ initialized to $\ket{0}_m$.
In other words,
\begin{equation}
\label{eqn:weighted_sum_definition}
\mathcal{S} \ket{a}_n \ket{0}_m = \ket{a}_n \Ket{\sum_{i=1}^{n} \omega_i a_i}_m,
\end{equation}
where
\begin{equation}
\label{eqn:weighted_sum_num_sum_qubits}
m = \left \lfloor \log_2 \left( \sum_{i=0}^{n} \omega_i \right) \right \rfloor + 1.
\end{equation}
The choice of $m$ ensures that the sum register $\ket{s}_m$ is large enough to hold the largest possible weighted sum, i.e. the sum of all weights.
Alternatively, we can write the weights in the form of a binary matrix $\Omega = \left(\Omega_{i,j} \right) \in \left\{ 0, 1 \right\}^{n \times n^*}$, where the $i$-th row in $\Omega$ is the binary representation of weight $\omega_i$ and $n^* = \max_{i=1}^{d} n_i$. 
We use the convention that less significant digits have smaller indices, so $\ket{s_1}$ and $\Omega_{i,1}$ are the least significant digits of the respective binary numbers.
Using this binary matrix representation, $\mathcal{S}$ is to add the $i$-th qubit $\ket{a_i}$ of the state register to the $j$-th qubit $\ket{s_j}$ of the sum register if and only if $\Omega_{i,j} = 1$.
Depending on the values of the weights, an additional quantum register may be necessary to temporarily store the carries during addition operations. 
We use $\ket{c_j}$ to denote the ancilla qubit used to store the carry from adding a digit to $\ket{s_j}$.
These ancilla qubits are initialized to $\ket{0}$ and will be reset to their initial states at the end of the computation.

\begin{figure}
\centering
\includegraphics[width=0.45 \textwidth]{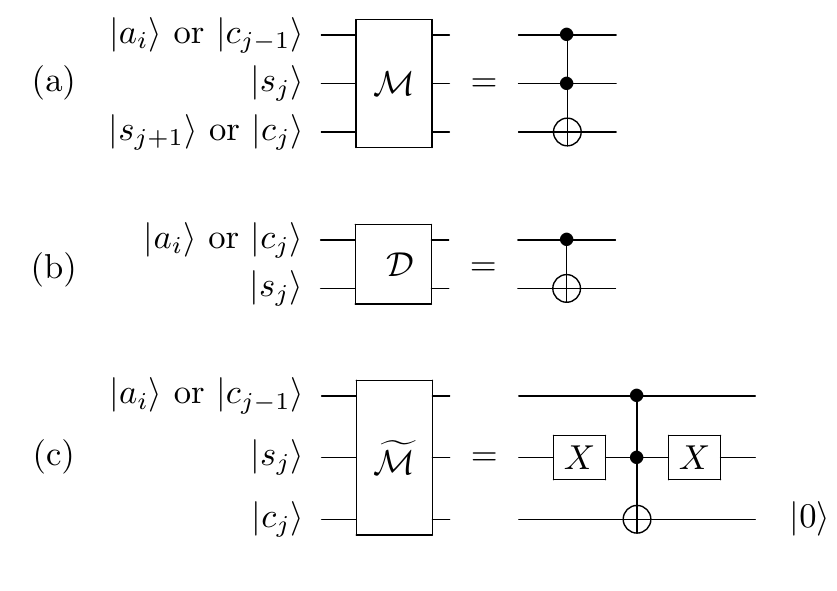}
\caption{Three component gates used to construct the weighted sum operator $\mathcal{S}$. 
(a) The carry operator $\mathcal{M}$ consisting of one Toffoli gate, which computes the carry from adding $\ket{a_i}$ (or $\ket{c_{j-1}}$) and $\ket{s_j}$ into $\ket{s_{j+1}}$ or $\ket{c_j}$. 
(b) The bit addition operator $\mathcal{D}$ consisting of one CNOT gate, which adds the state qubit $\ket{a_i}$ or the carry qubit from the previous digit $\ket{c_{j-1}}$ to the sum qubit $\ket{s_j}$.
(c) The carry reset operator $\widetilde{\mathcal{M}}$ consisting of two X gates and one Toffoli gate, which resets the carry qubit $\ket{c_j}$ back to $\ket{0}$. }
\label{fig:weighted_sum_basic_operations}
\end{figure}

Based on the above setup, we build quantum circuits for the weighted sum operator using three elementary gates: X (NOT), CNOT, and the Toffoli gate (CCNOT). 
These three gates suffice to build any Boolean function \cite{Vedral1995}.
Starting from the first column in $\Omega$, for each column $j$, we find all elements with $\Omega_{i,j} = 1$ and add the corresponding state qubit $\ket{a_i}$ to $\ket{s_j}$.
The addition of two qubits involves three operations detailed in Fig.~\ref{fig:weighted_sum_basic_operations}:
(a) computation of the carry using a Toffoli gate ($\mathcal{M}$), 
(b) computation of the current digit using a CNOT ($\mathcal{D}$),
(c) reset of the carry computation using two X gates and one Toffoli gate ($\widetilde{\mathcal{M}}$).
When adding $\ket{a_i}$ to the $j$-th qubit of the sum register, the computation starts by applying $\mathcal{M}$ and then $\mathcal{D}$ to $\ket{a_i}$, $\ket{s_j}$ and $\ket{c_j}$, which adds $\ket{a_i}$ to $\ket{s_j}$ and stores the carry into $\ket{c_j}$. 
Then, using the same two operations, it adds the carry $\ket{c_j}$ to the next sum qubit $\ket{s_{j+1}}$ with carry recorded in $\ket{c_{j+1}}$.
The process is iterated until all carries are handled.
Finally, it resets the carry qubits by applying $\widetilde{\mathcal{M}}$ in reverse order of the carry computation. 
We reset the carry qubits in order to reuse them in later computations if necessary.

In general, we need $\max(k-2, 0)$ carry qubits to compute the addition of $\ket{a_i}$ on $\ket{s_j}$, where $k \ge 1$ is the smallest integer satisfying
\begin{equation}
\label{eqn:weighted_sum_carries}
\prescript{}{k}{\bra{1}} \rho^s_{j,j+k-1} \ket{1}_k = 0,
\end{equation}
where $\rho^s_{j,j+k-1}$ is the density matrix corresponding to $\ket{s_j \dotsm s_{j+k-1}}$.
In the $k = 1$ case, i.e. $\ket{s_j} = 0$, the computation is reduced to ``copying'' $\ket{a_i}$ to $\ket{s_j}$ using the bit addition operator $\mathcal{D}$, and no carries would be produced.
For $k \ge 2$, Eq.~(\ref{eqn:weighted_sum_carries}) guarantees no carries from $\ket{s_{j+k-1}}$ and beyond. 
Therefore we can directly compute the carry from $\ket{s_{j+k-2}}$ into $\ket{s_{j+k-1}}$ without worrying about additional carries.
This eliminates the need for an ancilla qubit $\ket{c_{j+k-2}}$, and hence the number of carry qubits needed is $k-2$.
To further reduce the number of ancilla qubits, we can use any sum qubit $\ket{s_j} = \ket{0}$ during the computation.
In our case, since we are processing $\Omega$ column by column, all sum qubits more significant than $\ket{s_{j+k-1}}$ would be $\ket{0}$.
In other words, we have the last $m-(j+k-1)$ sum qubits usable as carry qubits in the computation described above.

As the weights are known at the time of building the circuit, the possible states that $\ket{s}_m$ can have before each addition of a state qubit $\ket{a_i}$ are also computable. 
Since we are adding $\ket{a_i}$ to $\ket{s}_m$ starting from the least significant bit, $k$ equals the bit length of the maximum possible sum on $\ket{s_j \dotsc s_m}$ after adding $\ket{a_i}$ to $\ket{s_j}$.
In other words, 
\begin{equation}
\label{eqn:weighted_sum_k}
k = \log_{2} \left \lfloor \sum_{\substack{u \le i, \, \mathrm{or}\\ v \le j}} \frac{\Omega_{u,v}}{2^{j-v}} \right \rfloor + 1.
\end{equation}
Therefore, the number of carry operations and additional ancilla qubits required for each addition of $\ket{a_i}$ can be determined.
The term in the $\left \lfloor \cdot \right \rfloor$ in Eq.~(\ref{eqn:weighted_sum_k}) is upper-bounded by
\begin{equation}
\label{eqn:weighted_sum_k_upper_bound_inner}
\sum_{\substack{u \le i, \, \mathrm{or}\\ v \le j}} \frac{\Omega_{u,v}}{2^v} 
\le \sum_{j=1}^{m} \frac{n_\mathrm{max}}{2^{j-1}}
< 2 n_\mathrm{max}
\le 2 n,
\end{equation}
where $n_\mathrm{max} = \max_{j=1}^{m} \sum_{i=1}{n} \Omega_{i, j}$ is the maximum number of $1$'s in a column of $\Omega$.
It immediately follows that the number of non-trivial carry operations (i.e. carry operations that requires $\widetilde{\mathcal{M}}$) required to add $\ket{a_i}$ to $\ket{s_j \dotsc s_m}$  is upper-bounded by
\begin{equation}
\label{eqn:weighted_sum_k_upper_bound}
k - 2 < \log_{2} \left \lfloor n_\mathrm{max} \right \rfloor
\le \log_{2} \left \lfloor n \right \rfloor,
\end{equation}
and the number of ancilla qubits required for the entire implementation of $\mathcal{S}$ is at most the upper bound for $k-2$, since we may use some sum qubits as carries.
In other words, the number of ancilla qubits required for $\mathcal{S}$ grows at most logarithmically with the number of state qubits $n$.

\subsection{Sum of multi-qubit integers}
The weighted sum operator $\mathcal{S}$ can be used to calculate the sum of $d$ multi-qubit positive integers on a quantum register.
To do that we first prepare the input register in the state
\begin{equation}
\label{eqn:weighted_sum_add_integers_state}
\ket{a}_n = \ket{a^{(1)}_1 \dotsc a^{(1)}_{n_1} \dotsc a^{(d)}_1 \dotsc a^{(d)}_{n_d}}, \quad n = \sum_{i=1}^{d} n_i,
\end{equation}
where $\ket{a^{(i)}_1 \dotsc a^{(i)}_{n_i}}, i \in [1, d]$ is the binary representation of the $i$-th integer to sum with $n_i$ qubits, least significant figure first.
Then we set the weights as
\begin{equation}
\label{eqn:weighted_sum_add_integers_weight}
\mathbf{\omega} = (2^0, \dotsc, 2^{n_1-1}, \dotsc, 2^{0}, \dotsc, 2^{n_d-1}),
\end{equation}
or equivalently,
\begin{equation}
\label{eqn:weighted_sum_add_integers_weight_matrix}
\Omega_{n \times n^*} = \left( I_{n_1 \times n^*}^T, \dotsc, I_{n_d \times n^*}^T \right)^T,
\end{equation}
where $I_{n_i \times n^*} = \left( I_{n_i}, 0_{n_i \times (n^* - n_i)} \right), i \in [1, d]$ and $I_{n_i}$ is the $n_i$-dimensional identity matrix.
Now if we build a weighted sum operator based on the weights in Eq.~(\ref{eqn:weighted_sum_add_integers_weight}) and apply it on the input state qubits in Eq.~(\ref{eqn:weighted_sum_add_integers_state}), we would have the sum of the $d$ integers in $\ket{s}_m$.

\begin{figure*}
\centering
\includegraphics[width=0.9 \textwidth]{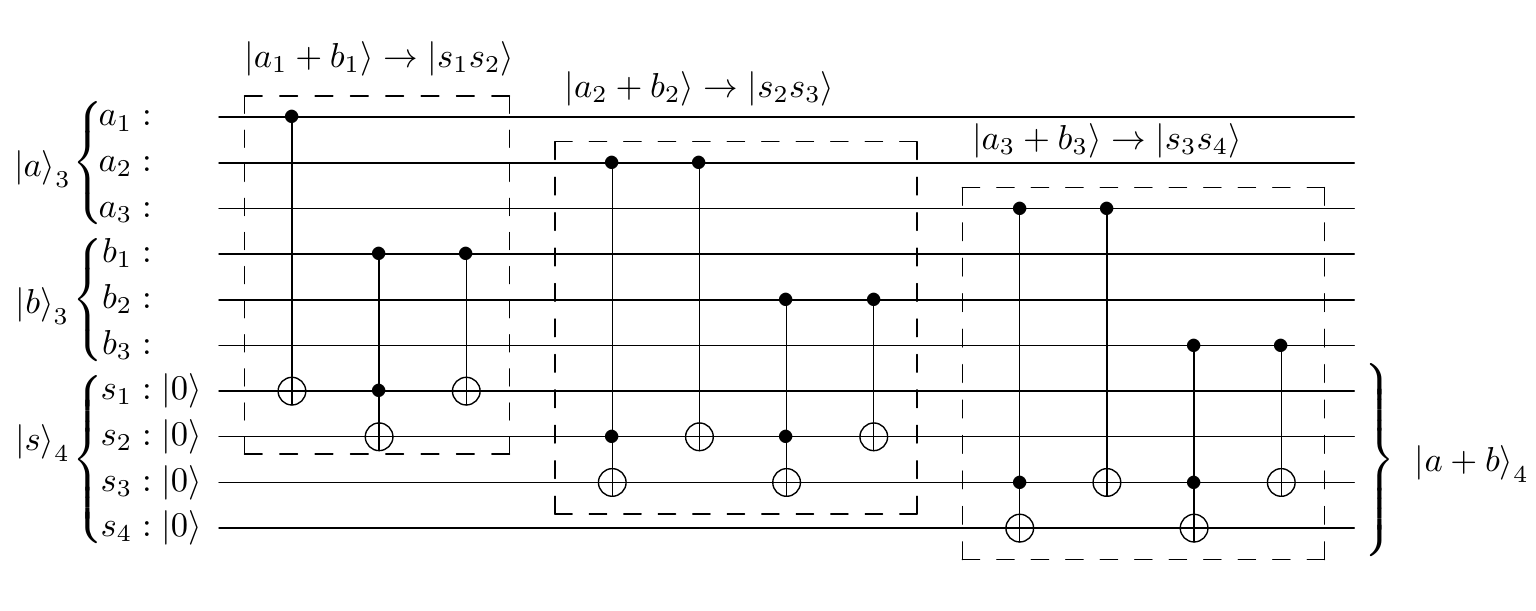}
\caption{A circuit computing the sum of binary numbers $\ket{a}_3$ and $\ket{b}_3$ into $\ket{s}_4$ implemented using the weighted sum operator with weights $\omega = (1, 2, 4, 1, 2, 4)$.} 
\label{fig:sum_of_integers_circuit}
\end{figure*}
Fig.~\ref{fig:sum_of_integers_circuit} shows an example circuit computing the sum of two $3$-digit binary numbers represented on a $6$-qubit quantum register $\ket{a}_3\ket{b}_3$, and storing the result into a $4$-qubit register $\ket{s}_4$.
The circuit is implemented by a weighted sum operator $\mathcal{S}$ with weights $\omega = (1, 2, 4, 1, 2, 4)$.
The addition of each qubit onto the sum qubits requires one carry gate ($\mathcal{M}$) followed by one addition gate ($\mathcal{D}$), except for the first bit $\ket{a_1}$ which does not have any carries before its addition.
This results in a total of $6$ CNOT ($\mathcal{D}$) gates and $5$ Toffoli ($\mathcal{M}$) gates.
The $11$ gates are grouped in three groups, as is shown in Fig.~\ref{fig:sum_of_integers_circuit} by dashed boxes.
Each group computes the sum of the bits $\ket{a_j}$ and $\ket{b_j}$ into $\ket{s_j}$ and the carry into $\ket{s_{j+1}}$.
Note that separate carry qubits are not required, therefore no carry reset operators $\widetilde{\mathcal{M}}$ are used.
In fact, using the above construction for $\mathcal{S}$, no extra carry qubits will be required for the addition of any two binary numbers.
In general, $\mathcal{S}$ requires at most $\left \lfloor \log_2 d \right \rfloor$ ancilla qubits for carrying operations, which directly comes from Eq.~(\ref{eqn:weighted_sum_k_upper_bound}).

\subsection{Weighted sum of multi-qubit integers}
In addition to summing up $d$ integers equally, a weight $w_i$ may also be added to each integer $a^{(i)}$. 
In that case, the weight matrix would be
\begin{equation}
\label{eqn:weighted_sum_add_integers_with_weights}
\Omega = \left( w_1 \cdot I_{n_1 \times n^*}^T, \dotsc, w_d \cdot I_{n_d \times n^*}^T \right)^T.
\end{equation}
In the case where $w_i$ are not integers, we can rescale the values represented on the quantum register by a common factor to make all weights integers. 
For example, if we are adding two numbers with weights $0.2$ and $0.8$, we could use integer weights of $w_1 = 1$ and $w_2 = 4$ instead, and reinterpret the resulting sum in postprocessing by dividing it by $5$.

\section{\label{sec:ga_circuit} Optimized Circuit for $\mathcal{QA}\ket{0}_3$}

In the following, we describe the circuit used for $\mathcal{QA}\ket{0}_3$ requiring only 18 CNOT gates.
We have that $\mathcal{Q} = -\mathcal{A}\mathcal{S}_0\mathcal{A}^\dagger\mathcal{S}_{\psi_0}$, where $\mathcal{S}_0 = 1 - 2\ket{0}\bra{0}$ and $\mathcal{S}_{\psi_0} = 1 - 2\ket{\psi_0}\ket{0}\bra{\psi_0}\bra{0}$ perform reflections on $\ket{0}_3$ and $\ket{\psi_0}_2\ket{0}$, respectively.
$S_{\psi_0}$ can be implemented up to a global phase using a single-qubit $Z$-gate on the last qubit, which is sufficient to differentiate between $\ket{\psi_0}\ket{0}$ and $\ket{\psi_1}\ket{1}$.
$\mathcal{S}_0$ is a bit more difficult and we use circuit synthesis for diagonal unitary matrices to achieve an efficient decomposition into gates \cite{Bullock2004}.
This construction lead to 16 CNOT gates for $\mathcal{Q}$ and 21 for $\mathcal{QA}$, which was still a bit too much to run on real hardware.

\begin{figure*}[htbp!]
\centering
\includegraphics[width=\textwidth]{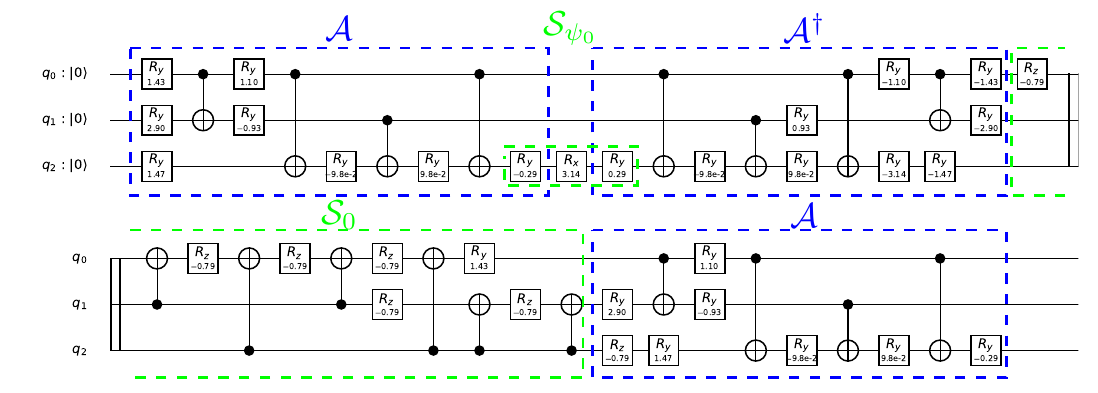}
\caption{\label{fig:qa_circuit} The optimized circuit for $\mathcal{QA}\ket{0}_3$ used for the experiments on real quantum hardware. It requires 18 CNOT gates and 33 single qubit gates. The initial spot price is assumed to be equal to 2. The dashed boxes indicate which parts are used for $\mathcal{A}$, $\mathcal{A}^{\dagger}$, $\mathcal{S}_{\psi_0}$, and $\mathcal{S}_0$. Note that due to the circuit optimization, some boxes are slightly overlapping.}
\end{figure*}

To further reduce the CNOT count, we look at the full circuit $\mathcal{QA}\ket{0}_3$ and we applied the following optimization steps.
The last part in $\mathcal{Q}$ is the application of $\mathcal{A}$.
As mentioned in Sec.~\ref{sec:hardware}, we can drop the very last CNOT gate and apply it in a classical postprocessing.
Furthermore, in $\mathcal{QA}\ket{0}_3$, we have $S_{\psi_0}$ between $\mathcal{A}$ and $\mathcal{A}^{\dagger}$, i.e. $\mathcal{A}^{\dagger} S_{\psi_0} \mathcal{A}$, where the uniformly controlled $Y$-rotations in $\mathcal{A}$ ($\mathcal{A}^{\dagger}$) are right before (after) $S_{\psi_0}$.
On the other hand, the $Z$-gate that implements $S_{\psi_0}$ can be decomposed into an $X$-rotation and a $Y$-rotation.
The $Y$-rotation can subsequently be absorbed into one of the uniformly controlled $Y$-rotations in $\mathcal{A}$ or $\mathcal{A}^{\dagger}$, changing the angles accordingly.
Since the remaining $X$-rotation commutes with the two neighboring CNOT gates from $\mathcal{A}$ and $\mathcal{A}^{\dagger}$, we can move the $X$-rotation so that the two CNOT gates cancel each other.
This reduces the CNOT gate count for $\mathcal{QA}\ket{0}_3$ to 18 and the resulting circuit is reported in Fig.~\ref{fig:qa_circuit}.

\bibliographystyle{plainnat}
\bibliography{option_pricing_review}

\begin{thebibliography}{55}%
\makeatletter
\providecommand \@ifxundefined [1]{%
 \@ifx{#1\undefined}
}%
\providecommand \@ifnum [1]{%
 \ifnum #1\expandafter \@firstoftwo
 \else \expandafter \@secondoftwo
 \fi
}%
\providecommand \@ifx [1]{%
 \ifx #1\expandafter \@firstoftwo
 \else \expandafter \@secondoftwo
 \fi
}%
\providecommand \natexlab [1]{#1}%
\providecommand \enquote  [1]{``#1''}%
\providecommand \bibnamefont  [1]{#1}%
\providecommand \bibfnamefont [1]{#1}%
\providecommand \citenamefont [1]{#1}%
\providecommand \href@noop [0]{\@secondoftwo}%
\providecommand \href [0]{\begingroup \@sanitize@url \@href}%
\providecommand \@href[1]{\@@startlink{#1}\@@href}%
\providecommand \@@href[1]{\endgroup#1\@@endlink}%
\providecommand \@sanitize@url [0]{\catcode `\\12\catcode `\$12\catcode
  `\&12\catcode `\#12\catcode `\^12\catcode `\_12\catcode `\%12\relax}%
\providecommand \@@startlink[1]{}%
\providecommand \@@endlink[0]{}%
\providecommand \url  [0]{\begingroup\@sanitize@url \@url }%
\providecommand \@url [1]{\endgroup\@href {#1}{\urlprefix }}%
\providecommand \urlprefix  [0]{URL }%
\providecommand \Eprint [0]{\href }%
\providecommand \doibase [0]{http://dx.doi.org/}%
\providecommand \selectlanguage [0]{\@gobble}%
\providecommand \bibinfo  [0]{\@secondoftwo}%
\providecommand \bibfield  [0]{\@secondoftwo}%
\providecommand \translation [1]{[#1]}%
\providecommand \BibitemOpen [0]{}%
\providecommand \bibitemStop [0]{}%
\providecommand \bibitemNoStop [0]{.\EOS\space}%
\providecommand \EOS [0]{\spacefactor3000\relax}%
\providecommand \BibitemShut  [1]{\csname bibitem#1\endcsname}%
\let\auto@bib@innerbib\@empty
\bibitem [{\citenamefont {Hull}(2006)}]{Hull}%
  \BibitemOpen
  \bibfield  {author} {\bibinfo {author} {\bibfnamefont {{John C.}}\
  \bibnamefont {Hull}},\ }\href {\doibase 10.1007/978-1-4419-9230-7_2} {\emph
  {\bibinfo {title} {Options, futures, and other derivatives}}},\ \bibinfo
  {edition} {6th}\ ed.\ (\bibinfo  {publisher} {Pearson Prentice Hall},\
  \bibinfo {address} {Upper Saddle River, NJ [u.a.]},\ \bibinfo {year}
  {2006})\BibitemShut {NoStop}%
\bibitem [{\citenamefont {Black}\ and\ \citenamefont
  {Scholes}(1973)}]{BlackScholes}%
  \BibitemOpen
  \bibfield  {author} {\bibinfo {author} {\bibfnamefont {Fischer}\ \bibnamefont
  {Black}}\ and\ \bibinfo {author} {\bibfnamefont {Myron}\ \bibnamefont
  {Scholes}},\ }\bibfield  {title} {\enquote {\bibinfo {title} {The pricing of
  options and corporate liabilities},}\ }\href {\doibase 10.1086/260062}
  {\bibfield  {journal} {\bibinfo  {journal} {Journal of Political Economy}\
  }\textbf {\bibinfo {volume} {81}},\ \bibinfo {pages} {637--654} (\bibinfo
  {year} {1973})}\BibitemShut {NoStop}%
\bibitem [{\citenamefont {Dupire}(1994)}]{Dupire1994}%
  \BibitemOpen
  \bibfield  {author} {\bibinfo {author} {\bibfnamefont {Bruno}\ \bibnamefont
  {Dupire}},\ }\bibfield  {title} {\enquote {\bibinfo {title} {Pricing with a
  smile},}\ }\href@noop {} {\bibfield  {journal} {\bibinfo  {journal} {Risk
  Magazine}\ ,\ \bibinfo {pages} {18--20}} (\bibinfo {year}
  {1994})}\BibitemShut {NoStop}%
\bibitem [{\citenamefont {Boyle}(1977)}]{Boyle1977}%
  \BibitemOpen
  \bibfield  {author} {\bibinfo {author} {\bibfnamefont {Phelim~P.}\
  \bibnamefont {Boyle}},\ }\bibfield  {title} {\enquote {\bibinfo {title}
  {{Options: A Monte Carlo approach}},}\ }\href {\doibase
  10.1016/0304-405X(77)90005-8} {\bibfield  {journal} {\bibinfo  {journal}
  {Journal of Financial Economics}\ }\textbf {\bibinfo {volume} {4}},\ \bibinfo
  {pages} {323--338} (\bibinfo {year} {1977})}\BibitemShut {NoStop}%
\bibitem [{\citenamefont {Glasserman}(2003)}]{Glasserman2003}%
  \BibitemOpen
  \bibfield  {author} {\bibinfo {author} {\bibfnamefont {Paul}\ \bibnamefont
  {Glasserman}},\ }\href {\doibase 10.1007/978-0-387-21617-1} {\emph {\bibinfo
  {title} {Monte Carlo Methods in Financial Engineering}}}\ (\bibinfo
  {publisher} {Springer-Verlag New York},\ \bibinfo {year} {2003})\ p.\
  \bibinfo {pages} {596}\BibitemShut {NoStop}%
\bibitem [{\citenamefont {Nielsen}\ and\ \citenamefont
  {Chuang}(2010)}]{Nielsen2010}%
  \BibitemOpen
  \bibfield  {author} {\bibinfo {author} {\bibfnamefont {Michael~A.}\
  \bibnamefont {Nielsen}}\ and\ \bibinfo {author} {\bibfnamefont {Isaac~L.}\
  \bibnamefont {Chuang}},\ }\href {\doibase 10.1017/CBO9780511976667} {\emph
  {\bibinfo {title} {Cambridge University Press}}}\ (\bibinfo {year} {2010})\
  p.\ \bibinfo {pages} {702}\BibitemShut {NoStop}%
\bibitem [{\citenamefont {Kandala}\ \emph {et~al.}(2017)\citenamefont
  {Kandala}, \citenamefont {Mezzacapo}, \citenamefont {Temme}, \citenamefont
  {Takita}, \citenamefont {Brink}, \citenamefont {Chow},\ and\ \citenamefont
  {Gambetta}}]{Kandala2017}%
  \BibitemOpen
  \bibfield  {author} {\bibinfo {author} {\bibfnamefont {Abhinav}\ \bibnamefont
  {Kandala}}, \bibinfo {author} {\bibfnamefont {Antonio}\ \bibnamefont
  {Mezzacapo}}, \bibinfo {author} {\bibfnamefont {Kristan}\ \bibnamefont
  {Temme}}, \bibinfo {author} {\bibfnamefont {Maika}\ \bibnamefont {Takita}},
  \bibinfo {author} {\bibfnamefont {Markus}\ \bibnamefont {Brink}}, \bibinfo
  {author} {\bibfnamefont {Jerry~M.}\ \bibnamefont {Chow}}, \ and\ \bibinfo
  {author} {\bibfnamefont {Jay~M.}\ \bibnamefont {Gambetta}},\ }\bibfield
  {title} {\enquote {\bibinfo {title} {Hardware-efficient variational quantum
  eigensolver for small molecules and quantum magnets},}\ }\href {\doibase
  10.1038/nature23879} {\bibfield  {journal} {\bibinfo  {journal} {Nature}\
  }\textbf {\bibinfo {volume} {549}},\ \bibinfo {pages} {242} (\bibinfo {year}
  {2017})}\BibitemShut {NoStop}%
\bibitem [{\citenamefont {Kandala}\ \emph {et~al.}(2019)\citenamefont
  {Kandala}, \citenamefont {Temme}, \citenamefont {D.~Corcoles}, \citenamefont
  {Mezzacapo}, \citenamefont {Chow},\ and\ \citenamefont
  {Gambetta}}]{Kandala2019}%
  \BibitemOpen
  \bibfield  {author} {\bibinfo {author} {\bibfnamefont {Abhinav}\ \bibnamefont
  {Kandala}}, \bibinfo {author} {\bibfnamefont {Kristan}\ \bibnamefont
  {Temme}}, \bibinfo {author} {\bibfnamefont {Antonio}\ \bibnamefont
  {D.~Corcoles}}, \bibinfo {author} {\bibfnamefont {Antonio}\ \bibnamefont
  {Mezzacapo}}, \bibinfo {author} {\bibfnamefont {Jerry~M.}\ \bibnamefont
  {Chow}}, \ and\ \bibinfo {author} {\bibfnamefont {Jay~M.}\ \bibnamefont
  {Gambetta}},\ }\bibfield  {title} {\enquote {\bibinfo {title} {Error
  mitigation extends the computational reach of a noisy quantum processor},}\
  }\href {\doibase 10.1038/s41586-019-1040-7} {\bibfield  {journal} {\bibinfo
  {journal} {Nature}\ }\textbf {\bibinfo {volume} {567}},\ \bibinfo {pages}
  {491--495} (\bibinfo {year} {2019})}\BibitemShut {NoStop}%
\bibitem [{\citenamefont {{Moll}}\ \emph {et~al.}(2018)\citenamefont {{Moll}},
  \citenamefont {{Barkoutsos}}, \citenamefont {{Bishop}}, \citenamefont
  {{Chow}}, \citenamefont {{Cross}}, \citenamefont {{Egger}}, \citenamefont
  {{Filipp}}, \citenamefont {{Fuhrer}}, \citenamefont {{Gambetta}},
  \citenamefont {{Ganzhorn}}, \citenamefont {{Kandala}}, \citenamefont
  {{Mezzacapo}}, \citenamefont {{M{\"u}ller}}, \citenamefont {{Riess}},
  \citenamefont {{Salis}}, \citenamefont {{Smolin}}, \citenamefont
  {{Tavernelli}},\ and\ \citenamefont {{Temme}}}]{Moll2018}%
  \BibitemOpen
  \bibfield  {author} {\bibinfo {author} {\bibfnamefont {N.}~\bibnamefont
  {{Moll}}}, \bibinfo {author} {\bibfnamefont {P.}~\bibnamefont
  {{Barkoutsos}}}, \bibinfo {author} {\bibfnamefont {L.~S.}\ \bibnamefont
  {{Bishop}}}, \bibinfo {author} {\bibfnamefont {J.~M.}\ \bibnamefont
  {{Chow}}}, \bibinfo {author} {\bibfnamefont {A.}~\bibnamefont {{Cross}}},
  \bibinfo {author} {\bibfnamefont {D.~J.}\ \bibnamefont {{Egger}}}, \bibinfo
  {author} {\bibfnamefont {S.}~\bibnamefont {{Filipp}}}, \bibinfo {author}
  {\bibfnamefont {A.}~\bibnamefont {{Fuhrer}}}, \bibinfo {author}
  {\bibfnamefont {J.~M.}\ \bibnamefont {{Gambetta}}}, \bibinfo {author}
  {\bibfnamefont {M.}~\bibnamefont {{Ganzhorn}}}, \bibinfo {author}
  {\bibfnamefont {A.}~\bibnamefont {{Kandala}}}, \bibinfo {author}
  {\bibfnamefont {A.}~\bibnamefont {{Mezzacapo}}}, \bibinfo {author}
  {\bibfnamefont {P.}~\bibnamefont {{M{\"u}ller}}}, \bibinfo {author}
  {\bibfnamefont {W.}~\bibnamefont {{Riess}}}, \bibinfo {author} {\bibfnamefont
  {G.}~\bibnamefont {{Salis}}}, \bibinfo {author} {\bibfnamefont
  {J.}~\bibnamefont {{Smolin}}}, \bibinfo {author} {\bibfnamefont
  {I.}~\bibnamefont {{Tavernelli}}}, \ and\ \bibinfo {author} {\bibfnamefont
  {K.}~\bibnamefont {{Temme}}},\ }\bibfield  {title} {\enquote {\bibinfo
  {title} {{Quantum optimization using variational algorithms on near-term
  quantum devices}},}\ }\href {\doibase 10.1088/2058-9565/aab822} {\bibfield
  {journal} {\bibinfo  {journal} {Quantum Science and Technology}\ }\textbf
  {\bibinfo {volume} {3}},\ \bibinfo {pages} {030503} (\bibinfo {year}
  {2018})}\BibitemShut {NoStop}%
\bibitem [{\citenamefont {Ganzhorn}\ \emph {et~al.}(2019)\citenamefont
  {Ganzhorn}, \citenamefont {Egger}, \citenamefont {Barkoutsos}, \citenamefont
  {Ollitrault}, \citenamefont {Salis}, \citenamefont {Moll}, \citenamefont
  {Roth}, \citenamefont {Fuhrer}, \citenamefont {Mueller}, \citenamefont
  {Woerner}, \citenamefont {Tavernelli},\ and\ \citenamefont
  {Filipp}}]{Ganzhorn2019}%
  \BibitemOpen
  \bibfield  {author} {\bibinfo {author} {\bibfnamefont {M.}~\bibnamefont
  {Ganzhorn}}, \bibinfo {author} {\bibfnamefont {D.J.}\ \bibnamefont {Egger}},
  \bibinfo {author} {\bibfnamefont {P.}~\bibnamefont {Barkoutsos}}, \bibinfo
  {author} {\bibfnamefont {P.}~\bibnamefont {Ollitrault}}, \bibinfo {author}
  {\bibfnamefont {G.}~\bibnamefont {Salis}}, \bibinfo {author} {\bibfnamefont
  {N.}~\bibnamefont {Moll}}, \bibinfo {author} {\bibfnamefont {M.}~\bibnamefont
  {Roth}}, \bibinfo {author} {\bibfnamefont {A.}~\bibnamefont {Fuhrer}},
  \bibinfo {author} {\bibfnamefont {P.}~\bibnamefont {Mueller}}, \bibinfo
  {author} {\bibfnamefont {S.}~\bibnamefont {Woerner}}, \bibinfo {author}
  {\bibfnamefont {I.}~\bibnamefont {Tavernelli}}, \ and\ \bibinfo {author}
  {\bibfnamefont {S.}~\bibnamefont {Filipp}},\ }\bibfield  {title} {\enquote
  {\bibinfo {title} {Gate-efficient simulation of molecular eigenstates on a
  quantum computer},}\ }\href {\doibase 10.1103/PhysRevApplied.11.044092}
  {\bibfield  {journal} {\bibinfo  {journal} {Phys. Rev. Applied}\ }\textbf
  {\bibinfo {volume} {11}},\ \bibinfo {pages} {044092} (\bibinfo {year}
  {2019})}\BibitemShut {NoStop}%
\bibitem [{\citenamefont {Harrow}\ \emph {et~al.}(2009)\citenamefont {Harrow},
  \citenamefont {Hassidim},\ and\ \citenamefont {Lloyd}}]{Harrow2009b}%
  \BibitemOpen
  \bibfield  {author} {\bibinfo {author} {\bibfnamefont {Aram~W.}\ \bibnamefont
  {Harrow}}, \bibinfo {author} {\bibfnamefont {Avinatan}\ \bibnamefont
  {Hassidim}}, \ and\ \bibinfo {author} {\bibfnamefont {Seth}\ \bibnamefont
  {Lloyd}},\ }\bibfield  {title} {\enquote {\bibinfo {title} {Quantum algorithm
  for linear systems of equations},}\ }\href {\doibase
  10.1103/PhysRevLett.103.150502} {\bibfield  {journal} {\bibinfo  {journal}
  {Phys. Rev. Lett.}\ }\textbf {\bibinfo {volume} {103}},\ \bibinfo {pages}
  {150502} (\bibinfo {year} {2009})}\BibitemShut {NoStop}%
\bibitem [{\citenamefont {Lloyd}\ \emph {et~al.}(2014)\citenamefont {Lloyd},
  \citenamefont {Mohseni},\ and\ \citenamefont {Rebentrost}}]{Lloyd2014}%
  \BibitemOpen
  \bibfield  {author} {\bibinfo {author} {\bibfnamefont {Seth}\ \bibnamefont
  {Lloyd}}, \bibinfo {author} {\bibfnamefont {Masoud}\ \bibnamefont {Mohseni}},
  \ and\ \bibinfo {author} {\bibfnamefont {Patrick}\ \bibnamefont
  {Rebentrost}},\ }\bibfield  {title} {\enquote {\bibinfo {title} {Quantum
  principal component analysis},}\ }\href {\doibase 10.1038/nphys3029}
  {\bibfield  {journal} {\bibinfo  {journal} {Nature Physics}\ }\textbf
  {\bibinfo {volume} {10}},\ \bibinfo {pages} {631--633} (\bibinfo {year}
  {2014})}\BibitemShut {NoStop}%
\bibitem [{\citenamefont {Biamonte}\ \emph {et~al.}(2017)\citenamefont
  {Biamonte}, \citenamefont {Wittek}, \citenamefont {Pancotti}, \citenamefont
  {Rebentrost}, \citenamefont {Wiebe},\ and\ \citenamefont
  {Lloyd}}]{Biamonte2017}%
  \BibitemOpen
  \bibfield  {author} {\bibinfo {author} {\bibfnamefont {Jacob}\ \bibnamefont
  {Biamonte}}, \bibinfo {author} {\bibfnamefont {Peter}\ \bibnamefont
  {Wittek}}, \bibinfo {author} {\bibfnamefont {Nicola}\ \bibnamefont
  {Pancotti}}, \bibinfo {author} {\bibfnamefont {Patrick}\ \bibnamefont
  {Rebentrost}}, \bibinfo {author} {\bibfnamefont {Nathan}\ \bibnamefont
  {Wiebe}}, \ and\ \bibinfo {author} {\bibfnamefont {Seth}\ \bibnamefont
  {Lloyd}},\ }\bibfield  {title} {\enquote {\bibinfo {title} {{Quantum machine
  learning}},}\ }\href {\doibase 10.1038/nature23474} {\bibfield  {journal}
  {\bibinfo  {journal} {Nature}\ }\textbf {\bibinfo {volume} {549}},\ \bibinfo
  {pages} {195--202} (\bibinfo {year} {2017})}\BibitemShut {NoStop}%
\bibitem [{\citenamefont {Havlicek}\ \emph {et~al.}(2019)\citenamefont
  {Havlicek}, \citenamefont {Corcoles}, \citenamefont {Temme}, \citenamefont
  {Harrow}, \citenamefont {Kandala}, \citenamefont {Chow},\ and\ \citenamefont
  {Gambetta}}]{Havlicek2019}%
  \BibitemOpen
  \bibfield  {author} {\bibinfo {author} {\bibfnamefont {Vojtech}\ \bibnamefont
  {Havlicek}}, \bibinfo {author} {\bibfnamefont {Antonio~D.}\ \bibnamefont
  {Corcoles}}, \bibinfo {author} {\bibfnamefont {Kristan}\ \bibnamefont
  {Temme}}, \bibinfo {author} {\bibfnamefont {Aram~W.}\ \bibnamefont {Harrow}},
  \bibinfo {author} {\bibfnamefont {Abhinav}\ \bibnamefont {Kandala}}, \bibinfo
  {author} {\bibfnamefont {Jerry~M.}\ \bibnamefont {Chow}}, \ and\ \bibinfo
  {author} {\bibfnamefont {Jay~M.}\ \bibnamefont {Gambetta}},\ }\bibfield
  {title} {\enquote {\bibinfo {title} {Supervised learning with
  quantum-enhanced feature spaces},}\ }\href {\doibase
  10.1038/s41586-019-0980-2} {\bibfield  {journal} {\bibinfo  {journal}
  {Nature}\ }\textbf {\bibinfo {volume} {567}},\ \bibinfo {pages} {209 -- 212}
  (\bibinfo {year} {2019})}\BibitemShut {NoStop}%
\bibitem [{\citenamefont {Orus}\ \emph {et~al.}(2019)\citenamefont {Orus},
  \citenamefont {Mugel},\ and\ \citenamefont {Lizaso}}]{Orus2019}%
  \BibitemOpen
  \bibfield  {author} {\bibinfo {author} {\bibfnamefont {Roman}\ \bibnamefont
  {Orus}}, \bibinfo {author} {\bibfnamefont {Samuel}\ \bibnamefont {Mugel}}, \
  and\ \bibinfo {author} {\bibfnamefont {Enrique}\ \bibnamefont {Lizaso}},\
  }\bibfield  {title} {\enquote {\bibinfo {title} {Quantum computing for
  finance: Overview and prospects},}\ }\href {\doibase
  10.1016/j.revip.2019.100028} {\bibfield  {journal} {\bibinfo  {journal}
  {Reviews in Physics}\ }\textbf {\bibinfo {volume} {4}},\ \bibinfo {pages}
  {100028} (\bibinfo {year} {2019})}\BibitemShut {NoStop}%
\bibitem [{\citenamefont {Rebentrost}\ and\ \citenamefont
  {Lloyd}(2018)}]{Rebentrost2018b}%
  \BibitemOpen
  \bibfield  {author} {\bibinfo {author} {\bibfnamefont {Patrick}\ \bibnamefont
  {Rebentrost}}\ and\ \bibinfo {author} {\bibfnamefont {Seth}\ \bibnamefont
  {Lloyd}},\ }\bibfield  {title} {\enquote {\bibinfo {title} {Quantum
  computational finance: quantum algorithm for portfolio optimization},}\
  }\href@noop {} {\  (\bibinfo {year} {2018})},\ \Eprint
  {http://arxiv.org/abs/1811.03975} {arXiv:1811.03975} \BibitemShut {NoStop}%
\bibitem [{\citenamefont {Woerner}\ and\ \citenamefont
  {Egger}(2019)}]{Woerner2019}%
  \BibitemOpen
  \bibfield  {author} {\bibinfo {author} {\bibfnamefont {Stefan}\ \bibnamefont
  {Woerner}}\ and\ \bibinfo {author} {\bibfnamefont {{Daniel~J.}}\ \bibnamefont
  {Egger}},\ }\bibfield  {title} {\enquote {\bibinfo {title} {Quantum risk
  analysis},}\ }\href {\doibase 10.1038/s41534-019-0130-6} {\bibfield
  {journal} {\bibinfo  {journal} {npj Quantum Information}\ }\textbf {\bibinfo
  {volume} {5}},\ \bibinfo {pages} {15} (\bibinfo {year} {2019})}\BibitemShut
  {NoStop}%
\bibitem [{\citenamefont {Rebentrost}\ \emph {et~al.}(2018)\citenamefont
  {Rebentrost}, \citenamefont {Gupt},\ and\ \citenamefont
  {Bromley}}]{Rebentrost2018}%
  \BibitemOpen
  \bibfield  {author} {\bibinfo {author} {\bibfnamefont {Patrick}\ \bibnamefont
  {Rebentrost}}, \bibinfo {author} {\bibfnamefont {Brajesh}\ \bibnamefont
  {Gupt}}, \ and\ \bibinfo {author} {\bibfnamefont {Thomas~R.}\ \bibnamefont
  {Bromley}},\ }\bibfield  {title} {\enquote {\bibinfo {title} {Quantum
  computational finance: Monte carlo pricing of financial derivatives},}\
  }\href {\doibase 10.1103/PhysRevA.98.022321} {\bibfield  {journal} {\bibinfo
  {journal} {Phys. Rev. A}\ }\textbf {\bibinfo {volume} {98}},\ \bibinfo
  {pages} {022321} (\bibinfo {year} {2018})}\BibitemShut {NoStop}%
\bibitem [{\citenamefont {Zoufal}\ \emph {et~al.}(2019)\citenamefont {Zoufal},
  \citenamefont {Lucchi},\ and\ \citenamefont {Woerner}}]{Zoufal2019}%
  \BibitemOpen
  \bibfield  {author} {\bibinfo {author} {\bibfnamefont {Christa}\ \bibnamefont
  {Zoufal}}, \bibinfo {author} {\bibfnamefont {Aur{\'e}lien}\ \bibnamefont
  {Lucchi}}, \ and\ \bibinfo {author} {\bibfnamefont {Stefan}\ \bibnamefont
  {Woerner}},\ }\bibfield  {title} {\enquote {\bibinfo {title} {Quantum
  generative adversarial networks for learning and loading random
  distributions},}\ }\href {\doibase 10.1038/s41534-019-0223-2} {\bibfield
  {journal} {\bibinfo  {journal} {npj Quantum Information}\ }\textbf {\bibinfo
  {volume} {5}},\ \bibinfo {pages} {1--9} (\bibinfo {year} {2019})}\BibitemShut
  {NoStop}%
\bibitem [{\citenamefont {Martin}\ \emph {et~al.}(2019)\citenamefont {Martin},
  \citenamefont {Candelas}, \citenamefont {Rodriguez-Rozas}, \citenamefont
  {D.~Martin-Guerrero}, \citenamefont {Chen}, \citenamefont {Lamata},
  \citenamefont {Orus}, \citenamefont {Solano},\ and\ \citenamefont
  {Sanz}}]{Martin2019}%
  \BibitemOpen
  \bibfield  {author} {\bibinfo {author} {\bibfnamefont {Ana}\ \bibnamefont
  {Martin}}, \bibinfo {author} {\bibfnamefont {Bruno}\ \bibnamefont
  {Candelas}}, \bibinfo {author} {\bibfnamefont {Angel}\ \bibnamefont
  {Rodriguez-Rozas}}, \bibinfo {author} {\bibfnamefont {Jose}\ \bibnamefont
  {D.~Martin-Guerrero}}, \bibinfo {author} {\bibfnamefont {Xi}~\bibnamefont
  {Chen}}, \bibinfo {author} {\bibfnamefont {Lucas}\ \bibnamefont {Lamata}},
  \bibinfo {author} {\bibfnamefont {Roman}\ \bibnamefont {Orus}}, \bibinfo
  {author} {\bibfnamefont {Enrique}\ \bibnamefont {Solano}}, \ and\ \bibinfo
  {author} {\bibfnamefont {Mikel}\ \bibnamefont {Sanz}},\ }\bibfield  {title}
  {\enquote {\bibinfo {title} {Towards pricing financial derivatives with an
  ibm quantum computer},}\ }\href@noop {} {\  (\bibinfo {year} {2019})},\
  \Eprint {http://arxiv.org/abs/1904.05803} {arXiv:1904.05803} \BibitemShut
  {NoStop}%
\bibitem [{\citenamefont {Brassard}\ \emph {et~al.}(2002)\citenamefont
  {Brassard}, \citenamefont {Hoyer}, \citenamefont {Mosca},\ and\ \citenamefont
  {Tapp}}]{Brassard2000}%
  \BibitemOpen
  \bibfield  {author} {\bibinfo {author} {\bibfnamefont {Gilles}\ \bibnamefont
  {Brassard}}, \bibinfo {author} {\bibfnamefont {Peter}\ \bibnamefont {Hoyer}},
  \bibinfo {author} {\bibfnamefont {Michele}\ \bibnamefont {Mosca}}, \ and\
  \bibinfo {author} {\bibfnamefont {Alain}\ \bibnamefont {Tapp}},\ }\bibfield
  {title} {\enquote {\bibinfo {title} {{Quantum Amplitude Amplification and
  Estimation}},}\ }\href {\doibase 10.1090/conm/305/05215} {\bibfield
  {journal} {\bibinfo  {journal} {Contemporary Mathematics}\ }\textbf {\bibinfo
  {volume} {305}} (\bibinfo {year} {2002}),\
  10.1090/conm/305/05215}\BibitemShut {NoStop}%
\bibitem [{\citenamefont {Suzuki}\ \emph {et~al.}(2020)\citenamefont {Suzuki},
  \citenamefont {Uno}, \citenamefont {Raymond}, \citenamefont {Tanaka},
  \citenamefont {Onodera},\ and\ \citenamefont {Yamamoto}}]{Suzuki2020}%
  \BibitemOpen
  \bibfield  {author} {\bibinfo {author} {\bibfnamefont {Yohichi}\ \bibnamefont
  {Suzuki}}, \bibinfo {author} {\bibfnamefont {Shumpei}\ \bibnamefont {Uno}},
  \bibinfo {author} {\bibfnamefont {Rudy}\ \bibnamefont {Raymond}}, \bibinfo
  {author} {\bibfnamefont {Tomoki}\ \bibnamefont {Tanaka}}, \bibinfo {author}
  {\bibfnamefont {Tamiya}\ \bibnamefont {Onodera}}, \ and\ \bibinfo {author}
  {\bibfnamefont {Naoki}\ \bibnamefont {Yamamoto}},\ }\bibfield  {title}
  {\enquote {\bibinfo {title} {Amplitude estimation without phase
  estimation},}\ }\href {\doibase 10.1007/s11128-019-2565-2} {\bibfield
  {journal} {\bibinfo  {journal} {Quantum Information Processing}\ }\textbf
  {\bibinfo {volume} {19}},\ \bibinfo {pages} {75} (\bibinfo {year}
  {2020})}\BibitemShut {NoStop}%
\bibitem [{\citenamefont {Rubinstein}(1981)}]{Rubinstein1981}%
  \BibitemOpen
  \bibfield  {author} {\bibinfo {author} {\bibfnamefont {Reuven~Y.}\
  \bibnamefont {Rubinstein}},\ }\href {\doibase 10.1002/9780470316511} {\emph
  {\bibinfo {title} {Simulation and the Monte Carlo Method}}},\ Wiley Series in
  Probability and Statistics\ (\bibinfo  {publisher} {Wiley},\ \bibinfo {year}
  {1981})\BibitemShut {NoStop}%
\bibitem [{\citenamefont {Abrams}\ and\ \citenamefont
  {Williams}(1999)}]{Abrams1999}%
  \BibitemOpen
  \bibfield  {author} {\bibinfo {author} {\bibfnamefont {Daniel~S}\
  \bibnamefont {Abrams}}\ and\ \bibinfo {author} {\bibfnamefont {Colin~P}\
  \bibnamefont {Williams}},\ }\bibfield  {title} {\enquote {\bibinfo {title}
  {{Fast quantum algorithms for numerical integrals and stochastic
  processes}},}\ }\href@noop {} {\  (\bibinfo {year} {1999})},\ \Eprint
  {http://arxiv.org/abs/quant-ph/9908083} {arxiv:quant-ph/9908083} \BibitemShut
  {NoStop}%
\bibitem [{\citenamefont {Montanaro}(2015)}]{Montanaro2017}%
  \BibitemOpen
  \bibfield  {author} {\bibinfo {author} {\bibfnamefont {Ashley}\ \bibnamefont
  {Montanaro}},\ }\bibfield  {title} {\enquote {\bibinfo {title} {Quantum
  speedup of monte carlo methods},}\ }\href {\doibase 10.1098/rspa.2015.0301}
  {\bibfield  {journal} {\bibinfo  {journal} {Proceedings of the Royal Society
  of London A: Mathematical, Physical and Engineering Sciences}\ }\textbf
  {\bibinfo {volume} {471}} (\bibinfo {year} {2015}),\
  10.1098/rspa.2015.0301}\BibitemShut {NoStop}%
\bibitem [{\citenamefont {Kitaev}(1995)}]{Kitaev1995}%
  \BibitemOpen
  \bibfield  {author} {\bibinfo {author} {\bibfnamefont {A.~Yu.}\ \bibnamefont
  {Kitaev}},\ }\bibfield  {title} {\enquote {\bibinfo {title} {{Quantum
  measurements and the Abelian Stabilizer Problem}},}\ }\href@noop {} {\
  (\bibinfo {year} {1995})},\ \Eprint
  {http://arxiv.org/abs/arXiv:quant-ph/9511026} {arXiv:quant-ph/9511026}
  \BibitemShut {NoStop}%
\bibitem [{\citenamefont {Cleve}\ \emph {et~al.}(1998)\citenamefont {Cleve},
  \citenamefont {Ekert}, \citenamefont {Macchiavello},\ and\ \citenamefont
  {Mosca}}]{Cleve1998}%
  \BibitemOpen
  \bibfield  {author} {\bibinfo {author} {\bibfnamefont {R.}~\bibnamefont
  {Cleve}}, \bibinfo {author} {\bibfnamefont {A.}~\bibnamefont {Ekert}},
  \bibinfo {author} {\bibfnamefont {C.}~\bibnamefont {Macchiavello}}, \ and\
  \bibinfo {author} {\bibfnamefont {M.}~\bibnamefont {Mosca}},\ }\bibfield
  {title} {\enquote {\bibinfo {title} {Quantum algorithms revisited},}\ }\href
  {\doibase 10.1098/rspa.1998.0164} {\bibfield  {journal} {\bibinfo  {journal}
  {Proceedings of the Royal Society of London. Series A: Mathematical, Physical
  and Engineering Sciences}\ }\textbf {\bibinfo {volume} {454}},\ \bibinfo
  {pages} {339--354} (\bibinfo {year} {1998})}\BibitemShut {NoStop}%
\bibitem [{\citenamefont {Glasserman}\ \emph {et~al.}(2000)\citenamefont
  {Glasserman}, \citenamefont {Heidelberger},\ and\ \citenamefont
  {Shahabuddin}}]{Glasserman2000}%
  \BibitemOpen
  \bibfield  {author} {\bibinfo {author} {\bibfnamefont {Paul}\ \bibnamefont
  {Glasserman}}, \bibinfo {author} {\bibfnamefont {Philip}\ \bibnamefont
  {Heidelberger}}, \ and\ \bibinfo {author} {\bibfnamefont {Perwez}\
  \bibnamefont {Shahabuddin}},\ }\bibfield  {title} {\enquote {\bibinfo {title}
  {{Efficient Monte Carlo Methods for Value-at-Risk}},}\ }in\ \href@noop {}
  {\emph {\bibinfo {booktitle} {Mastering Risk}}},\ Vol.~\bibinfo {volume} {2}\
  (\bibinfo {year} {2000})\ pp.\ \bibinfo {pages} {5--18}\BibitemShut {NoStop}%
\bibitem [{\citenamefont {Merton}(1973)}]{Merton1973}%
  \BibitemOpen
  \bibfield  {author} {\bibinfo {author} {\bibfnamefont {Robert~C.}\
  \bibnamefont {Merton}},\ }\bibfield  {title} {\enquote {\bibinfo {title}
  {Theory of rational option pricing},}\ }\href {\doibase 10.2307/3003143}
  {\bibfield  {journal} {\bibinfo  {journal} {The Bell Journal of Economics and
  Management Science}\ }\textbf {\bibinfo {volume} {4}},\ \bibinfo {pages}
  {141--183} (\bibinfo {year} {1973})}\BibitemShut {NoStop}%
\bibitem [{\citenamefont {Grover}\ and\ \citenamefont
  {Rudolph}(2002)}]{Grover2002}%
  \BibitemOpen
  \bibfield  {author} {\bibinfo {author} {\bibfnamefont {Lov}\ \bibnamefont
  {Grover}}\ and\ \bibinfo {author} {\bibfnamefont {Terry}\ \bibnamefont
  {Rudolph}},\ }\bibfield  {title} {\enquote {\bibinfo {title} {{Creating
  superpositions that correspond to efficiently integrable probability
  distributions}},}\ }\href@noop {} {\  (\bibinfo {year} {2002})},\ \Eprint
  {http://arxiv.org/abs/quant-ph/0208112} {arXiv:quant-ph/0208112} \BibitemShut
  {NoStop}%
\bibitem [{\citenamefont {Koshiyama}\ \emph {et~al.}(2019)\citenamefont
  {Koshiyama}, \citenamefont {Firoozye},\ and\ \citenamefont
  {Treleaven}}]{Koshiyama2019}%
  \BibitemOpen
  \bibfield  {author} {\bibinfo {author} {\bibfnamefont {Adriano}\ \bibnamefont
  {Koshiyama}}, \bibinfo {author} {\bibfnamefont {Nick}\ \bibnamefont
  {Firoozye}}, \ and\ \bibinfo {author} {\bibfnamefont {Philip}\ \bibnamefont
  {Treleaven}},\ }\bibfield  {title} {\enquote {\bibinfo {title} {Generative
  adversarial networks for financial trading strategies fine-tuning and
  combination},}\ }\href@noop {} {\  (\bibinfo {year} {2019})},\ \Eprint
  {http://arxiv.org/abs/arXiv:1901.01751} {arXiv:1901.01751} \BibitemShut
  {NoStop}%
\bibitem [{\citenamefont {Horvath}\ \emph {et~al.}(2019)\citenamefont
  {Horvath}, \citenamefont {Muguruza},\ and\ \citenamefont
  {Tomas}}]{Horvath2019}%
  \BibitemOpen
  \bibfield  {author} {\bibinfo {author} {\bibfnamefont {Blanka}\ \bibnamefont
  {Horvath}}, \bibinfo {author} {\bibfnamefont {Aitor}\ \bibnamefont
  {Muguruza}}, \ and\ \bibinfo {author} {\bibfnamefont {Mehdi}\ \bibnamefont
  {Tomas}},\ }\bibfield  {title} {\enquote {\bibinfo {title} {Deep learning
  volatility},}\ }\href {\doibase 10.2139/ssrn.3322085} {\bibfield  {journal}
  {\bibinfo  {journal} {SSRN Electronic Journal}\ } (\bibinfo {year} {2019}),\
  10.2139/ssrn.3322085}\BibitemShut {NoStop}%
\bibitem [{\citenamefont {Plesch}\ and\ \citenamefont
  {Brukner}(2011)}]{Plesch2010}%
  \BibitemOpen
  \bibfield  {author} {\bibinfo {author} {\bibfnamefont {Martin}\ \bibnamefont
  {Plesch}}\ and\ \bibinfo {author} {\bibfnamefont {\ifmmode
  \check{C}\else~\v{C}\fi{}aslav}\ \bibnamefont {Brukner}},\ }\bibfield
  {title} {\enquote {\bibinfo {title} {Quantum-state preparation with universal
  gate decompositions},}\ }\href {\doibase 10.1103/PhysRevA.83.032302}
  {\bibfield  {journal} {\bibinfo  {journal} {Phys. Rev. A}\ }\textbf {\bibinfo
  {volume} {83}},\ \bibinfo {pages} {032302} (\bibinfo {year}
  {2011})}\BibitemShut {NoStop}%
\bibitem [{\citenamefont {Goodfellow}\ \emph {et~al.}(2014)\citenamefont
  {Goodfellow}, \citenamefont {Pouget-Abadie}, \citenamefont {Mirza},
  \citenamefont {Xu}, \citenamefont {Warde-Farley}, \citenamefont {Ozair},
  \citenamefont {Courville},\ and\ \citenamefont {Bengio}}]{Goodfellow2014}%
  \BibitemOpen
  \bibfield  {author} {\bibinfo {author} {\bibfnamefont {Ian}\ \bibnamefont
  {Goodfellow}}, \bibinfo {author} {\bibfnamefont {Jean}\ \bibnamefont
  {Pouget-Abadie}}, \bibinfo {author} {\bibfnamefont {Mehdi}\ \bibnamefont
  {Mirza}}, \bibinfo {author} {\bibfnamefont {Bing}\ \bibnamefont {Xu}},
  \bibinfo {author} {\bibfnamefont {David}\ \bibnamefont {Warde-Farley}},
  \bibinfo {author} {\bibfnamefont {Sherjil}\ \bibnamefont {Ozair}}, \bibinfo
  {author} {\bibfnamefont {Aaron}\ \bibnamefont {Courville}}, \ and\ \bibinfo
  {author} {\bibfnamefont {Yoshua}\ \bibnamefont {Bengio}},\ }\bibfield
  {title} {\enquote {\bibinfo {title} {Generative adversarial nets},}\ }in\
  \href@noop {} {\emph {\bibinfo {booktitle} {Advances in Neural Information
  Processing Systems 27}}},\ \bibinfo {editor} {edited by\ \bibinfo {editor}
  {\bibfnamefont {Z.}~\bibnamefont {Ghahramani}}, \bibinfo {editor}
  {\bibfnamefont {M.}~\bibnamefont {Welling}}, \bibinfo {editor} {\bibfnamefont
  {C.}~\bibnamefont {Cortes}}, \bibinfo {editor} {\bibfnamefont {N.~D.}\
  \bibnamefont {Lawrence}}, \ and\ \bibinfo {editor} {\bibfnamefont {K.~Q.}\
  \bibnamefont {Weinberger}}}\ (\bibinfo  {publisher} {Curran Associates,
  Inc.},\ \bibinfo {year} {2014})\ pp.\ \bibinfo {pages}
  {2672--2680}\BibitemShut {NoStop}%
\bibitem [{\citenamefont {Barenco}\ \emph {et~al.}(1995)\citenamefont
  {Barenco}, \citenamefont {Bennett}, \citenamefont {Cleve}, \citenamefont
  {Divincenzo}, \citenamefont {Margolus}, \citenamefont {Shor}, \citenamefont
  {Sleator}, \citenamefont {Smolin},\ and\ \citenamefont
  {Weinfurter}}]{Barenco1995}%
  \BibitemOpen
  \bibfield  {author} {\bibinfo {author} {\bibfnamefont {Adriano}\ \bibnamefont
  {Barenco}}, \bibinfo {author} {\bibfnamefont {Charles~H.}\ \bibnamefont
  {Bennett}}, \bibinfo {author} {\bibfnamefont {Richard}\ \bibnamefont
  {Cleve}}, \bibinfo {author} {\bibfnamefont {David~P.}\ \bibnamefont
  {Divincenzo}}, \bibinfo {author} {\bibfnamefont {Norman}\ \bibnamefont
  {Margolus}}, \bibinfo {author} {\bibfnamefont {Peter}\ \bibnamefont {Shor}},
  \bibinfo {author} {\bibfnamefont {Tycho}\ \bibnamefont {Sleator}}, \bibinfo
  {author} {\bibfnamefont {John~A.}\ \bibnamefont {Smolin}}, \ and\ \bibinfo
  {author} {\bibfnamefont {Harald}\ \bibnamefont {Weinfurter}},\ }\bibfield
  {title} {\enquote {\bibinfo {title} {{Elementary gates for quantum
  computation}},}\ }\href {\doibase 10.1103/PhysRevA.52.3457} {\bibfield
  {journal} {\bibinfo  {journal} {Phys. Rev. A}\ }\textbf {\bibinfo {volume}
  {52}},\ \bibinfo {pages} {3457--3467} (\bibinfo {year} {1995})}\BibitemShut
  {NoStop}%
\bibitem [{\citenamefont {Cuccaro}\ \emph {et~al.}(2004)\citenamefont
  {Cuccaro}, \citenamefont {Draper}, \citenamefont {Kutin},\ and\ \citenamefont
  {Moulton}}]{Cuccaro2004}%
  \BibitemOpen
  \bibfield  {author} {\bibinfo {author} {\bibfnamefont {Steven~A}\
  \bibnamefont {Cuccaro}}, \bibinfo {author} {\bibfnamefont {Thomas~G}\
  \bibnamefont {Draper}}, \bibinfo {author} {\bibfnamefont {Samuel~A}\
  \bibnamefont {Kutin}}, \ and\ \bibinfo {author} {\bibfnamefont
  {David~Petrie}\ \bibnamefont {Moulton}},\ }\bibfield  {title} {\enquote
  {\bibinfo {title} {{A new quantum ripple-carry addition circuit}},}\
  }\href@noop {} {\  (\bibinfo {year} {2004})},\ \Eprint
  {http://arxiv.org/abs/quant-ph/0410184} {arxiv:quant-ph/0410184} \BibitemShut
  {NoStop}%
\bibitem [{\citenamefont {Aleksandrowicz}\ \emph {et~al.}(2019)\citenamefont
  {Aleksandrowicz}, \citenamefont {Alexander}, \citenamefont {Barkoutsos},
  \citenamefont {Bello}, \citenamefont {Ben-Haim}, \citenamefont {Bucher},
  \citenamefont {Cabrera-Hern{\'a}ndez}, \citenamefont {Carballo-Franquis},
  \citenamefont {Chen}, \citenamefont {Chen}, \citenamefont {Chow},
  \citenamefont {C{\'o}rcoles-Gonzales}, \citenamefont {Cross}, \citenamefont
  {Cross}, \citenamefont {Cruz-Benito}, \citenamefont {Culver}, \citenamefont
  {Gonz{\'a}lez}, \citenamefont {Torre}, \citenamefont {Ding}, \citenamefont
  {Dumitrescu}, \citenamefont {Duran}, \citenamefont {Eendebak}, \citenamefont
  {Everitt}, \citenamefont {Sertage}, \citenamefont {Frisch}, \citenamefont
  {Fuhrer}, \citenamefont {Gambetta}, \citenamefont {Gago}, \citenamefont
  {Gomez-Mosquera}, \citenamefont {Greenberg}, \citenamefont {Hamamura},
  \citenamefont {Havlicek}, \citenamefont {Hellmers}, \citenamefont {Herok},
  \citenamefont {Horii}, \citenamefont {Hu}, \citenamefont {Imamichi},
  \citenamefont {Itoko}, \citenamefont {Javadi-Abhari}, \citenamefont
  {Kanazawa}, \citenamefont {Karazeev}, \citenamefont {Krsulich}, \citenamefont
  {Liu}, \citenamefont {Luh}, \citenamefont {Maeng}, \citenamefont {Marques},
  \citenamefont {Mart{\'\i}n-Fern{\'a}ndez}, \citenamefont {McClure},
  \citenamefont {McKay}, \citenamefont {Meesala}, \citenamefont {Mezzacapo},
  \citenamefont {Moll}, \citenamefont {Rodr{\'\i}guez}, \citenamefont
  {Nannicini}, \citenamefont {Nation}, \citenamefont {Ollitrault},
  \citenamefont {O'Riordan}, \citenamefont {Paik}, \citenamefont {P{\'e}rez},
  \citenamefont {Phan}, \citenamefont {Pistoia}, \citenamefont {Prutyanov},
  \citenamefont {Reuter}, \citenamefont {Rice}, \citenamefont {Davila},
  \citenamefont {Rudy}, \citenamefont {Ryu}, \citenamefont {Sathaye},
  \citenamefont {Schnabel}, \citenamefont {Schoute}, \citenamefont {Setia},
  \citenamefont {Shi}, \citenamefont {Silva}, \citenamefont {Siraichi},
  \citenamefont {Sivarajah}, \citenamefont {Smolin}, \citenamefont {Soeken},
  \citenamefont {Takahashi}, \citenamefont {Tavernelli}, \citenamefont
  {Taylor}, \citenamefont {Taylour}, \citenamefont {Trabing}, \citenamefont
  {Treinish}, \citenamefont {Turner}, \citenamefont {Vogt-Lee}, \citenamefont
  {Vuillot}, \citenamefont {Wildstrom}, \citenamefont {Wilson}, \citenamefont
  {Winston}, \citenamefont {Wood}, \citenamefont {Wood}, \citenamefont
  {W{\"o}rner}, \citenamefont {Akhalwaya},\ and\ \citenamefont
  {Zoufal}}]{Qiskit}%
  \BibitemOpen
  \bibfield  {author} {\bibinfo {author} {\bibfnamefont {Gadi}\ \bibnamefont
  {Aleksandrowicz}}, \bibinfo {author} {\bibfnamefont {Thomas}\ \bibnamefont
  {Alexander}}, \bibinfo {author} {\bibfnamefont {Panagiotis}\ \bibnamefont
  {Barkoutsos}}, \bibinfo {author} {\bibfnamefont {Luciano}\ \bibnamefont
  {Bello}}, \bibinfo {author} {\bibfnamefont {Yael}\ \bibnamefont {Ben-Haim}},
  \bibinfo {author} {\bibfnamefont {David}\ \bibnamefont {Bucher}}, \bibinfo
  {author} {\bibfnamefont {Francisco~Jose}\ \bibnamefont
  {Cabrera-Hern{\'a}ndez}}, \bibinfo {author} {\bibfnamefont {Jorge}\
  \bibnamefont {Carballo-Franquis}}, \bibinfo {author} {\bibfnamefont {Adrian}\
  \bibnamefont {Chen}}, \bibinfo {author} {\bibfnamefont {Chun-Fu}\
  \bibnamefont {Chen}}, \bibinfo {author} {\bibfnamefont {Jerry~M.}\
  \bibnamefont {Chow}}, \bibinfo {author} {\bibfnamefont {Antonio~D.}\
  \bibnamefont {C{\'o}rcoles-Gonzales}}, \bibinfo {author} {\bibfnamefont
  {Abigail~J.}\ \bibnamefont {Cross}}, \bibinfo {author} {\bibfnamefont
  {Andrew}\ \bibnamefont {Cross}}, \bibinfo {author} {\bibfnamefont {Juan}\
  \bibnamefont {Cruz-Benito}}, \bibinfo {author} {\bibfnamefont {Chris}\
  \bibnamefont {Culver}}, \bibinfo {author} {\bibfnamefont {Salvador De
  La~Puente}\ \bibnamefont {Gonz{\'a}lez}}, \bibinfo {author} {\bibfnamefont
  {Enrique De~La}\ \bibnamefont {Torre}}, \bibinfo {author} {\bibfnamefont
  {Delton}\ \bibnamefont {Ding}}, \bibinfo {author} {\bibfnamefont {Eugene}\
  \bibnamefont {Dumitrescu}}, \bibinfo {author} {\bibfnamefont {Ivan}\
  \bibnamefont {Duran}}, \bibinfo {author} {\bibfnamefont {Pieter}\
  \bibnamefont {Eendebak}}, \bibinfo {author} {\bibfnamefont {Mark}\
  \bibnamefont {Everitt}}, \bibinfo {author} {\bibfnamefont {Ismael~Faro}\
  \bibnamefont {Sertage}}, \bibinfo {author} {\bibfnamefont {Albert}\
  \bibnamefont {Frisch}}, \bibinfo {author} {\bibfnamefont {Andreas}\
  \bibnamefont {Fuhrer}}, \bibinfo {author} {\bibfnamefont {Jay}\ \bibnamefont
  {Gambetta}}, \bibinfo {author} {\bibfnamefont {Borja~Godoy}\ \bibnamefont
  {Gago}}, \bibinfo {author} {\bibfnamefont {Juan}\ \bibnamefont
  {Gomez-Mosquera}}, \bibinfo {author} {\bibfnamefont {Donny}\ \bibnamefont
  {Greenberg}}, \bibinfo {author} {\bibfnamefont {Ikko}\ \bibnamefont
  {Hamamura}}, \bibinfo {author} {\bibfnamefont {Vojtech}\ \bibnamefont
  {Havlicek}}, \bibinfo {author} {\bibfnamefont {Joe}\ \bibnamefont
  {Hellmers}}, \bibinfo {author} {\bibfnamefont {{\L}ukasz}\ \bibnamefont
  {Herok}}, \bibinfo {author} {\bibfnamefont {Hiroshi}\ \bibnamefont {Horii}},
  \bibinfo {author} {\bibfnamefont {Shaohan}\ \bibnamefont {Hu}}, \bibinfo
  {author} {\bibfnamefont {Takashi}\ \bibnamefont {Imamichi}}, \bibinfo
  {author} {\bibfnamefont {Toshinari}\ \bibnamefont {Itoko}}, \bibinfo {author}
  {\bibfnamefont {Ali}\ \bibnamefont {Javadi-Abhari}}, \bibinfo {author}
  {\bibfnamefont {Naoki}\ \bibnamefont {Kanazawa}}, \bibinfo {author}
  {\bibfnamefont {Anton}\ \bibnamefont {Karazeev}}, \bibinfo {author}
  {\bibfnamefont {Kevin}\ \bibnamefont {Krsulich}}, \bibinfo {author}
  {\bibfnamefont {Peng}\ \bibnamefont {Liu}}, \bibinfo {author} {\bibfnamefont
  {Yang}\ \bibnamefont {Luh}}, \bibinfo {author} {\bibfnamefont {Yunho}\
  \bibnamefont {Maeng}}, \bibinfo {author} {\bibfnamefont {Manoel}\
  \bibnamefont {Marques}}, \bibinfo {author} {\bibfnamefont {Francisco~Jose}\
  \bibnamefont {Mart{\'\i}n-Fern{\'a}ndez}}, \bibinfo {author} {\bibfnamefont
  {Douglas~T.}\ \bibnamefont {McClure}}, \bibinfo {author} {\bibfnamefont
  {David}\ \bibnamefont {McKay}}, \bibinfo {author} {\bibfnamefont {Srujan}\
  \bibnamefont {Meesala}}, \bibinfo {author} {\bibfnamefont {Antonio}\
  \bibnamefont {Mezzacapo}}, \bibinfo {author} {\bibfnamefont {Nikolaj}\
  \bibnamefont {Moll}}, \bibinfo {author} {\bibfnamefont {Diego~Moreda}\
  \bibnamefont {Rodr{\'\i}guez}}, \bibinfo {author} {\bibfnamefont {Giacomo}\
  \bibnamefont {Nannicini}}, \bibinfo {author} {\bibfnamefont {Paul}\
  \bibnamefont {Nation}}, \bibinfo {author} {\bibfnamefont {Pauline}\
  \bibnamefont {Ollitrault}}, \bibinfo {author} {\bibfnamefont {Lee~James}\
  \bibnamefont {O'Riordan}}, \bibinfo {author} {\bibfnamefont {Hanhee}\
  \bibnamefont {Paik}}, \bibinfo {author} {\bibfnamefont {Jes{\'u}s}\
  \bibnamefont {P{\'e}rez}}, \bibinfo {author} {\bibfnamefont {Anna}\
  \bibnamefont {Phan}}, \bibinfo {author} {\bibfnamefont {Marco}\ \bibnamefont
  {Pistoia}}, \bibinfo {author} {\bibfnamefont {Viktor}\ \bibnamefont
  {Prutyanov}}, \bibinfo {author} {\bibfnamefont {Max}\ \bibnamefont {Reuter}},
  \bibinfo {author} {\bibfnamefont {Julia}\ \bibnamefont {Rice}}, \bibinfo
  {author} {\bibfnamefont {Abd{\'o}n~Rodr{\'\i}guez}\ \bibnamefont {Davila}},
  \bibinfo {author} {\bibfnamefont {Raymond Harry~Putra}\ \bibnamefont {Rudy}},
  \bibinfo {author} {\bibfnamefont {Mingi}\ \bibnamefont {Ryu}}, \bibinfo
  {author} {\bibfnamefont {Ninad}\ \bibnamefont {Sathaye}}, \bibinfo {author}
  {\bibfnamefont {Chris}\ \bibnamefont {Schnabel}}, \bibinfo {author}
  {\bibfnamefont {Eddie}\ \bibnamefont {Schoute}}, \bibinfo {author}
  {\bibfnamefont {Kanav}\ \bibnamefont {Setia}}, \bibinfo {author}
  {\bibfnamefont {Yunong}\ \bibnamefont {Shi}}, \bibinfo {author}
  {\bibfnamefont {Adenilton}\ \bibnamefont {Silva}}, \bibinfo {author}
  {\bibfnamefont {Yukio}\ \bibnamefont {Siraichi}}, \bibinfo {author}
  {\bibfnamefont {Seyon}\ \bibnamefont {Sivarajah}}, \bibinfo {author}
  {\bibfnamefont {John~A.}\ \bibnamefont {Smolin}}, \bibinfo {author}
  {\bibfnamefont {Mathias}\ \bibnamefont {Soeken}}, \bibinfo {author}
  {\bibfnamefont {Hitomi}\ \bibnamefont {Takahashi}}, \bibinfo {author}
  {\bibfnamefont {Ivano}\ \bibnamefont {Tavernelli}}, \bibinfo {author}
  {\bibfnamefont {Charles}\ \bibnamefont {Taylor}}, \bibinfo {author}
  {\bibfnamefont {Pete}\ \bibnamefont {Taylour}}, \bibinfo {author}
  {\bibfnamefont {Kenso}\ \bibnamefont {Trabing}}, \bibinfo {author}
  {\bibfnamefont {Matthew}\ \bibnamefont {Treinish}}, \bibinfo {author}
  {\bibfnamefont {Wes}\ \bibnamefont {Turner}}, \bibinfo {author}
  {\bibfnamefont {Desiree}\ \bibnamefont {Vogt-Lee}}, \bibinfo {author}
  {\bibfnamefont {Christophe}\ \bibnamefont {Vuillot}}, \bibinfo {author}
  {\bibfnamefont {Jonathan~A.}\ \bibnamefont {Wildstrom}}, \bibinfo {author}
  {\bibfnamefont {Jessica}\ \bibnamefont {Wilson}}, \bibinfo {author}
  {\bibfnamefont {Erick}\ \bibnamefont {Winston}}, \bibinfo {author}
  {\bibfnamefont {Christopher}\ \bibnamefont {Wood}}, \bibinfo {author}
  {\bibfnamefont {Stephen}\ \bibnamefont {Wood}}, \bibinfo {author}
  {\bibfnamefont {Stefan}\ \bibnamefont {W{\"o}rner}}, \bibinfo {author}
  {\bibfnamefont {Ismail~Yunus}\ \bibnamefont {Akhalwaya}}, \ and\ \bibinfo
  {author} {\bibfnamefont {Christa}\ \bibnamefont {Zoufal}},\ }\href {\doibase
  10.5281/zenodo.2562110} {\enquote {\bibinfo {title} {Qiskit: An open-source
  framework for quantum computing},}\ } (\bibinfo {year} {2019})\BibitemShut
  {NoStop}%
\bibitem [{\citenamefont {Vedral}\ \emph {et~al.}(1996)\citenamefont {Vedral},
  \citenamefont {Barenco},\ and\ \citenamefont {Ekert}}]{Vedral1995}%
  \BibitemOpen
  \bibfield  {author} {\bibinfo {author} {\bibfnamefont {Vlatko}\ \bibnamefont
  {Vedral}}, \bibinfo {author} {\bibfnamefont {Adriano}\ \bibnamefont
  {Barenco}}, \ and\ \bibinfo {author} {\bibfnamefont {Artur}\ \bibnamefont
  {Ekert}},\ }\bibfield  {title} {\enquote {\bibinfo {title} {Quantum networks
  for elementary arithmetic operations},}\ }\href {\doibase
  10.1103/PhysRevA.54.147} {\bibfield  {journal} {\bibinfo  {journal} {Phys.
  Rev. A}\ }\textbf {\bibinfo {volume} {54}},\ \bibinfo {pages} {147--153}
  (\bibinfo {year} {1996})}\BibitemShut {NoStop}%
\bibitem [{\citenamefont {Draper}(2000)}]{Draper2000}%
  \BibitemOpen
  \bibfield  {author} {\bibinfo {author} {\bibfnamefont {Thomas~G}\
  \bibnamefont {Draper}},\ }\bibfield  {title} {\enquote {\bibinfo {title}
  {{Addition on a Quantum Computer}},}\ }\href@noop {} {\  (\bibinfo {year}
  {2000})},\ \Eprint {http://arxiv.org/abs/quant-ph/0008033}
  {arXiv:quant-ph/0008033} \BibitemShut {NoStop}%
\bibitem [{\citenamefont {Draper}\ \emph {et~al.}(2006)\citenamefont {Draper},
  \citenamefont {Kutin}, \citenamefont {Rains},\ and\ \citenamefont
  {Svore}}]{Draper2004}%
  \BibitemOpen
  \bibfield  {author} {\bibinfo {author} {\bibfnamefont {Thomas~G}\
  \bibnamefont {Draper}}, \bibinfo {author} {\bibfnamefont {Samuel~A}\
  \bibnamefont {Kutin}}, \bibinfo {author} {\bibfnamefont {Eric~M}\
  \bibnamefont {Rains}}, \ and\ \bibinfo {author} {\bibfnamefont {Krysta~M}\
  \bibnamefont {Svore}},\ }\bibfield  {title} {\enquote {\bibinfo {title} {{A
  logarithmic-depth quantum carry-lookahead adder}},}\ }\href@noop {}
  {\bibfield  {journal} {\bibinfo  {journal} {Quantum Information and
  Computation}\ }\textbf {\bibinfo {volume} {6}},\ \bibinfo {pages} {351--369}
  (\bibinfo {year} {2006})},\ \Eprint {http://arxiv.org/abs/quant-ph/0406142}
  {arXiv:quant-ph/0406142} \BibitemShut {NoStop}%
\bibitem [{\citenamefont {Tarmast}(2001)}]{Tarmast2001}%
  \BibitemOpen
  \bibfield  {author} {\bibinfo {author} {\bibfnamefont {Ghasem}\ \bibnamefont
  {Tarmast}},\ }\bibfield  {title} {\enquote {\bibinfo {title} {{Multivariate
  Log-Normal Distribution}},}\ }\href@noop {} {\bibfield  {journal} {\bibinfo
  {journal} {International Statistical Institute Proceedings: 53rd Session,
  Seoul}\ } (\bibinfo {year} {2001})}\BibitemShut {NoStop}%
\bibitem [{\citenamefont {Gobet}(2009)}]{Gobet2009}%
  \BibitemOpen
  \bibfield  {author} {\bibinfo {author} {\bibfnamefont {Emmanuel}\
  \bibnamefont {Gobet}},\ }\bibfield  {title} {\enquote {\bibinfo {title}
  {Advanced monte carlo methods for barrier and related exotic options},}\
  }\href {\doibase 10.1016/S1570-8659(08)00012-4} {\ \bibinfo {series}
  {Handbook of Numerical Analysis},\ \textbf {\bibinfo {volume} {15}},\
  \bibinfo {pages} {497 -- 528} (\bibinfo {year} {2009})}\BibitemShut {NoStop}%
\bibitem [{\citenamefont {Shevchenko}\ and\ \citenamefont
  {Moral}(2014)}]{Shevchenko2014}%
  \BibitemOpen
  \bibfield  {author} {\bibinfo {author} {\bibfnamefont {Pavel~V.}\
  \bibnamefont {Shevchenko}}\ and\ \bibinfo {author} {\bibfnamefont
  {Pierre~Del}\ \bibnamefont {Moral}},\ }\bibfield  {title} {\enquote {\bibinfo
  {title} {Valuation of barrier options using sequential monte carlo},}\ }\href
  {\doibase 10.2139/ssrn.2529539} {\bibfield  {journal} {\bibinfo  {journal}
  {Journal of Computational Finance}\ }\textbf {\bibinfo {volume} {20}},\
  \bibinfo {pages} {107--135} (\bibinfo {year} {2014})}\BibitemShut {NoStop}%
\bibitem [{\citenamefont {\v{Z}nidari\v{c}}\ \emph {et~al.}(2008)\citenamefont
  {\v{Z}nidari\v{c}}, \citenamefont {Giraud},\ and\ \citenamefont
  {Georgeot}}]{Znidaric2008}%
  \BibitemOpen
  \bibfield  {author} {\bibinfo {author} {\bibfnamefont {M.}~\bibnamefont
  {\v{Z}nidari\v{c}}}, \bibinfo {author} {\bibfnamefont {O.}~\bibnamefont
  {Giraud}}, \ and\ \bibinfo {author} {\bibfnamefont {B.}~\bibnamefont
  {Georgeot}},\ }\bibfield  {title} {\enquote {\bibinfo {title} {Optimal number
  of controlled-not gates to generate a three-qubit state},}\ }\href {\doibase
  10.1103/PhysRevA.77.032320} {\bibfield  {journal} {\bibinfo  {journal}
  {Physical Review A}\ }\textbf {\bibinfo {volume} {77}},\ \bibinfo {pages}
  {032320} (\bibinfo {year} {2008})}\BibitemShut {NoStop}%
\bibitem [{\citenamefont {Shende}\ \emph {et~al.}(2006)\citenamefont {Shende},
  \citenamefont {Bullock},\ and\ \citenamefont {Markov}}]{shende2006}%
  \BibitemOpen
  \bibfield  {author} {\bibinfo {author} {\bibfnamefont {Vivek~V}\ \bibnamefont
  {Shende}}, \bibinfo {author} {\bibfnamefont {Stephen~S}\ \bibnamefont
  {Bullock}}, \ and\ \bibinfo {author} {\bibfnamefont {Igor~L}\ \bibnamefont
  {Markov}},\ }\bibfield  {title} {\enquote {\bibinfo {title} {Synthesis of
  quantum-logic circuits},}\ }\href {\doibase 10.1109/TCAD.2005.855930}
  {\bibfield  {journal} {\bibinfo  {journal} {IEEE Transactions on
  Computer-Aided Design of Integrated Circuits and Systems}\ }\textbf {\bibinfo
  {volume} {25}},\ \bibinfo {pages} {1000--1010} (\bibinfo {year}
  {2006})}\BibitemShut {NoStop}%
\bibitem [{\citenamefont {Iten}\ \emph {et~al.}(2019)\citenamefont {Iten},
  \citenamefont {Reardon-Smith}, \citenamefont {Mondada}, \citenamefont
  {Redmond}, \citenamefont {Kohli},\ and\ \citenamefont {Colbeck}}]{iten2019}%
  \BibitemOpen
  \bibfield  {author} {\bibinfo {author} {\bibfnamefont {Raban}\ \bibnamefont
  {Iten}}, \bibinfo {author} {\bibfnamefont {Oliver}\ \bibnamefont
  {Reardon-Smith}}, \bibinfo {author} {\bibfnamefont {Luca}\ \bibnamefont
  {Mondada}}, \bibinfo {author} {\bibfnamefont {Ethan}\ \bibnamefont
  {Redmond}}, \bibinfo {author} {\bibfnamefont {Ravjot~Singh}\ \bibnamefont
  {Kohli}}, \ and\ \bibinfo {author} {\bibfnamefont {Roger}\ \bibnamefont
  {Colbeck}},\ }\bibfield  {title} {\enquote {\bibinfo {title} {Introduction to
  {UniversalQCompiler}},}\ }\href@noop {} {\  (\bibinfo {year} {2019})},\
  \Eprint {http://arxiv.org/abs/1904.01072} {arXiv:1904.01072} \BibitemShut
  {NoStop}%
\bibitem [{\citenamefont {Dewes}\ \emph {et~al.}(2012)\citenamefont {Dewes},
  \citenamefont {Ong}, \citenamefont {Schmitt}, \citenamefont {Lauro},
  \citenamefont {Boulant}, \citenamefont {Bertet}, \citenamefont {Vion},\ and\
  \citenamefont {Esteve}}]{Dewes2012}%
  \BibitemOpen
  \bibfield  {author} {\bibinfo {author} {\bibfnamefont {A.}~\bibnamefont
  {Dewes}}, \bibinfo {author} {\bibfnamefont {F.~R.}\ \bibnamefont {Ong}},
  \bibinfo {author} {\bibfnamefont {V.}~\bibnamefont {Schmitt}}, \bibinfo
  {author} {\bibfnamefont {R.}~\bibnamefont {Lauro}}, \bibinfo {author}
  {\bibfnamefont {N.}~\bibnamefont {Boulant}}, \bibinfo {author} {\bibfnamefont
  {P.}~\bibnamefont {Bertet}}, \bibinfo {author} {\bibfnamefont
  {D.}~\bibnamefont {Vion}}, \ and\ \bibinfo {author} {\bibfnamefont
  {D.}~\bibnamefont {Esteve}},\ }\bibfield  {title} {\enquote {\bibinfo {title}
  {Characterization of a two-transmon processor with individual single-shot
  qubit readout},}\ }\href {\doibase 10.1103/PhysRevLett.108.057002} {\bibfield
   {journal} {\bibinfo  {journal} {Phys. Rev. Lett.}\ }\textbf {\bibinfo
  {volume} {108}},\ \bibinfo {pages} {057002} (\bibinfo {year}
  {2012})}\BibitemShut {NoStop}%
\bibitem [{\citenamefont {Temme}\ \emph {et~al.}(2017)\citenamefont {Temme},
  \citenamefont {Bravyi},\ and\ \citenamefont {Gambetta}}]{Temme2017}%
  \BibitemOpen
  \bibfield  {author} {\bibinfo {author} {\bibfnamefont {Kristan}\ \bibnamefont
  {Temme}}, \bibinfo {author} {\bibfnamefont {Sergey}\ \bibnamefont {Bravyi}},
  \ and\ \bibinfo {author} {\bibfnamefont {Jay~M.}\ \bibnamefont {Gambetta}},\
  }\bibfield  {title} {\enquote {\bibinfo {title} {Error mitigation for
  short-depth quantum circuits},}\ }\href {\doibase
  10.1103/PhysRevLett.119.180509} {\bibfield  {journal} {\bibinfo  {journal}
  {Phys. Rev. Lett.}\ }\textbf {\bibinfo {volume} {119}},\ \bibinfo {pages}
  {180509} (\bibinfo {year} {2017})}\BibitemShut {NoStop}%
\bibitem [{\citenamefont {Dumitrescu}\ \emph {et~al.}(2018)\citenamefont
  {Dumitrescu}, \citenamefont {McCaskey}, \citenamefont {Hagen}, \citenamefont
  {Jansen}, \citenamefont {Morris}, \citenamefont {Papenbrock}, \citenamefont
  {Pooser}, \citenamefont {Dean},\ and\ \citenamefont
  {Lougovski}}]{Dumitrescu2018}%
  \BibitemOpen
  \bibfield  {author} {\bibinfo {author} {\bibfnamefont {E.~F.}\ \bibnamefont
  {Dumitrescu}}, \bibinfo {author} {\bibfnamefont {A.~J.}\ \bibnamefont
  {McCaskey}}, \bibinfo {author} {\bibfnamefont {G.}~\bibnamefont {Hagen}},
  \bibinfo {author} {\bibfnamefont {G.~R.}\ \bibnamefont {Jansen}}, \bibinfo
  {author} {\bibfnamefont {T.~D.}\ \bibnamefont {Morris}}, \bibinfo {author}
  {\bibfnamefont {T.}~\bibnamefont {Papenbrock}}, \bibinfo {author}
  {\bibfnamefont {R.~C.}\ \bibnamefont {Pooser}}, \bibinfo {author}
  {\bibfnamefont {D.~J.}\ \bibnamefont {Dean}}, \ and\ \bibinfo {author}
  {\bibfnamefont {P.}~\bibnamefont {Lougovski}},\ }\bibfield  {title} {\enquote
  {\bibinfo {title} {Cloud quantum computing of an atomic nucleus},}\ }\href
  {\doibase 10.1103/PhysRevLett.120.210501} {\bibfield  {journal} {\bibinfo
  {journal} {Phys. Rev. Lett.}\ }\textbf {\bibinfo {volume} {120}},\ \bibinfo
  {pages} {210501} (\bibinfo {year} {2018})}\BibitemShut {NoStop}%
\bibitem [{\citenamefont {Broadie}\ and\ \citenamefont
  {Glasserman}(1996)}]{Broadie1996}%
  \BibitemOpen
  \bibfield  {author} {\bibinfo {author} {\bibfnamefont {Mark}\ \bibnamefont
  {Broadie}}\ and\ \bibinfo {author} {\bibfnamefont {Paul}\ \bibnamefont
  {Glasserman}},\ }\bibfield  {title} {\enquote {\bibinfo {title} {Estimating
  security price derivatives using simulation},}\ }\href {\doibase
  10.1287/mnsc.42.2.269} {\bibfield  {journal} {\bibinfo  {journal} {Management
  Science}\ }\textbf {\bibinfo {volume} {42}} (\bibinfo {year} {1996}),\
  10.1287/mnsc.42.2.269}\BibitemShut {NoStop}%
\bibitem [{\citenamefont {Fowler}\ \emph {et~al.}(2012)\citenamefont {Fowler},
  \citenamefont {Mariantoni}, \citenamefont {Martinis},\ and\ \citenamefont
  {Cleland}}]{Fowler2012}%
  \BibitemOpen
  \bibfield  {author} {\bibinfo {author} {\bibfnamefont {Austin~G.}\
  \bibnamefont {Fowler}}, \bibinfo {author} {\bibfnamefont {Matteo}\
  \bibnamefont {Mariantoni}}, \bibinfo {author} {\bibfnamefont {John~M.}\
  \bibnamefont {Martinis}}, \ and\ \bibinfo {author} {\bibfnamefont
  {Andrew~N.}\ \bibnamefont {Cleland}},\ }\bibfield  {title} {\enquote
  {\bibinfo {title} {Surface codes: Towards practical large-scale quantum
  computation},}\ }\href {\doibase 10.1103/PhysRevA.86.032324} {\bibfield
  {journal} {\bibinfo  {journal} {Phys. Rev. A}\ }\textbf {\bibinfo {volume}
  {86}},\ \bibinfo {pages} {032324} (\bibinfo {year} {2012})}\BibitemShut
  {NoStop}%
\bibitem [{\citenamefont {Dob\ifmmode \check{s}\else
  \v{s}\fi{}\'{\i}\ifmmode~\check{c}\else \v{c}\fi{}ek}\ \emph
  {et~al.}(2007)\citenamefont {Dob\ifmmode \check{s}\else
  \v{s}\fi{}\'{\i}\ifmmode~\check{c}\else \v{c}\fi{}ek}, \citenamefont
  {Johansson}, \citenamefont {Shumeiko},\ and\ \citenamefont
  {Wendin}}]{Dobsicek2006}%
  \BibitemOpen
  \bibfield  {author} {\bibinfo {author} {\bibfnamefont {Miroslav}\
  \bibnamefont {Dob\ifmmode \check{s}\else
  \v{s}\fi{}\'{\i}\ifmmode~\check{c}\else \v{c}\fi{}ek}}, \bibinfo {author}
  {\bibfnamefont {G\"oran}\ \bibnamefont {Johansson}}, \bibinfo {author}
  {\bibfnamefont {Vitaly}\ \bibnamefont {Shumeiko}}, \ and\ \bibinfo {author}
  {\bibfnamefont {G\"oran}\ \bibnamefont {Wendin}},\ }\bibfield  {title}
  {\enquote {\bibinfo {title} {Arbitrary accuracy iterative quantum phase
  estimation algorithm using a single ancillary qubit: A two-qubit
  benchmark},}\ }\href {\doibase 10.1103/PhysRevA.76.030306} {\bibfield
  {journal} {\bibinfo  {journal} {Phys. Rev. A}\ }\textbf {\bibinfo {volume}
  {76}},\ \bibinfo {pages} {030306} (\bibinfo {year} {2007})}\BibitemShut
  {NoStop}%
\bibitem [{\citenamefont {O'Loan}(2010)}]{OLoan2010}%
  \BibitemOpen
  \bibfield  {author} {\bibinfo {author} {\bibfnamefont {C.~J.}\ \bibnamefont
  {O'Loan}},\ }\bibfield  {title} {\enquote {\bibinfo {title} {{Iterative phase
  estimation}},}\ }\href {\doibase 10.1088/1751-8113/43/1/015301} {\bibfield
  {journal} {\bibinfo  {journal} {Journal of Physics A: Mathematical and
  Theoretical}\ }\textbf {\bibinfo {volume} {43}} (\bibinfo {year} {2010}),\
  10.1088/1751-8113/43/1/015301}\BibitemShut {NoStop}%
\bibitem [{\citenamefont {Svore}\ \emph {et~al.}(2014)\citenamefont {Svore},
  \citenamefont {Hastings},\ and\ \citenamefont {Freedman}}]{Svore2013}%
  \BibitemOpen
  \bibfield  {author} {\bibinfo {author} {\bibfnamefont {Krysta~M}\
  \bibnamefont {Svore}}, \bibinfo {author} {\bibfnamefont {Matthew~B}\
  \bibnamefont {Hastings}}, \ and\ \bibinfo {author} {\bibfnamefont {Michael}\
  \bibnamefont {Freedman}},\ }\bibfield  {title} {\enquote {\bibinfo {title}
  {{Faster phase estimation}},}\ }\href@noop {} {\bibfield  {journal} {\bibinfo
   {journal} {Quantum Information {\&} Computation}\ }\textbf {\bibinfo
  {volume} {14}},\ \bibinfo {pages} {306--328} (\bibinfo {year} {2014})},\
  \Eprint {http://arxiv.org/abs/1304.0741} {arXiv:1304.0741} \BibitemShut
  {NoStop}%
\bibitem [{\citenamefont {Bullock}\ and\ \citenamefont
  {Markov}(2004)}]{Bullock2004}%
  \BibitemOpen
  \bibfield  {author} {\bibinfo {author} {\bibfnamefont {S.S.}\ \bibnamefont
  {Bullock}}\ and\ \bibinfo {author} {\bibfnamefont {I.L.}\ \bibnamefont
  {Markov}},\ }\bibfield  {title} {\enquote {\bibinfo {title} {Smaller circuits
  for arbitrary n-qubit diagonal computations},}\ }\href@noop {} {\bibfield
  {journal} {\bibinfo  {journal} {Quantum Information \& Computation}\ }\textbf
  {\bibinfo {volume} {4}},\ \bibinfo {pages} {027--047} (\bibinfo {year}
  {2004})},\ \Eprint {http://arxiv.org/abs/quant-ph/0303039}
  {arXiv:quant-ph/0303039} \BibitemShut {NoStop}%
\end{thebibliography}%

\end{document}